\documentclass[useAMS,usenatbib]{mnras}
\usepackage{graphicx}
\usepackage{float}
\usepackage{mathtools}
\usepackage{color}
\usepackage{multicol}
\usepackage{tabularx}
\usepackage{sidecap}
\usepackage[T1]{fontenc}
\usepackage{hyperref}
\usepackage{textcomp}
\usepackage{pdflscape}

\usepackage{longtable}

\def\Ha{H$\alpha$}

\def\CoII{Co\,{\sc ii}}
 \def\NiII{Ni\,{\sc ii}}

\def\OI{O\,{\sc i}}

\def\TiII{Ti\,{\sc ii}}

\def\FeII{Fe\,{\sc ii}}

\def\to{$\rightarrow$}
\def\msun{M$_{\odot}$}
\def\rsun{R$_{\odot}$}
\def\lesssim{\mathrel{\hbox{\rlap{\hbox{\lower4pt\hbox{$\sim$}}}\hbox{$<$}}}}
\def\gtrsim{\mathrel{\hbox{\rlap{\hbox{\lower4pt\hbox{$\sim$}}}\hbox{$>$}}}}

\title[SN~II with LCOGT]
{The Diversity of Type II Supernova vs. The Similarity in Their Progenitors}
\author[Valenti et al.]{S. Valenti$^{1}$
\thanks{e--mail: stfn.valenti@gmail.com},  D.~A. Howell$^{2,3}$, M. D. Stritzinger$^{4}$, M. L. Graham$^{5}$,  G. Hosseinzadeh$^{2,3}$, 
\and 
I. Arcavi$^{2,6}$,  L. Bildsten$^{6}$,   A. Jerkstrand$^{7}$,  C. McCully$^{2,3}$,  A. Pastorello$^{8}$, A. L. Piro$^{9}$,
\and
 D. Sand$^{10}$,  S.~J. Smartt$^{7}$,    G. Terreran$^{7,8}$,  C. Baltay$^{11}$, S. Benetti$^{8}$,   P. Brown$^{12}$,
\and
  A. V. Filippenko$^{5}$, M. Fraser$^{13}$,    D. Rabinowitz$^{11}$, M. Sullivan$^{14}$, F. Yuan$^{15,16}$
 \\
$^{1 ~}$ Department of Physics, University of California, Davis, CA 95616, USA \\
$^{2 ~}$ Las Cumbres Observatory Global Telescope Network, 6740 Cortona Dr., Suite 102, Goleta, CA 93117, USA \\
$^{3 ~}$ Department of Physics, University of California, Santa Barbara, Broida Hall, Mail Code 9530, Santa Barbara, CA 93106-9530, USA \\
$^{4~}$ Department of Physics and Astronomy, Aarhus University, Ny Munkegade 120, DK-8000 Aarhus C, Denmark \\
$^{5 ~}$ Department of Astronomy, University of California, Berkeley, CA 94720-3411, USA \\
$^{6 ~}$ Kavli Institute for Theoretical Physics, Kohn Hall, University of California, Santa Barbara, CA 93106-4030, USA\\
$^{7 ~}$ Astrophysics Research Centre,  School of Mathematics and Physics, Queens University Belfast, Belfast BT7 1NN, UK \\
$^{8 ~}$ INAF Osservatorio Astronomico di Padova, Vicolo dell'Osservatorio 5, 35122 Padova, Italy \\
$^{9~}$ Carnegie Observatories, 813 Santa Barbara Street, Pasadena, CA 91101, USA \\
$^{10}$ Physics Department, Texas Tech University, Lubbock, TX , 79409, USA\\
$^{11}$ Department of Physics, Yale University, New Haven, CT 06520-8120, USA \\
$^{12}$ George P. and Cynthia Woods Mitchell Institute for Fundamental Physics \&{} Astronomy, \\
Texas A. \&{} M. University, Department of Physics and Astronomy, 4242 TAMU, College Station, TX 77843, USA \\
$^{13}$ Institute of Astronomy, University of Cambridge, Madingley Road, Cambridge, CB3 0HA, UK\\
$^{14}$ School of Physics and Astronomy, University of Southampton, Southampton, SO17 1BJ, UK \\
$^{15}$ Research School of Astronomy and Astrophysics, Australian National University, Canberra, ACT 2611, Australia \\
$^{16}$  ARC Centre of Excellence for All-Sky Astrophysics (CAASTRO) \\
}

\begin{document}

\date{Accepted .....; Received ....; in original form ....}


\maketitle

\begin{abstract}
High-quality collections  of Type II  supernova (SN) light  curves are
scarce  because they  evolve for  hundreds of  days, making  follow-up
observations  time   consuming  and  often  extending   over  multiple
observing seasons.  In  light of these difficulties,  the diversity of
SNe~II  is not  fully  understood.  Here  we  present ultraviolet  and
optical  photometry  of  12  SNe~II   monitored  by  the  Las  Cumbres
Observatory Global  Telescope Network  (LCOGT) during  2013--2014, and
compare  them with  previously studied  SNe having  well-sampled light
curves.   We explore  SN~II  diversity by  searching for  correlations
between the slope of the  linear light-curve decay after maximum light
(historically  used to  divide  SNe~II  into IIL  and  IIP) and  other
measured physical properties. While SNe~IIL are found to be on average
more luminous than  SNe~IIP, SNe~IIL do not appear  to synthesize more
$^{56}$Ni than SNe~IIP.  Finally, optical nebular spectra obtained for
several SNe  in our sample are  found to be consistent  with models of
red supergiant progenitors in  the 12--16~\msun{} range. Consequently,
SNe~IIL  appear  not  to  account  for  the  deficit  of  massive  red
supergiants as SN~II progenitors.
\end{abstract}


\section{Introduction}
\label{parintroduction}
The majority of core-collapse supernovae (SNe) are associated with the
deaths  of  hydrogen-rich  massive stars  with  zero-age-main-sequence
masses  $M_{\rm   ZAMS}$  \textgreater~8~\msun{}.   The   most  common
hydrogen-rich Type  II SNe exhibit a  $\sim 100$ day plateau  in their
light curves (SNe~IIP), and the  physics behind their emission is well
understood.   In particular,  the energy  source powering  the plateau
phase  is the  energy  deposited  by the  shock  wave  soon after  the
explosion.  The  recombination  process  of the  ionized  hydrogen  is
responsible for the photosphere moving  inward and allowing the stored
energy      to      be       radiated      during      this      phase
\citep{Popov1993,2009ApJ...703.2205K,Dessart2010,Bersten2013a}.    The
recombination wave travels through the  hydrogen-rich ejecta as the SN
expands homologously and cools \citep{Falk1973,Chevalier1976}.

However, a  fraction of SNe~II  have a  linear decline in  their light
curves     after     a     rapid     rise     to     maximum     light
\citep[SNe~IIL;][]{Barbon1979,  Filippenko1997a}.   After  either  the
plateau or linear decline phase in  SNe II there is a subsequent rapid
drop,  usually around  100-140 days  after explosion  for SNe  IIP and
80-100 days for  SNe IIL \citep{Valenti2015}.  This  difference may be
due to a smaller ejecta  mass in SNe IIL.  \cite{Blinnikov1993} argued
that the fast light-curve decline of SNe IIL is the result of both the
small amount  (1--2 \msun{}) of  hydrogen envelope in a  progenitor of
relatively larger  radius (a  few 1000 \rsun{}).  These values  can be
compared with  the progenitors of  SNe IIP, which are  red supergiants
(RSGs) that span radii of 10--1600 \rsun{} \citep{Levesque2005} at the
moment of  explosion and should have  an ejected mass of  $\sim$ 6--10
\msun{}.    Recent    hydrodynamical   and   radiative-transfer
  calculations have  also confirmed  that a  smaller ejected  mass may
  produce  a short  plateau, but  cannot reproduce  the luminosity  of
  SNe~IIL  \citep{Morozova2015,Moriya2015}.    A  larger   radius,  a
different hydrogen distribution,  or extra energy sources  may play an
important role in shaping the light curves of SNe~IIL.

RSGs have been  definitively identified as the  progenitors of SNe~IIP
(e.g.,       \citealt{Maund2005};       \citealt{2011MNRAS.417.1417F};
\citealt{VanDyk2012}),  with   masses  estimated  from   the  measured
luminosities of  the progenitor stars to  be in the range  8--17 \msun
\citep{2009ARA&A..47...63S}.    The   fact   that   more-massive   RSG
progenitors  have   not  been  identified  has   been  highlighted  as
significant    and    termed    the   ``red    supergiant    problem''
\citep{Smartt2009}.
  
The separation,  if any,  between SNe~IIP  and IIL  is an  open issue.
Some  authors  suggest  that   they  form  a  continuous  distribution
\citep{Patat1994,   Anderson2014,  Sanders2015,   Valenti2015},  while
others    favour    their    separation    into    distinct    classes
\citep{Arcavi2012,Faran2014,Faran2014a}.   If  one   of  the  keys  to
differentiating  SNe~IIP from  IIL is  the amount  of hydrogen  in the
envelope, a  clear dichotomy may  suggest the presence of  a specific,
unknown ingredient  in the evolution  of massive stars  that sometimes
strips them of discrete (large for SNe~IIL, small for SNe~IIP) amounts
of  hydrogen.    Despite  an increasing  number  of  SNe~II  with
  published light curves  and the growing evidence that  SNe~II form a
  continuum  of light-curve  properties,  it  remains unknown  whether
  their progenitors span a range of stellar populations.

In this  paper, we  present comprehensive optical  light curves  of 12
SNe~II that have been monitored  by the Las Cumbras Observatory Global
Telescope  (LCOGT,  \citealt{Brown2013a})  Network  between  2013  and
2014. The  LCOGT network of  nine 1~m  and two 2~m  robotic telescopes
enables  photometric coverage  with long  temporal baselines  and fast
response times to cover the most time-critical phases of SN evolution.
Many  of these  SNe~II have  also  been followed  at ultraviolet  (UV)
wavelengths    with   the    UVOT   camera    onboard   {\it    Swift}
\citep{Roming2005}.   When possible,  we  will compare  our data  with
those   previously  published   and  publicly   available  (see   next
section).  The  main  goal  of  this paper  is  to  focus  on  generic
properties  of our  sample to  gain an  expanded understanding  of the
SN~IIP/IIL diversity.

Various authors  use different values  of the linear decay  rate after
maximum brightness to  separate SNe IIP and SNe IIL.   {\it The reader
  should be  aware that in this  paper, we will not  define a specific
  decline-rate value  to distinguish  SNe~IIP and SNe~IIL;  rather, we
  will use  IIP to refer  to SNe~II  with flatter linear  decays after
  maximum light (IIP-like SNe) and IIL  to refer to SNe~II with faster
  decline rates  after maximum light (IIL-like  SNe).}  When possible,
we will  colour-code the sample of  objects using the decline  rate of
the $V$-band  light curve ($s_{V50}$; see  Section \ref{sec:slope} for
the definition)  to emphasize that we  are not dividing SNe~II  to IIP
and IIL {\it a priori}.

In Section~\ref{sec:sample}, we present  our sample of new SNe as
  well as SNe from the literature  that are used in the analysis.  In
Section~\ref{sec:data} we  present our  photometric data,  and Section
\ref{sec:slope}  describes  the  main parameters  characterizing  them
(listed  also  in  the  Appendix  \ref{apefigures}).   We  repeat  the
analysis  in   Section~\ref{sec:bolo}  using   pseudobolometric  light
curves.   In  Section~\ref{sec:early}   we  discuss  early-time  data,
focusing on the UV differences between  SNe~IIP and IIL.  We derive in
Section~\ref{sec:nickel} the  amount of  $^{56}$Ni produced  in SNe~II
and   compare  our   results   with  progenitor   studies  of   SNe~II
\citep{Smartt2009}.   In  Section~\ref{sec:nebular} we  constrain  the
mass of the progenitor for a few SNe in our sample using their nebular
spectra, and Section \ref{sec:discussion} summarizes our results.

\section{Supernova Sample}
\label{sec:sample}
\begin{table*}
\caption{LCOGT Type II Supernovae}
\scriptsize
\begin{tabular}{lcccccccc}
\hline
SN                &   Discovery    &    Ref-discovery$^{*}$   &  Classification$^{\dagger}$     & Last                  & Photometry     & N  points   &  $\alpha$(J2000)   &  $\delta$(J2000)   \\
                    &                     &                                       &       SN~II            & nondetection$^{\diamond}$   & coverage (days)    &                  &   (h:m:s)          &  ($^\circ$:$'$:$''$)      \\
\hline 
    2013fs            & 2013 Oct. 07.46   &  a                    & young,  aa     &   2013 Oct. 5.24, o     &  106  &  371  &   23:19:44.67 & $+$10:11:04.5    \\ 
    2013bu           & 2013 Apr. 21.76   &  b                   &  young,  bb    &  2013 Apr. 12.8,  p    &  130  &  163    &   22:37:05.60 & $+$34:24:31.9   \\ 
ASASSN-14ha      & 2014 Sep. 10.29   &  ASASSN, c     &  young,  cc     & 2014  Sep.  08.3, c     & 180   &  372   &     04:20:01.41 & $-$54:56:17.0  \\ 
ASASSN-14dq      & 2014 Jul. 8.48      &  ASASSN, d     &  SN~II  ,  dd    & 2014 Jun.  27.49, d  &  341 &  219  & 21:57:59.97 & $+$24:16:08.1    \\ 
ASASSN-14gm     & 2014 Sep. 2.47    &  ASASSN, e      &  young , ee     &  2014 Aug. 30.48, e   & 318   &  327  &  00:59:47.83 & $-$07:34:19.3     \\ 
    2014cy            & 2014 Aug. 31.0   &   f                    & SN~II   ,  ff      &  2013 Aug.  29.3, LOSS  &  160  &  162   & 23:44:16.03 & $+$10:46:12.5     \\ 
ASASSN-14kg      & 2014 Nov. 17.36 &  ASASSN, g      &  young , gg     &  2014 Nov. 11.38, g    & 100   &  99   &  01:44:38.38 & $+$35:48:20.5     \\ 
    2013ai            & 2013 Mar. 1.66    &  h                    &  young, hh      &    --  &  83  & 163    &  06:16:18.35  & $-$21:22:32.9     \\ 
    2014dw          & 2014 Nov. 6.59    &  i                     &  young,  ii       &    --    &  141  &  149   &  11:10:48.41  & $-$37:27:02.2     \\ 
  LSQ13dpa         & 2013 Dec. 18.28  &  LSQ, l             &  young,  ll       &  2013 Dec 14.2, l   &  189  &  179   &  11:01:12.91  & $-$05:50:52.4     \\ 
   LSQ14gv          & 2014 Jan. 17.3     &  LSQ,  m           &  young, mm   &  2014  Jan. 13.17, m    & 97   &  113   &  10:54:11.71  & $-$15:01:30.0     \\ 
  2015W       & 2015 Jan. 12.17   &  LOSS, n &  SN~II , nn      &  2014 Dec. 14, n    &  119  &  221   &  06:57:43.03 & $+$13:34:45.7     \\ 
\hline
\end{tabular}
\label{tabsummary0}
\justify \noindent $(*)$ a, \citet{Nakano2013}; b, \citet{Itagaki2013}; c, \citet{Kiyota2014}; d, \citet{Stanek2014};  e, \citet{Holoien2014}; f, \citet{Nishimura2014}; g, \citet{Nicolas2014}; h, \citet{Conseil2013a}; i, \citet{Parker2015}; l, \citet{Hsiao2013}; m, \citet{Ergon2014a}; n, \citet{Kim2015}.\\
\noindent $(\dagger)$  aa, \citet{Childress2013a}; bb, \citet{Ochner2013}; cc, \citet{Arcavi2014}; dd, \citet{Arcavi2014a}; ee, \citet{Elias-Rosa2014};
ff, \citet{Morell2014}; gg, \citet{Falco2014}; hh, \citet{Conseil2013}; ii, \citet{Arcavi2014b}; ll, \citet{Hsiao2013}; mm, \citet{Ergon2014a}; 
nn, \citet{Tomasella2015}.
\noindent $(\diamond)$  o, Yaron et al. (in preparation); p, \citet{Itagaki2013}.  
\end{table*}

\begin{table*}
\centering
\caption{Sample of Type II Supernovae}
\scriptsize
\begin{tabular}{ccccccccc}
\hline
Object$^{+}$   &   Distance    &   Ref.$^{*}$       &    Explosion    &
m Ref.$^{*}$ &  $E(B-V)$  & Ref.$^{*}$       &  $E(B-V)^{x}$   &   Host      \\
     &   modulus     &   ($\mu$)         &      epoch      & (explosion)  &  (host)  & ($E(B-V)_{\rm host})$   &   M.W.          &   galaxy      \\
\hline \\     SN 2013ai  &     32.20       0.15    &   a  &    2456348.00  ~~   5.0     &   w   &               0.15           0.08      &     w     &    0.08     &      NGC 2207   \\ 
    SN 2013bu  &     30.79       0.08    &   b  &    2456399.80  ~~   4.5     &   w   &               0.00           0.00      &     w     &    0.08     &      NGC 7331   \\ 
    SN 2013fs  &     33.45       0.15    &   a  &    2456571.12  ~~   0.5     &   ll   &               0.00           0.00      &     w     &    0.04     &      NGC 7610   \\ 
  lsq13dpa      &     35.08       0.15    &   a  &    2456642.70  ~~   2.0     &   w   &               0.00           0.00      &     w     &    0.04     &   LCSBS 1492O   \\ 
    SN 2014cy  &     31.87       0.15    &   a  &    2456900.00  ~~   1.0     &   w   &               0.00           0.00      &     w     &    0.05     &      NGC 7742   \\ 
    SN 2014dw  &     32.46       0.15    &   a  &    2456958.00  ~~   10.0     &   w   &               0.11           0.06      &     w     &    0.11     &      NGC 3568   \\ 
   lsq14gv        &     35.15       0.15    &   a  &    2456674.80  ~~   2.0     &   w   &               0.00           0.00      &     w     &    0.06      &   2MASX J10541092-1501228   \\ 
ASASSN-14dq  &     33.26       0.15    &   a  &    2456841.50  ~~   5.5     &   w   &               0.00           0.00      &     w     &    0.07     &     UGC 11860   \\ 
ASASSN-14gm  &     31.74       0.15    &   a  &    2456901.00  ~~   1.5     &   w   &               0.00           0.00      &     w     &    0.10     &      NGC 337   \\ 
ASASSN-14kg  &     33.83       0.15    &   a  &    2456970.00  ~~   3.0     &   w   &               0.25           0.12      &     w     &    0.04     &   CGCG 521-075   \\ 
ASASSN-14ha  &     29.53       0.50    &   a  &    2456910.50  ~~   1.5     &   w   &               0.00           0.00      &     w     &    0.01     &      NGC 1566   \\ 
    SN 2015W   &     33.74       0.15    &   a  &    2457025.00  ~~   10.0     &   w   &               0.00           0.00      &     w     &    0.15     &     UGC 3617   \\ 
     \hline
    SN 2013ab  &     31.90       0.08    &   d  &    2456340.00  ~~   1.0     &   d   &               0.02           0.07      &     d     &    0.02     &     1430+101   \\ 
    SN 2013by  &     30.81       0.15    &   a  &    2456404.00  ~~   2.0     &   x   &               0.00           0.00      &     x     &    0.23     &   ESO 138-G010   \\ 
    SN 2013ej  &     29.79       0.20    &   c  &    2456515.10  ~~   1.0     &   y   &               0.00           0.00      &     y     &    0.06     &   M74   \\ 
     SN 2014G  &     31.90       0.15    &   a  &    2456668.35  ~~   1.0     &   w   &               0.20           0.00      &     kk     &    0.01     &      NGC 3448   \\ 
            --       &   --                            &  --      &          --                       &                    --                            &      --       &  --                     &      --      &   --       \\  
\hline
\end{tabular}
\label{tabsummary}
\justify $^{+}$ Table continues in the online version with objects from literature. \\
$^{*}$ a: NED , b: \citealt{Kanbur2003}, c: \citealt{Fraser2013a}, d: \citealt{Bose2015}, e: \citealt{Brown2010}, f: \citealt{Ferrarese1996}, g: \citealt{1992ApJ...395..366S}, h: \citealt{Poznanski2009}, i: \citealt{Jones2009a}, j: \citealt{Leonard2003}, k: \citealt{Wang2006}, l: \citealt{2006MNRAS.372.1735T}, m: \citealt{Anderson2014}, n: \citealt{Takats2012}, o: \citealt{Bose2014}, p: \citealt{Freedman2001}, q: \citealt{Takats2013}, r: \citealt{Takats2015}, s: \citealt{Gall2015}, t: \citealt{OlivaresE.2010}, u: \citealt{Mould2008}, v: \citealt{Rest2014}, w: This paper, x: \citealt{Valenti2015}, z: \citealt{Benetti1994}, aa: \citealt{Faran2014a}, bb: \citealt{Inserra2013}, cc: \citealt{Taddia2013}, dd: \citealt{Spiro2014}, ee: \citealt{Faran2014}, ff: \citealt{Pozzo2006}, gg: \citealt{DeVaucouleurs1981}, hh: \citealt{Leonard2002b}, jj: \citealt{2004MNRAS.347...74P}, kk: Terreran in preparation, ll: 
Yaron in preparation, mm: \citealt{Gandhi2013}, nn: \citealt{2010MNRAS.404..981M}, oo: \citealt{2009MNRAS.394.2266P}, pp: \citealt{2011MNRAS.417..261I}, qq: \citealt{Tomasella2013}, rr: \citealt{Barbarino2015}, ss: \citealt{2012MNRAS.422.1122I}, tt: \citealt{2011MNRAS.417.1417F}, uu: \citealt{Gal-Yam2011b}, vv: \citealt{Elias-Rosa2011}, ww: \citealt{Dessart2008}, xx: \citealt{Dall'Ora2014}, zz: \citealt{Poznanski2015}, ab: \citealt{Elias-Rosa2010}, ac: \citealt{Zwitter2004}, ad: \citealt{Quimby2007}, ae: \citealt{Andrews2011}.   \\
\end{table*}

Between  September 2012 and  July 2013, LCOGT deployed  eight 1~m
  telescopes around the world.  Commissioning of the telescopes lasted
  until  March  2014  when  the LCOGT  network  of  telescopes  become
  operational \citep{Brown2013a}.   It was  at this time  we initiated
  our programme to obtain detailed follow-up observations of SNe~II.  To
  maximize the  potential of  the LCOGT network  to contribute  to our
  understanding of SNe~II,  we elected to monitor  all objects located
  within  40~Mpc and  having  constraints on  the  explosion epoch  of
  within one  week or less,  or a  blue featureless spectrum  which is
  typical  of  young SNe~II.   Objects  selected  for monitoring  were
  discovered by various transient search  surveys (see below), and our
  main  criterion for  selection  was the  potential  to collect  data
  extending  from soon  after explosion  to at  least the  end of  the
  plateau phase.

The  sample consists of 16  bright SNe~II discovered in  2013 and
  2014.  Data for  12 of the objects are presented  here for the first
  time,     while    photometry     of    4     objects    (SN~2013ej,
  \citealt{Valenti2014},   Yuan   et   al.,   in   preparation;   SN~2013by,
  \citealt{Valenti2015};   SN~2014G,  Terreran   et  al.,   in  preparation;
  SN~2013ab, \citealt{Bose2015}) are presented  in papers cited herein
  though their data are included in our analysis below.

Information   characterizing   our  SNe~II   is   reported   in
  Table~\ref{tabsummary0}, including details concerning the discovery,
  classification, duration of monitoring, and coordinates. Most of the
  objects were discovered  1--4 days after explosion  and observed for
  more  than  100  days,  covering  the  end  of  the  plateau  phase.
  Unfortunately, the  end of  the plateau phase  was not  included for
  SN~2013ai and ASASSN-14kg as they disappeared into the daytime sky.

Four  objects  within  the  sample  were  discovered  by  the  All-Sky
Automated      Survey       for      SuperNovae      (ASASSN\footnote{
  \protect\href{http://www.astronomy.ohio-state.edu/~assassin/index.shtml}
               {www.astronomy.ohio-state.edu/$\sim$assassin/index.shtml}
},  \citealt{Shappee2014}),  two by  the  La  Silla Quest  Variability
Survey\footnote{LSQ,   \protect\href{http://hep.yale.edu/lasillaquest}
  {http://hep.yale.edu/lasillaquest}}  \citep{Baltay2013}, and  one by
the   Lick   Observatory  Supernova   Search   \citep[LOSS;][]{Li2000,
  Filippenko2001} with the Katzman Automatic Imaging Telescope (KAIT).
In  addition, five  objects  were reported  on  the Transient  Objects
Confirmation  Page  (TOCP)  of  the Central  Bureau  for  Astronomical
Telegrams                                         (CBAT\footnote{CBAT,
  \protect\href{http://www.cbat.eps.harvard.edu/}
               {http://www.cbat.eps.harvard.edu/}}).  More details are
available in Appendix \ref{appsample}.

We will compare our sample of 16 SNe~II with data of other SNe~II from
the  literature.   We  consider  all  SNe~II  with  dense  photometric
follow-up  observations  that  are presented  by  \cite{Anderson2014},
\cite{Faran2014},    \cite{Faran2014a},   \cite{Poznanski2015},    and
\cite{Spiro2014}, as well as  with several recent well-observed SNe~II
\citep{Tomasella2013, Dall'Ora2014,  Takats2013, Takats2015, Gall2015,
  2009MNRAS.394.2266P, Inserra2013, Leonard2002}.
  
We stress  that the number  of objects  taken from the  literature and
used for comparison in the following sections changes depending on the
exact  data available  for  each object.  For  example, including  our
sample, 107  SNe~II have been  observed during the linear  decay phase
allowing for an estimate of the  linear decline rate over 50 days, but
only  48 of  these objects  have $V$-band  observations that  enable a
measurement of the drop at the  end of the linear decay.  Moreover, at
early phases  and in the  UV the number of  objects with good  data is
even smaller (see Section~\ref{sec:early}).

\section{Observations}
\label{sec:data}

The  LCOGT  images were  reduced  using  {\tt lcogtsnpipe},  a  custom
pipeline  developed  by  one  of  us  (S.V.).  The  pipeline  computes
magnitudes with the point-spread-function  (PSF) fitting technique and
a low-order  polynomial fit  removes the host-galaxy  background.  The
template-subtraction          technique           (using          {\tt
  HOTPANTS} \footnote{http://www.astro.washington.edu/users/becker/v2.0/
  \newline
  hotpants.html})  was  used  in  the case  of  SN~2013bu,  SN~2014cy,
SN~2014dw, LSQ14gv,  and LSQ13dpa, as  in those cases  the host-galaxy
contamination was particularly relevant.  The absolute calibration was
performed                                                        using
APASS             \footnote{\protect\href{https://www.aavso.org/apass}
  {https://www.aavso.org/apass}}, which  implies that $B$ and  $V$ are
calibrated to the  Landolt system (Vega), while $g$, $r$,  and $i$ are
calibrated to  the Sloan  system (AB  magnitudes).  Details  about the
pipeline  can  be  found  in  Appendix  \ref{sec:pipe}.   {\it  Swift}
aperture photometry was computed following the prescriptions described
by   \cite{Brown2009a},   but   using  the   updated   zeropoints   of
\cite{Breeveld2010}.   {\it Swift}  spectra  were  reduced using  {\tt
  uvotpy}\footnote{https://github.com/PaulKuin/uvotpy}.  LSQ  data for
LSQ14gv  and   LSQ13dpa  were  reduced  following   the  prescriptions
described by \cite{Firth2014}.

\begin{figure*}
\begin{center}
 \includegraphics[width=16.cm,height=20.cm]{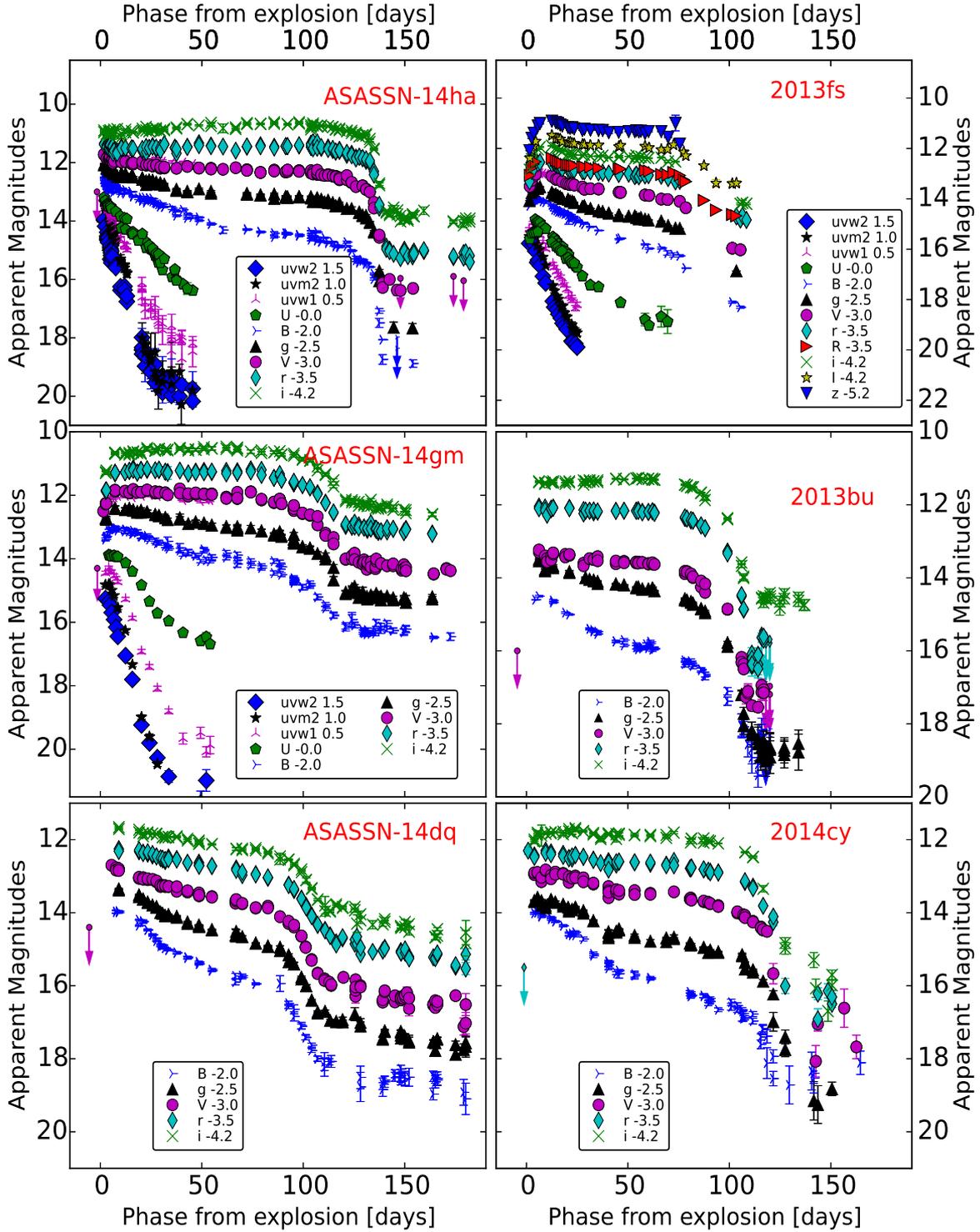}
 \caption{Multiband UV and optical light curves of SNe~II observed by LCOGT.}
  \label{fig:lc}
\end{center}
\end{figure*}

\begin{figure*}
\begin{center}
 \includegraphics[width=16.cm,height=20.cm]{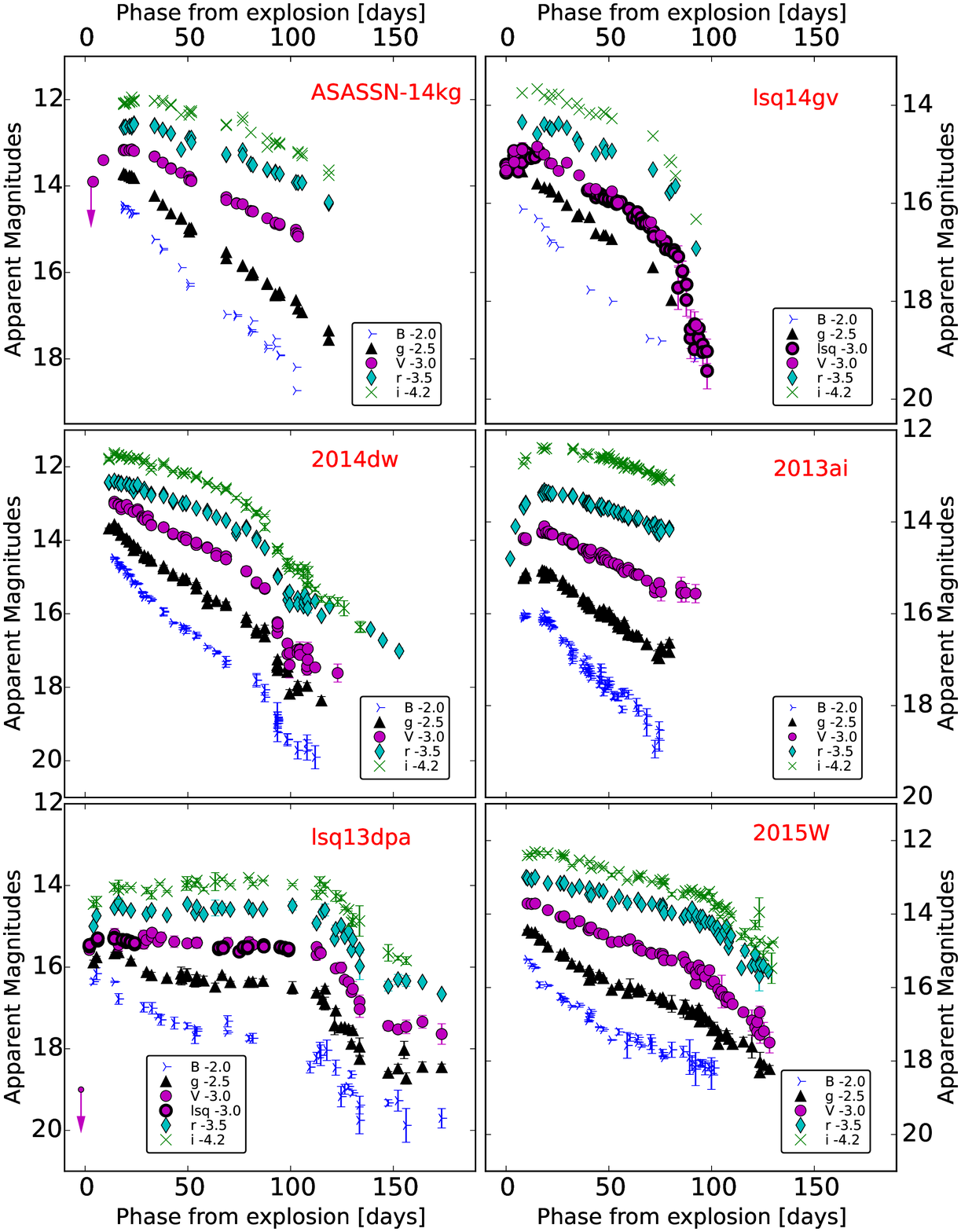}
 \caption{Multiband light curves of SNe~II observed by LCOGT. Data from La Silla Quest are marked 
 with a black border. The wide La Silla Quest filter ($r+g$) is close, but not identical, to Landolt $V$.}
  \label{fig:lc2}
\end{center}
\end{figure*}

We also  obtained six nebular spectra  of three of our  targets at the
Keck and  Gemini North  Observatories. SNe~II are  faint during
  their nebular phase; hence, it was not possible to obtain spectra of
  the entire sample.  These spectra together with nebular spectra from
  the literature  are used  in Section~\ref{sec:nebular}  to constrain
  the zero-age-main-sequence mass of the progenitors.

The  Keck   spectra  are  of   ASASSN-14dq  (2015  June   16.47,  2015
Oct.  10.34),  ASASSN-14gm  (2015  Oct.  10.39),  and  SN~2015W  (2015
Oct. 10.60) with the Low Resolution Imaging Spectrometer (LRIS; Oke et
al. 1995).  We  used the 1.0$\arcsec$ slit rotated  to the parallactic
angle   to   minimize   the    effects   of   atmospheric   dispersion
(\citealt{Filippenko1982};     in     addition,    LRIS     has     an
atmospheric-dispersion   corrector).   In   our  LRIS   configuration,
coverage  in  the  blue  with  the 600/4000  grism  extends  over  the
wavelength   range  3200--5600~\AA\   with   a   dispersion  of   0.63
\AA\ pixel$^{-1}$  and a full  width at half-maximum  intensity (FWHM)
resolution of  $\sim 4$ \AA. We  used the 5600 \AA\  dichroic, and our
coverage  in   the  red  with   the  400/8500  grating   extends  over
5600--10,200 \AA\  with a dispersion  of 1.16 \AA\ pixel$^{-1}$  and a
resolution of FWHM $\approx 7$ \AA.  At Gemini, we obtained a spectrum
of ASASSN-14gm (2015  July 27.62) and a spectrum  of ASASSN-14dq (2015
Oct.   09)  using GMOS.   The  Gemini  spectra  were observed  with  a
1.5$\arcsec$ slit rotated  to the parallactic angle  using grisms B600
and R400. Details  of the spectra and their reduction  can be found in
Appendix \ref{sec:specrredu}.

The final light curves of our  sample of SNe~II are plotted in Figures
\ref{fig:lc}  and  \ref{fig:lc2}.   The  corresponding  photometry  is
provided in the Appendix \ref{sec:specrredu}.

Table~\ref{tabsummary} gives more information  on our sample including
distance  modulus,  explosion  epoch  and  reddening.  To  obtain  the
distance  modulus,  we  use  (in  order  of  preference)  the  Cepheid
distance,   the  expanding   photosphere  method   distance,  or   the
Hubble-flow  distance  (using a  Hubble  constant  of 73  km  s$^{-1}$
Mpc$^{-1}$ and  corrected for  Virgo infall).   The only  exception is
ASASSN-14ha, for  which the distance modulus  ranges from
28.7 to  31.64 mag;  we decided  to use the  mean distance  modulus of
$29.53  \pm 0.5$  mag  (from  NED). This  choice  makes ASASSN-14ha  a
low-luminosity SN, similar to SN~2005cs \citep{Pastorello2005}. Giving
the extremely low velocities \citep{Arcavi2014b} and the long plateau,
our choice seems  appropriate; however, the reader should  be aware of
the large uncertainty in this value.

The explosion  epoch ($t_{0}$) is  assumed to be the  midpoint between
the last nondetection and the first detection.  The range between
  these two epochs has been used for the explosion epoch uncertainty.
If  the last  nondetection is  not  very constraining  and the  object
appears to have been discovered  quite young, we estimate an explosion
epoch by  comparing its spectra to  those of similar objects   and
  using the range of epochs of similar spectra for the explosion epoch
  uncertainty.

Galactic  reddening  values were  obtained  from  NED as  provided  by
\citet{Schlafly2011}. To estimate the host-galaxy  reddening we
  used the equivalent width (EW) of \ion{Na}{i}~D lines.  Gas and dust
  are often  mixed, and some authors  have claimed the existence  of a
  statistical correlation  between the  EW of the  \ion{Na}{i}~D lines
  and                 the                colour                 excess
  \citep{1997A&A...318..269M,2003fthp.conf..200T,
    2012MNRAS.426.1465P}.  We  use spectra  collected with  the FLOYDS
  spectrograph, by  the Public  ESO Spectroscopic Survey  of Transient
  Objects (PESSTO;  \citealt{Smartt2015}) and by the  Asiago Transient
  Classification Programme \citep{Tomasella2014}.

 Host-galaxy \ion{Na}{i}~D  lines are not  detected in the  spectra of
 SN~2013fs, SN~2013bu,  SN~2014cy, LSQ14gv,  ASASSN-14dq, ASASSN-14gm,
 ASASSN-14ha, and SN~2015W.  In the spectra of SN~2014dw both Galactic
 and  host-galaxy  \ion{Na}{i}~D lines  are  present  with similar  EW
 values, so  we set  the host-galaxy reddening  equal to  the Galactic
 reddening.   PESSTO  spectra  of   SN~2013ai  also  show  host-galaxy
 \ion{Na}{i}~D  absorption  lines that  have  an  EW value  twice  the
 Galactic  component.  Thus,  in  this case  we  set  the  host-galaxy
 reddening to  twice the Galactic reddening\footnote{Here  we assume a
   similar dust-to-gas ratio for NGC 2207 as in the Milky Way.}.

The  FLOYDS spectrum  of  ASASSN-14kg also  shows \ion{Na}{i}~D  lines
attributed  to the  host  galaxy with  an EW  of  1.5~\AA.  Using  the
equation from \cite{2003fthp.conf..200T}, this  implies a color excess
of $E(B-V)_{\rm host}  = 0.25 \pm 0.12$ mag. Using  this reddening the
color of ASASSN-14kg  is consistent with that of  several other SNe~II
(corrected for reddening). A larger  value would make ASASSN-14kg very
blue.  Note  that in  the case  of these  three SNe  we adopt  for the
host-galaxy reddening  a conservative  error of  50\% of  the inferred
value.  The only SN for which we do not have optical spectra available
to   search    for   the   presence   of    \ion{Na}{i}~D   lines   is
LSQ13dpa. However, the  light curve of this  SN is not red,  and it is
consistent  with  the  light  curves  of  reddening-corrected  SNe~II,
suggesting that it too is minimally reddened.

The distance modulus, explosion epoch, and reddening of objects in the
literature    are    reported    in     the    online    version    of
Table~\ref{tabsummary}, together with their references.

\section{SNe~II: Light-Curve Parameters}
\label{sec:slope}

\begin{figure}
 \begin{center}
  \includegraphics[width=8.4cm,height=6cm]{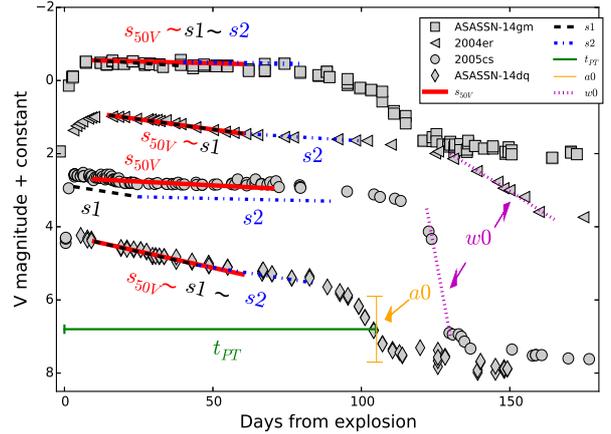}
  \caption{A sample of SN~II light curves with 
  the parameters in Section 3 indicated.}
  \label{fig:parameters}
\end{center}
\end{figure}

In  this  section we  introduce  several  observational parameters  to
characterize  the light  curves. The  light-curve properties  are then
compared  with those  of  objects from  the  literature.  A  graphical
visualization   of   the   various  light-curve   parameters   is   in
Figure~\ref{fig:parameters}.
\begin{itemize}
\item  $s_{50V}$:  decline  rate  of   the  $V$-band  light  curve  in
  mag  per  50 days,  calculated  soon  after the  light  curve
  reaches the plateau or the linear  decay phase.  SNe~II can take
  1--10 days  from explosion to reach peak  brightness; 
hence, we calculated the decline rate starting from  maximum
brightness and extending over 50 days.

\item  $t_{\rm PT}$:  this  is  an indication  of  the  length of  the
  plateau.   It  is  measured  from the  explosion  ($t_{0}$)  to  the
  midpoint  between the  end of  the plateau  phase and  start of  the
  radioactive decay-powered ($^{56}$Co $\rightarrow$ $^{56}$Fe) linear
  decline phase.
\end{itemize}

The parameter $t_{\rm  PT}$ is obtained by fitting the  SN light curve
in magnitudes around the fall from the plateau with the function
\begin{equation}
\label{eq:tpt}
y(t) = \frac{-a0}{1+e^{(t-t_{\rm PT})/w0}}+ ( p0\times{} t ) + m0 .
\end{equation}
The fit is computed using a Markov Chain Monte Carlo using the package
{\tt  emcee}  \citep{Foreman-Mackey2013}.   The   first  term  of  the
equation is  a Fermi-Dirac function,  which provides a  description of
the    transition     between    the    plateau     and    radioactive
phases\footnote{This function  has no direct physical  meaning, but it
  provides a  good fit to the  transition phase in the  light curve.}.
$t$ is  the time  from explosion  in days. The  parameter $a0$  is the
depth of  the drop,  while $w0$  measures the slope  of the  drop (see
Fig. \ref{fig:parameters}).   The second term  of the equation  adds a
slope to the  Fermi-Dirac function to reproduce  the light-curve slope
before and after the drop.  The light curve after the drop is close to
the slope  of the $^{56}$Co  radioactive decay, while before  the drop
the     light-curve    slope     may     be    more     heterogeneous.
\cite{OlivaresE.2010} and \cite{Anderson2014}  used a similar function
to estimate $t_{\rm PT}$ that  includes three additional parameters to
reproduce  the first  month of  the light  curve. However,  as we  are
mainly  interested  in constraining  the  time  ($t_{\rm PT}$),  depth
($a0$), and shape ($w0$) of the drop, a function with fewer parameters
is more  appropriate.  This allows us  to limit our fit  to the region
close  to the  drop. The  parameter $p0$,  which constrains  the slope
before and after the drop, is fitted {\it a priori} to the light curve
after the drop and kept fixed  during the fit of the other parameters.
For a few  objects, the light curve does not  extend sufficiently long
after the end of the drop to measure $p0$. For those objects, we fixed
$p0$ to the  median value of all the other  measurements ($p0 = 0.012$
mag per day).

Some SNe~II exhibit a change in the slope of the light curve at 10--20
days after  maximum. It is only  after this time that  the photosphere
starts to  recede deeper  than the  outer few tenths  of a  solar mass
\citep{Dessart2010}.  However, for most SNe~II the change in the slope
is not clearly  visible in a single  band, and the light  curve can be
reproduced by a single slope  until the end of hydrogen recombination.
To account  for this,  while studying an  expanded sample  of $V$-band
light   curves,   \cite{Anderson2014}    introduced   two   additional
parameters:
\begin{itemize}
\item $s1$:  the decline  rate of  the initial,  steeper slope  of the
  light curve. 
\item $s2$:  the decline rate  of the  second, shallower slope  of the
  light curve.
\end{itemize}

To be consistent with our definition of $s_{50V}$, the slopes $s1$ and
$s2$  are  also  in  units  of mag  per  50  days.\footnote{Note  that
  \citet{Anderson2014} instead  used mag  per 100 days.}   For objects
whose change in  slope is not clearly visible, $s1  \approx s2 \approx
s_{50V}$    (see   Fig.     \ref{fig:parameters},   ASASSN-14gm    and
ASASSN-14dq).  In several cases the  change in slope is visible 30--40
days     after     explosion      (see,     e.g.,     SN~2004er     in
Fig. \ref{fig:parameters}),  and $s1 \approx s_{50V}$.   In some other
cases, the change in slope is  visible around 20 days after explosion,
leading $s_{50V}$  to take on a  value between those of  $s1$ and $s2$
(e.g., SN~2005cs  in Fig. \ref{fig:parameters}).  While  $s1$ and $s2$
may  highlight  differences  among   SNe~II,  they  are  difficult  to
constrain independently without dense, multiband light-curve coverage.
For  a single-band  analysis, we  will mainly  use $s_{50V}$,  $t_{\rm
  PT}$, $a0$,  and $w0$.   We will  later show  pseudobolometric light
curves  that  exhibit more  pronounced  changes  in the  slope.   When
measuring the slopes of the pseudobolometric light curves, we will use
capitol letters ($S1$  and $S2$) to distinguish  these parameters from
the single V-band slopes parameters ($s1$ and $s2$).  Finally, we
  point out  that the  end of  the plateau is  usually defined  as the
  point  where  the  light  curve  starts to  drop  from  the  plateau
  \citep[OPTd; ][]{Anderson2014}. However, the  end of the plateau for
  the majority of  the objects occurs $\sim 20$ days  prior to $t_{\rm
    PT}$,  depending on  how much  the light  curve deviates  from the
  linear decay.  Given this arbitrary  definition, we use $t_{\rm PT}$
  as a measure  of the length of the plateau.   All of the parameters
described above are reported in Appendix \ref{apefigures}.

In  Section  \ref{sec:rise}, we  will  also  introduce two  additional
parameters: the  \emph{rise time} and the  \emph{magnitude at maximum}
of the light curve.  The rise time of SNe~II is relatively fast (1--10
days)  compared  to  the  rise  times  of  other  types  of  SNe,  and
determining it  accurately requires good constraints  on the explosion
epoch and decent  photometric coverage at early  phases.  Some SNe~IIP
do not actually  reach a maximum in their  light-curve evolution until
the end of the plateau, especially in red passbands.  For this reason,
we will refer to the light-curve maximum as the last point after which
the light curve  does not increase by  more than 0.1 mag  per day (see
also \citealt{Gall2015} and \citealt{Rubin2015}).

Starting from  our sample  of SNe  and SNe  from the  literature, when
possible,  we  measured  $s_{50V}$,  the  \emph{rise  time},  and  the
\emph{magnitude at  maximum light} in $V$,  $R$ (or $r$), and  $I$ (or
$i$).  Given  the limited number of  light curves in $R$  (or $r$) and
$I$  (or $i$)  that cover  the drop  of the  linear decay,  we measure
$t_{\rm PT}$, $a0$, and $w0$ only in $V$ and with the pseudobolometric
light curve.  This  gives us 45 and 30 objects  with information after
maximum  light  and  at  the  moment  of  the  drop  in  $V$  (see Fig.4 ) 
 and  the pseudobolometric light curve (respectively).
 
\begin{figure*}
\begin{center}
  \includegraphics[width=17.cm,height=20.cm]{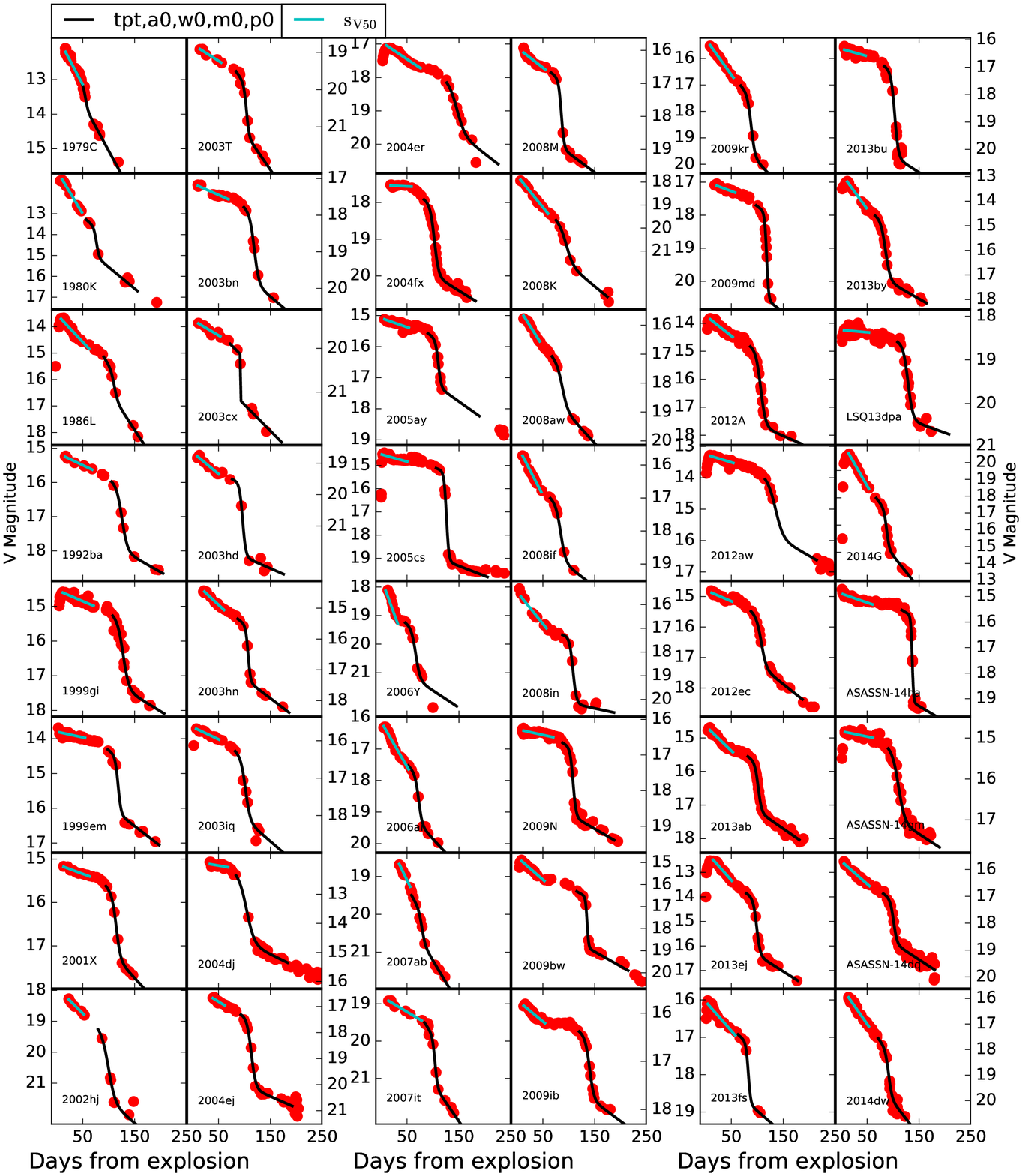}
  \caption{$V$-band light curves and their fits.}
  \label{fig:Vfit}
\end{center}
\end{figure*}
\begin{figure*}
 \begin{center}
  \includegraphics[width=8.cm,height=8cm]{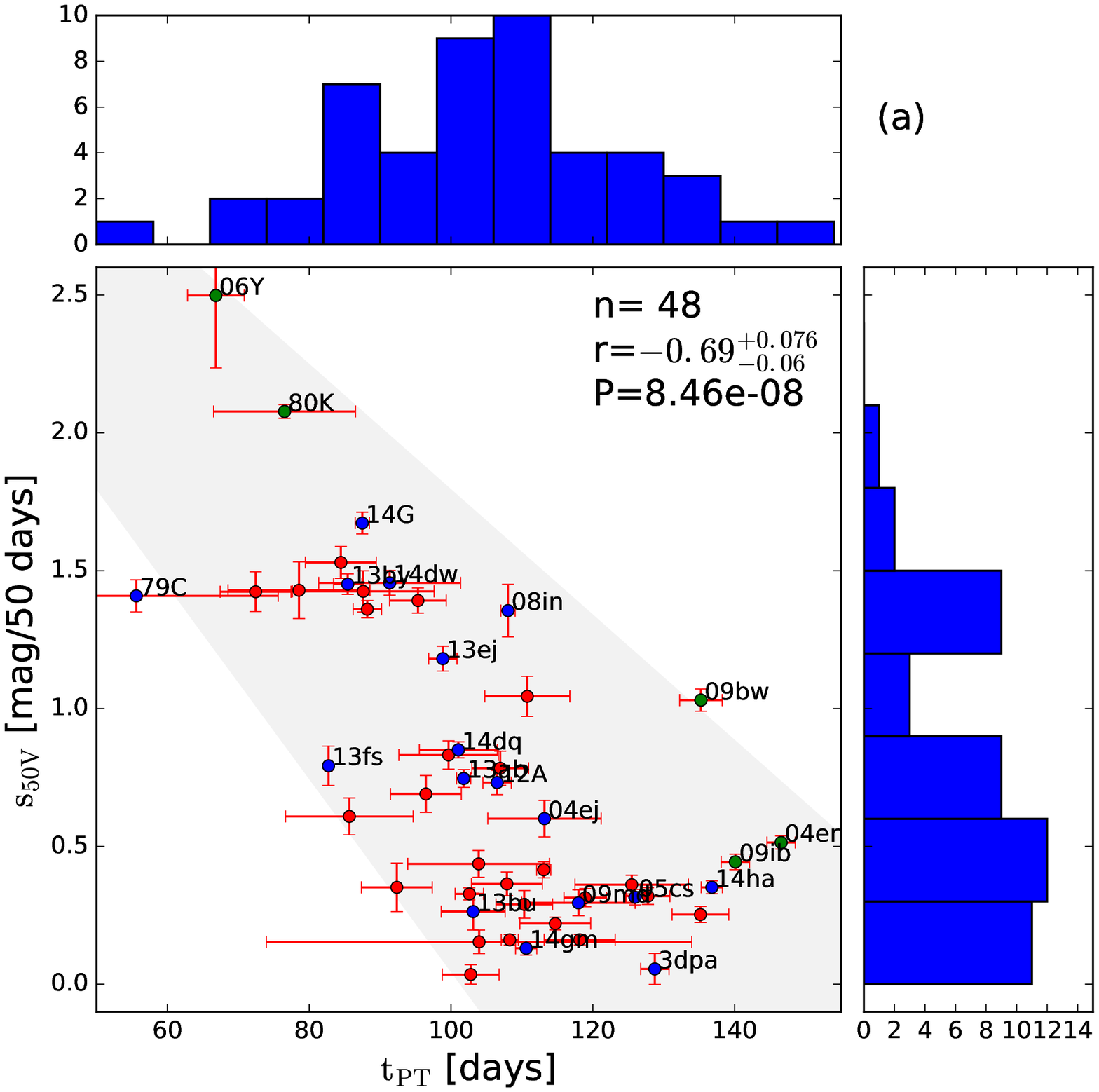}
  \includegraphics[width=8.cm,height=8cm]{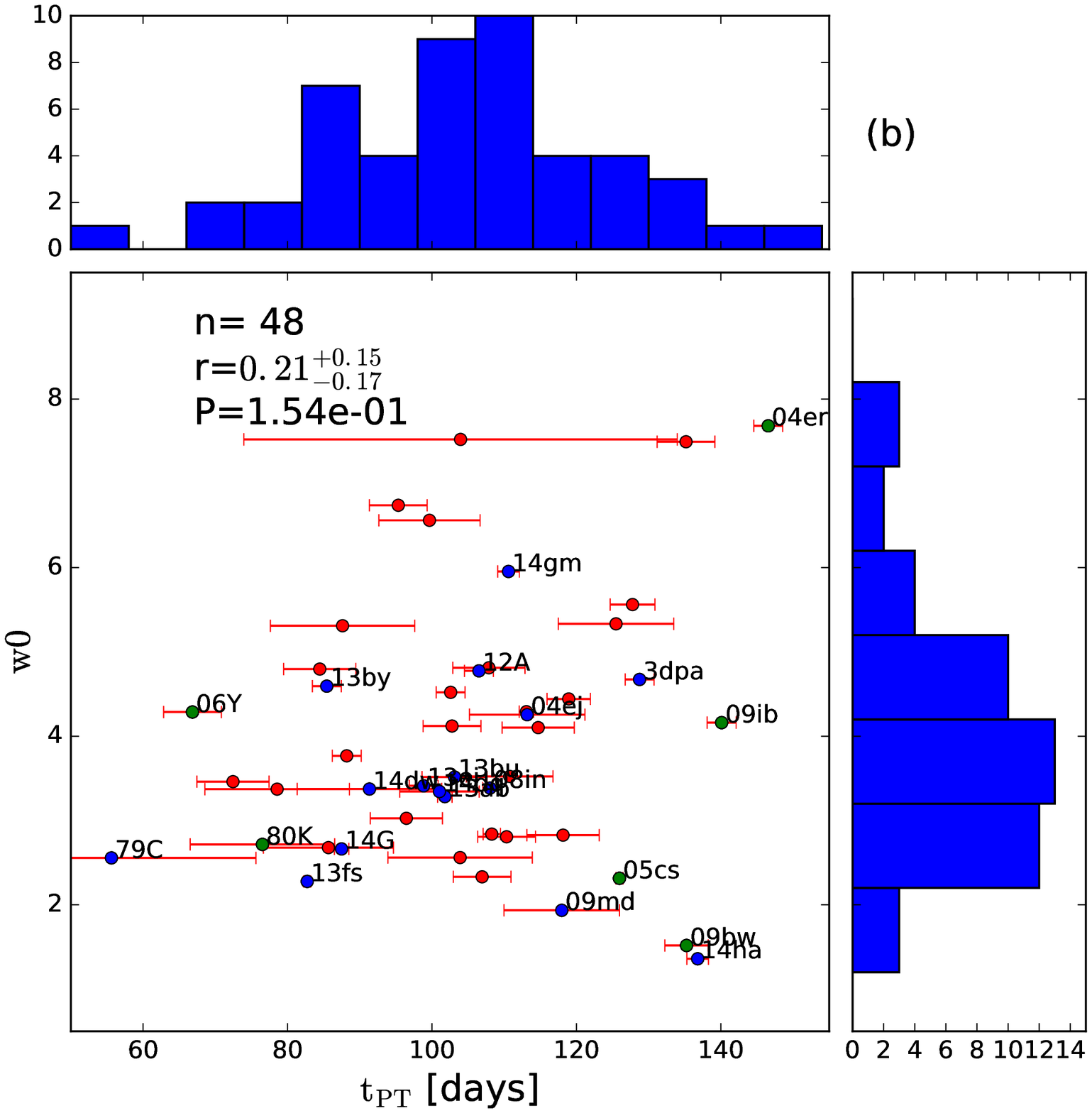}
  \includegraphics[width=8.cm,height=8cm]{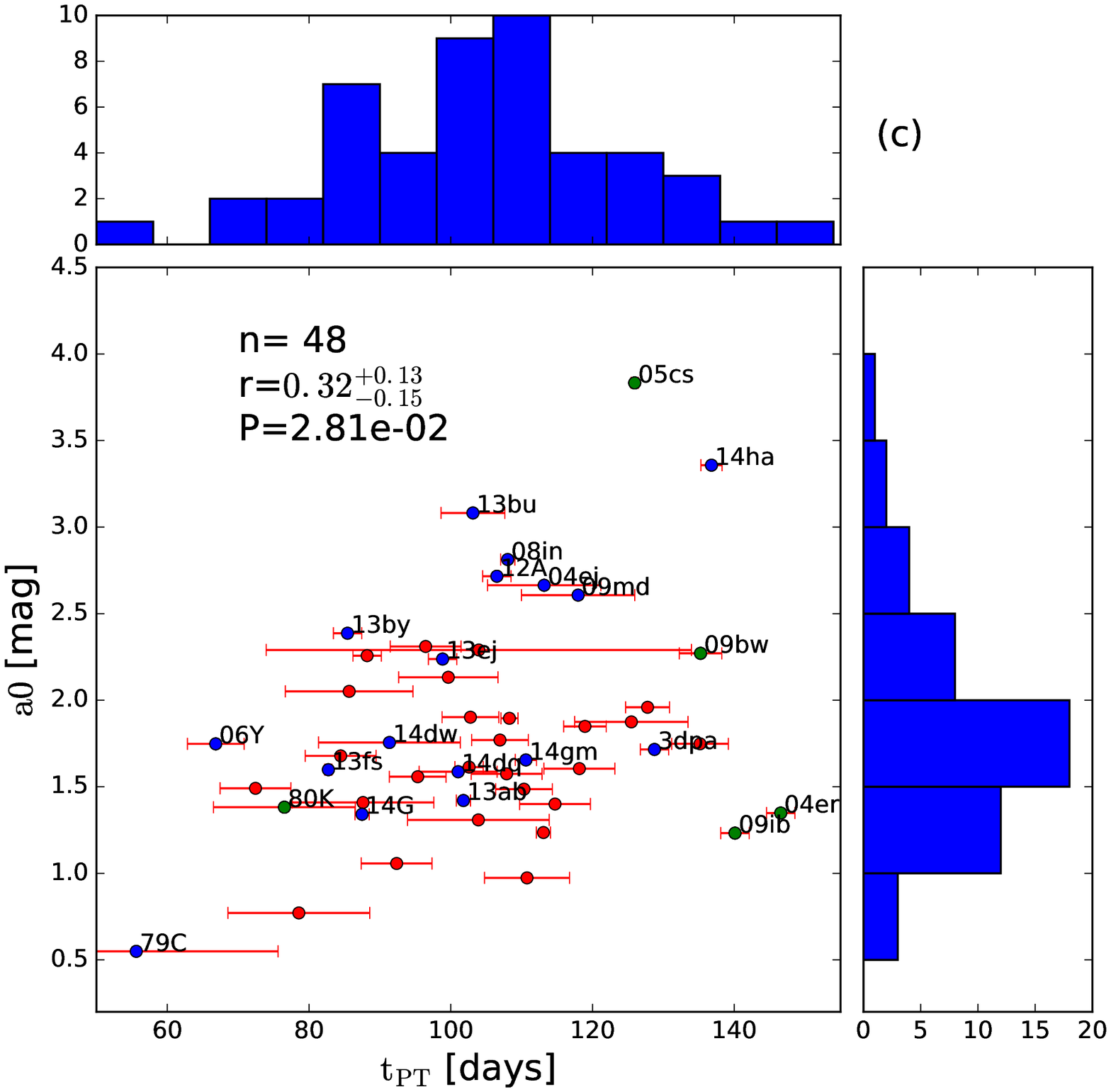}
  \includegraphics[width=8.cm,height=8cm]{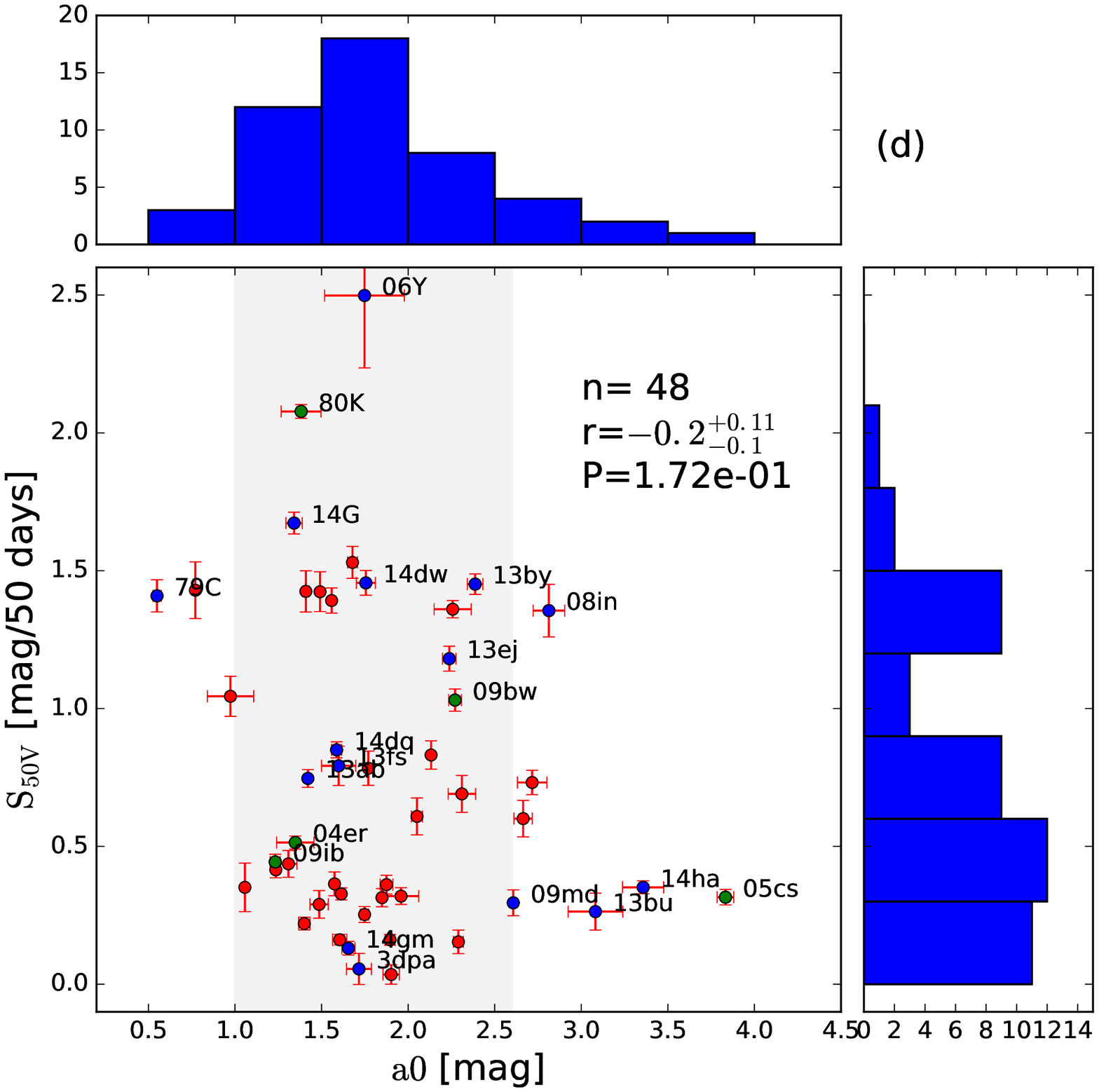}
  \caption{Correlation  among  $s_{50V}$,  $t_{\rm PT}$,  $w0$, and  $a0$,  as
    obtained  from $V$-band light  curves.  Our  sample is  marked in
    blue; some other SNe are highlighted in green. For each panel $n$ is
    the number of events,  $r$ is the Pearson $s$ correlation
    coefficient, and $P$ is the probability of detecting a correlation by
    chance. }
  \label{fig:tpt}
\end{center}
\end{figure*}

 
Figure \ref{fig:tpt}  shows several  relations among  $s_{50V}$, $a0$,
$t_{\rm  PT}$,  and $w0$.   We  used  a  Pearson  test to  search  for
correlations  between the  measured  values of  these parameters,  and
further  employed   Monte  Carlo  bootstrapping  to   investigate  the
reliability of the test.  This is the same approach as previously used
by  \cite{Anderson2014}.  No  clear  clustering for  the two  distinct
SN~IIL and SN~IIP subtypes is  found.  A moderate correlation is found
between $s_{50V}$ and $t_{\rm PT}$, with faster-declining SNe~II found
to  have  shorter  duration  plateaus.   \cite{Anderson2014}  found  a
similar relation  between $s1$  and OPTd.   However, their  result was
heavily dependent on the extreme case of SN~2006Y. Four of our objects
(SN~2013ej, SN~2013by,  SN~2013fs, and SN~2014G) show  a steep decline
and a  short $t_{\rm PT}$,  making this correlation more  robust.  The
length of  the plateau ($t_{\rm PT}$)  is not found to  correlate with
the slope of  the drop ($w0$; see Fig.   \ref{fig:tpt}b) or the
depth of the  drop ($a0$; see Fig.   \ref{fig:tpt}c).  The mean
length of the plateau is 100 days, while the wide range of the plateau
lengths is 60  to 150 days.  This is  similar to the distribution
  of OPTd values reported by \cite{Anderson2014} (if we add 20 days to
  account for the different parameters  OPTd and $t_{\rm PT}$), but it
  differs from results presented by \cite{Poznanski2013}.

An interesting result is that the  vast majority of our SNe~II (42 out
of 45) have  a drop ($a0$) between  1 and 2.6 mag,  independent of the
slope  of the  post-maximum decline  (see Fig.~\ref{fig:tpt}d).
The  similar drop  ($a0$) suggests  that SNe~II  with a  more luminous
plateau should  also be more luminous  on the radioactive tail.   As a
consequence, given that  the plateau luminosity depends  mainly on the
kinetic   energy   of   the  explosion   \protect\citep[$L_{\rm   SN}$
  $\propto{}$   $E^{5/6}$;  ][]{2009ApJ...703.2205K},   more-energetic
explosions should also produce more  nickel. This is in agreement with
the relation  between expansion velocity  and nickel mass  reported by
several authors \citep{Hamuy2003,Spiro2014,Valenti2015,Pejcha2015a}.

\begin{figure}
\begin{center}
  \includegraphics[width=8.cm,height=6cm]{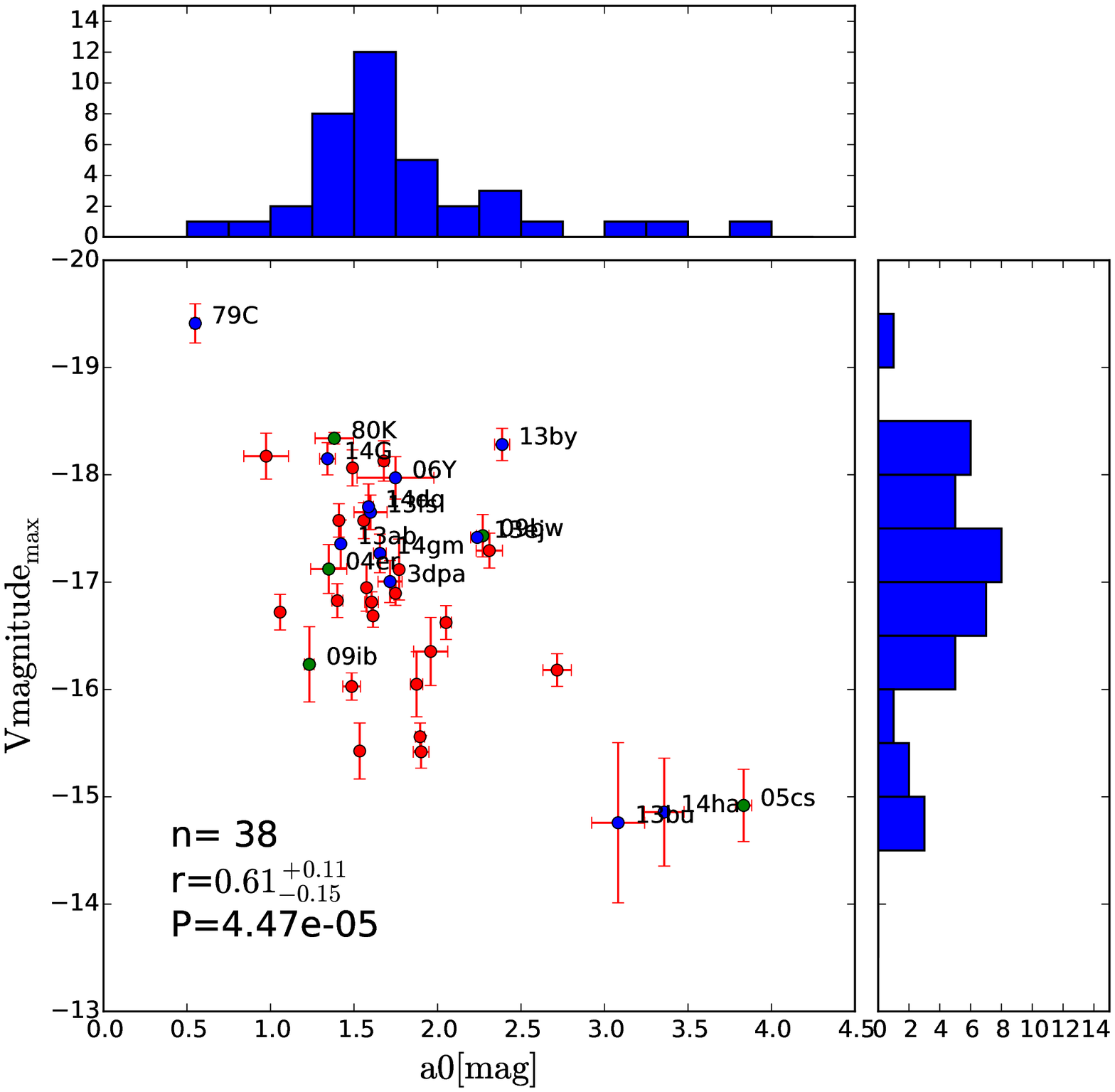}
  \includegraphics[width=8.cm,height=6cm]{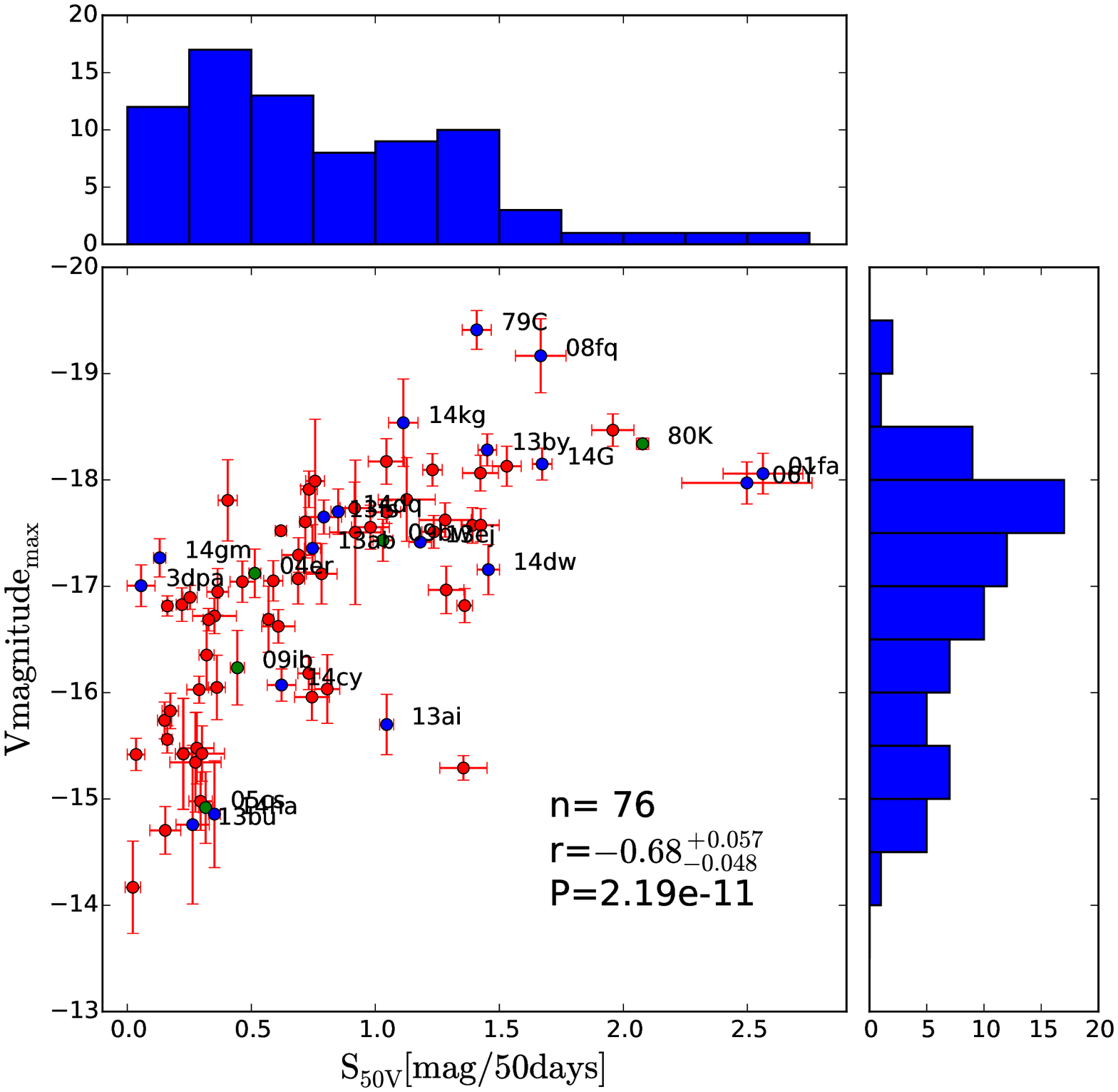}
  \caption{Correlation between  absolute magnitude and $a0$ (top panel)
and $s_{50V}$ (lower panel)  as obtained from $V$-band light curves. }
  \label{fig:drop}
\end{center}
\end{figure}

Three objects show a significant  high $a0$: SN~2005cs, SN~2013bu, and
ASASSN-14ha.   Interestingly, these  three objects  are low-luminosity
SNe~II.  The relative number of  SNe~II with a significant drop (3/45)
is probably  a lower  limit since  we do  not have  a large  number of
low-luminosity SNe~II in our sample.   The peculiar behaviour of these
three  SNe is  also  clear when  we correlate  with  the magnitude  at
maximum (see Fig.~\ref{fig:drop}); $a0$  and magnitude at maximum seem
to correlate, but  the correlation is mainly driven by  the large drop
of the three faint SNe and the  small drop of SN 1979C. In the latter,
the  drop  (if present  at  all)  is  around  50 days  after  maximum.
Finally,  in the  lower panel  of Figure  \ref{fig:drop}, we  show and
confirm the \cite{Anderson2014} result  of a clear correlation between
$V$-band slope ($s_{50V}$) and absolute magnitude.

\section{Pseudobolometric Light Curves}
\label{sec:bolo}
To date, 17 SNe~II have  multiband photometric coverage extending from
near the time  of explosion to the  radioactive-decay tail.
The addition  of the current  set nearly
doubles   the  published   sample.   We   now  proceed   to  construct
pseudobolometric light curves (see Fig.7) (from $U$/$B$  to $I$) for the 30 SNe~II
followed beyond the beginning of the radioactive-decay tail.

\begin{figure}
 \begin{center}
  \includegraphics[width=8.4cm,height=6cm]{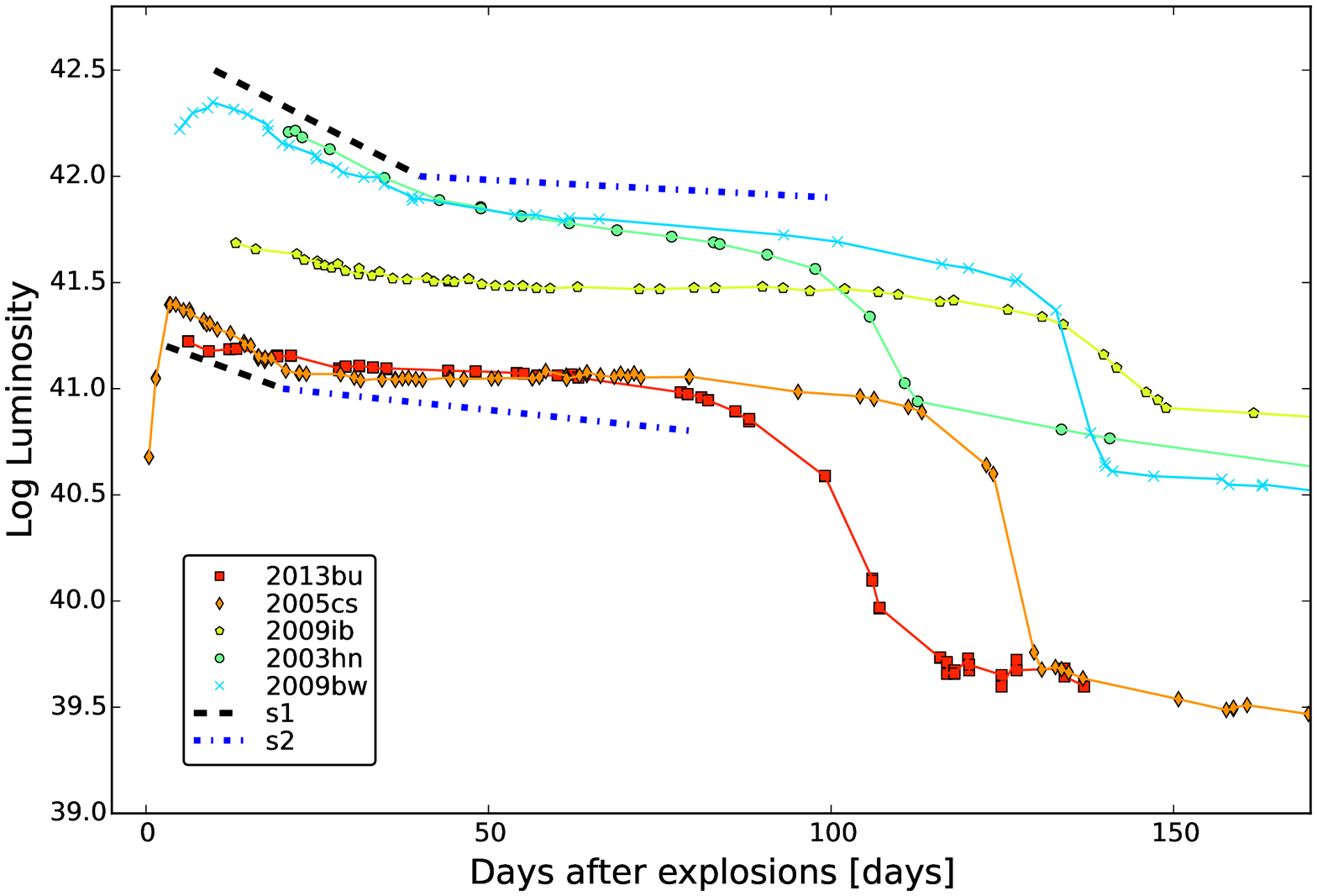}
  \caption{A sample of SN~II pseudobolometric light curves.}
  \label{fig:bolo}
\end{center}
\end{figure}

\begin{figure}
 \begin{center}
  \includegraphics[width=8.4cm,height=6cm]{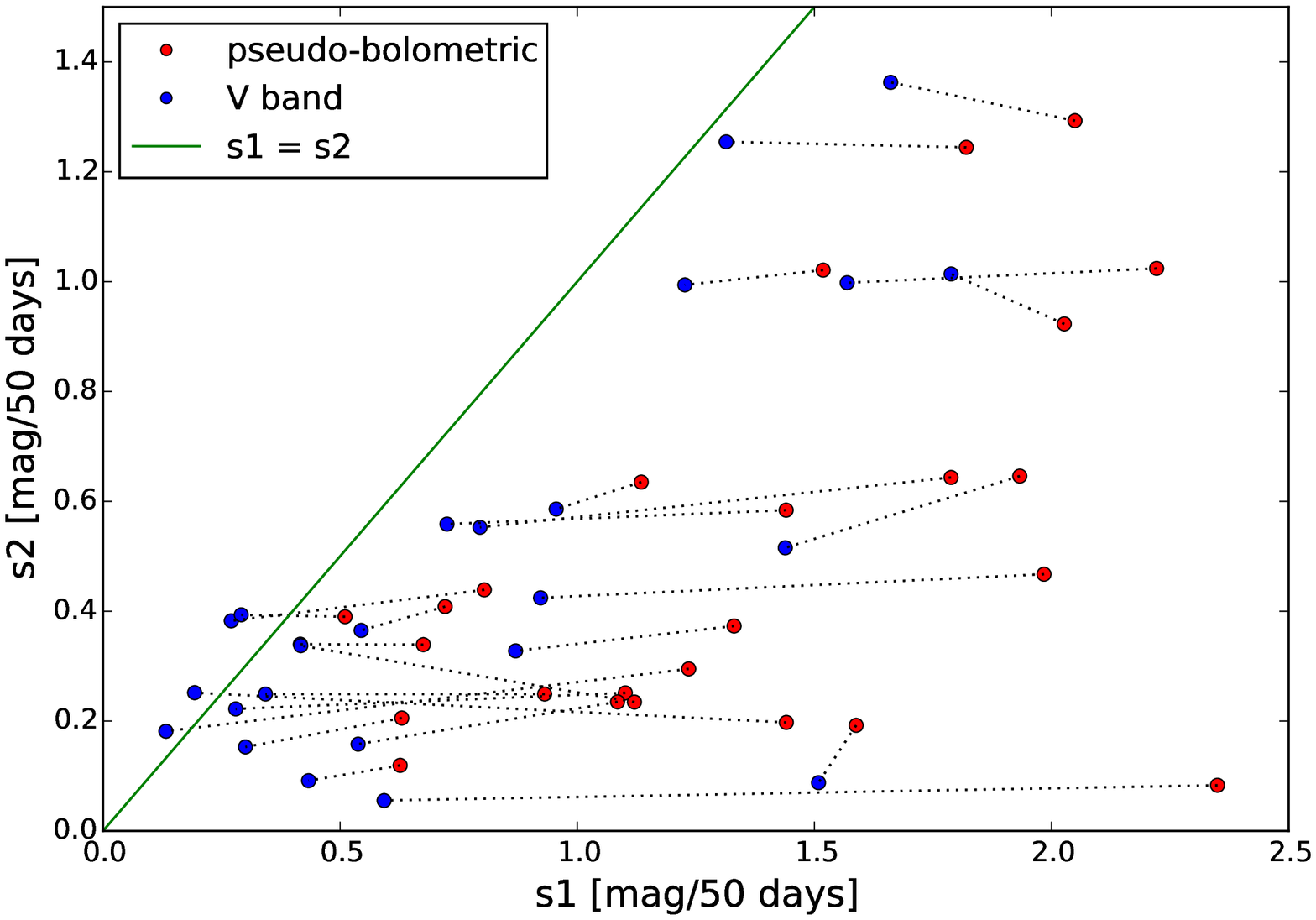}
  \caption{Parameters $s1$ vs. $s2$ computed using the $V$ band (blue) and the
    pseudobolometric light curve (red) for the sample of 22 SNe with
    multiband coverage.  While $s2$ remains nearly constant, $s1$
    increases if we use the integrated flux.}
  \label{fig:s1s2}
\end{center}
\end{figure}

The transition between the early fast  slope $s1$ and the shallow late
slope $s2$,  which is  not always  visible in a  single band,  is more
evident in the pseudobolometric  curves.  In Figure \ref{fig:s1s2}, we
show  $s1$  and $s2$  measured  both  for the  $V$  band  and for  the
pseudobolometric  light  curve  (the  dotted  lines  connect  $V$-band
measurements  with pseudobolometric  measurements).  When  integrating
over  the optical  range, the  light  curve becomes  steeper at  early
phases and changes  less in the second part of  the plateau.  At early
times, the SN is expanding and  the temperature is dropping, hence, at
early time,  the bluest  bands give large  contributions to  the light
curve.   Later,  hydrogen  recombination starts  and  the  temperature
remains close  to the recombination temperature  ($\sim 6000$~K).  The
pseudobolometric light curve better describes this transition phase.

We have repeated the light-curve analysis using pseudobolometric light
curves of 30 SNe~II with the function

\begin{equation}
\label{eq:tpt2}
y(t) = \frac{-A0}{1+e^{(t-T_{\rm PT})/W0}}+ ( P0\times{} t ) + M0 .
\end{equation}

The parameters $T_{\rm PT}$, $A0$, $W0$, $P0$, and $M0$ are equivalent
to  the parameters  $t_{\rm  PT}$,  $a0$, $w0$,  $p0$,  and $m0$,  but
fitting  Log\,(luminosity)  as  a  function of  time  instead  of  the
$V$-magnitude light curves.

As a first-order  approximation to the true  bolometric luminosity, we
integrated  the  SN  emission  in  the  optical  bands  ($(U)BVRI$  or
$(U)BVgri$, depending  on the  available bands).  All  magnitudes were
converted to fluxes and integrated  using Simpson's Rule \citep[see ][
  for more  details on  this method]{2008MNRAS.383.1485V}.   Given the
good light-curve coverage,  we measured the parameters  $S1$ and $S2$,
along with $A0$, $W0$, and $T_{\rm PT}$. Pseudobolometric light curves
and   the    main   computed   parameters   are    shown   in   Figure
\ref{fig:bolofit}.

\begin{figure*}
 \begin{center}
   \includegraphics[width=8.cm,height=8cm]{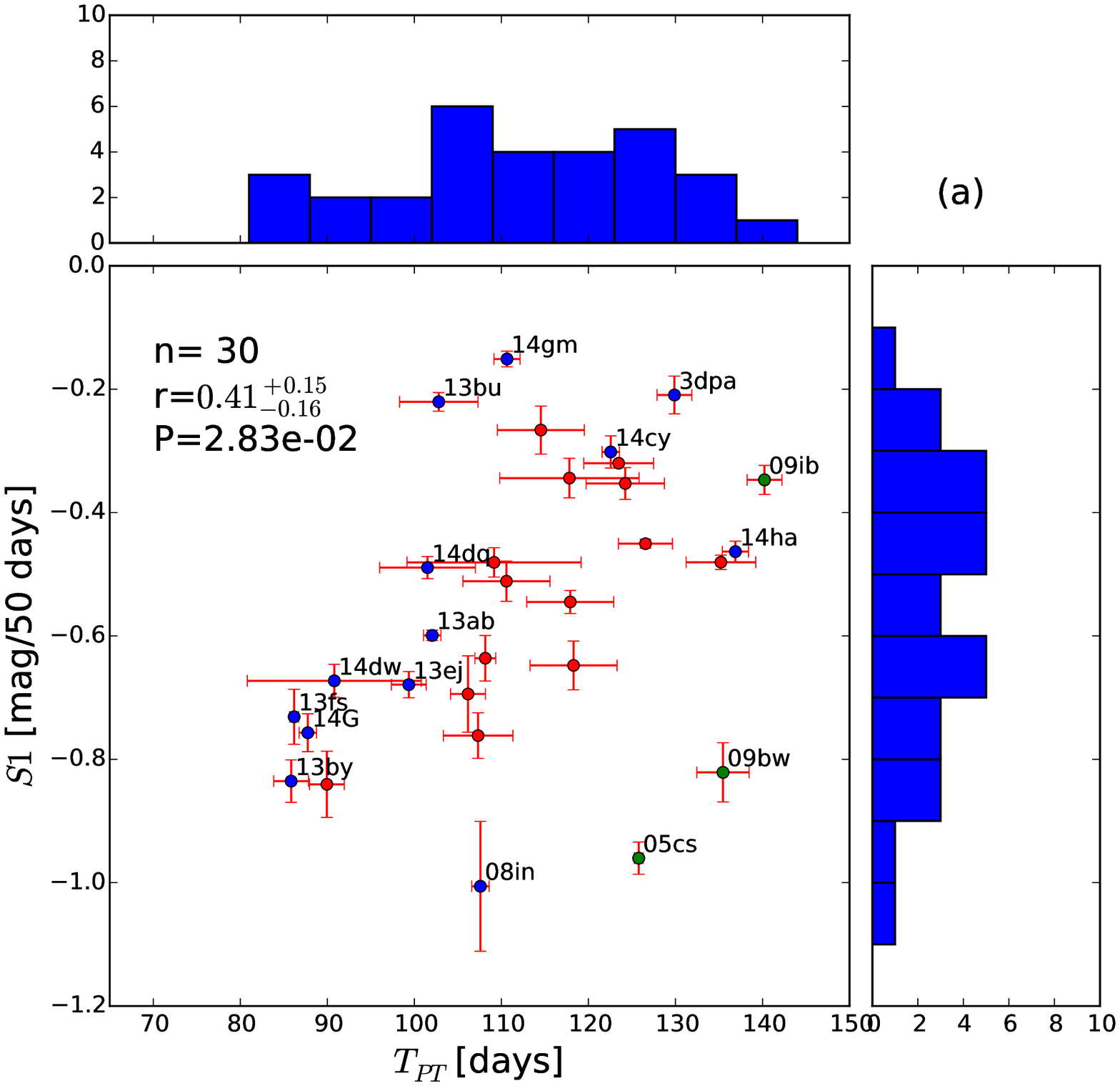}
   \includegraphics[width=8.cm,height=8cm]{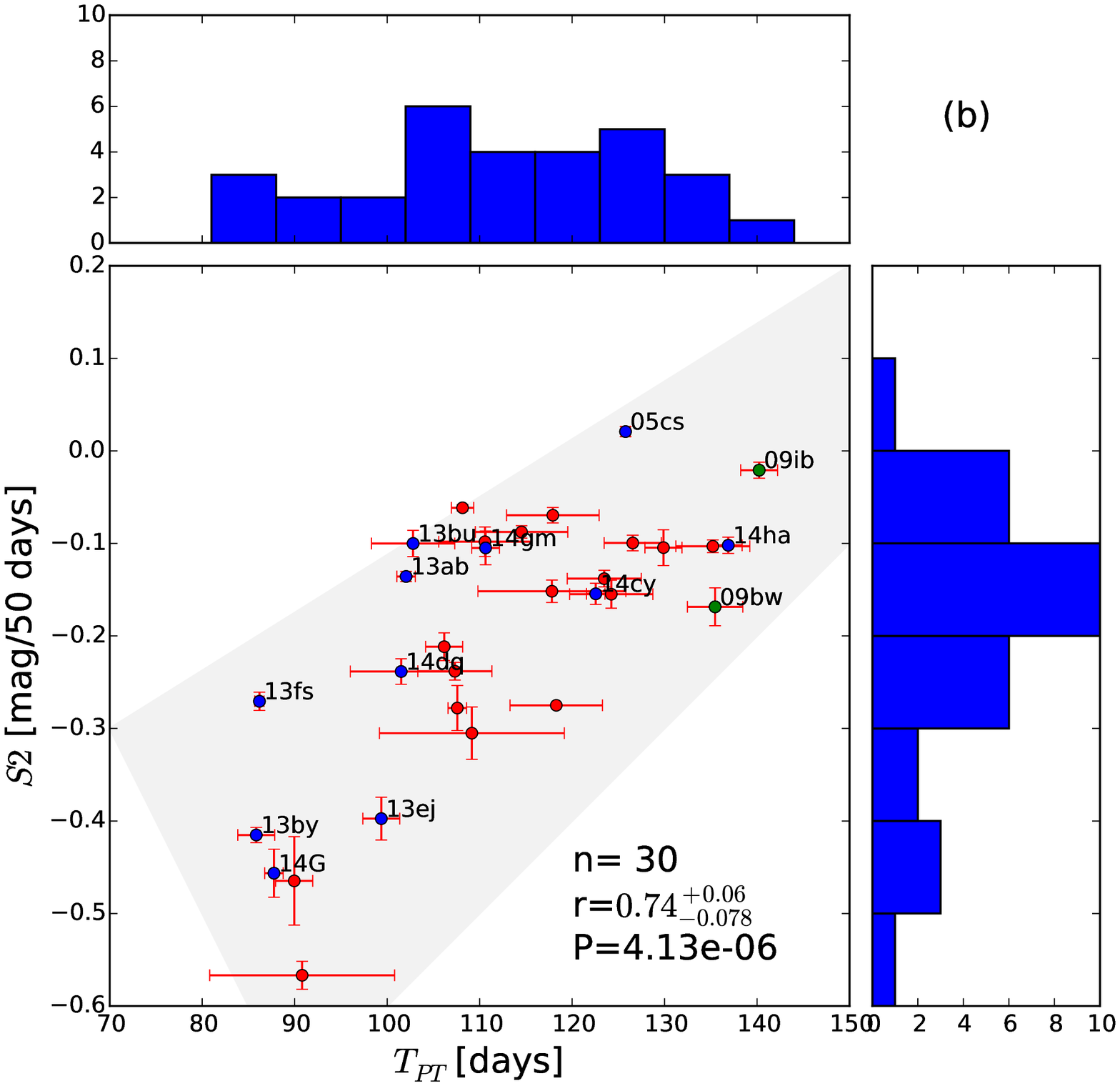}
   \includegraphics[width=8.cm,height=8cm]{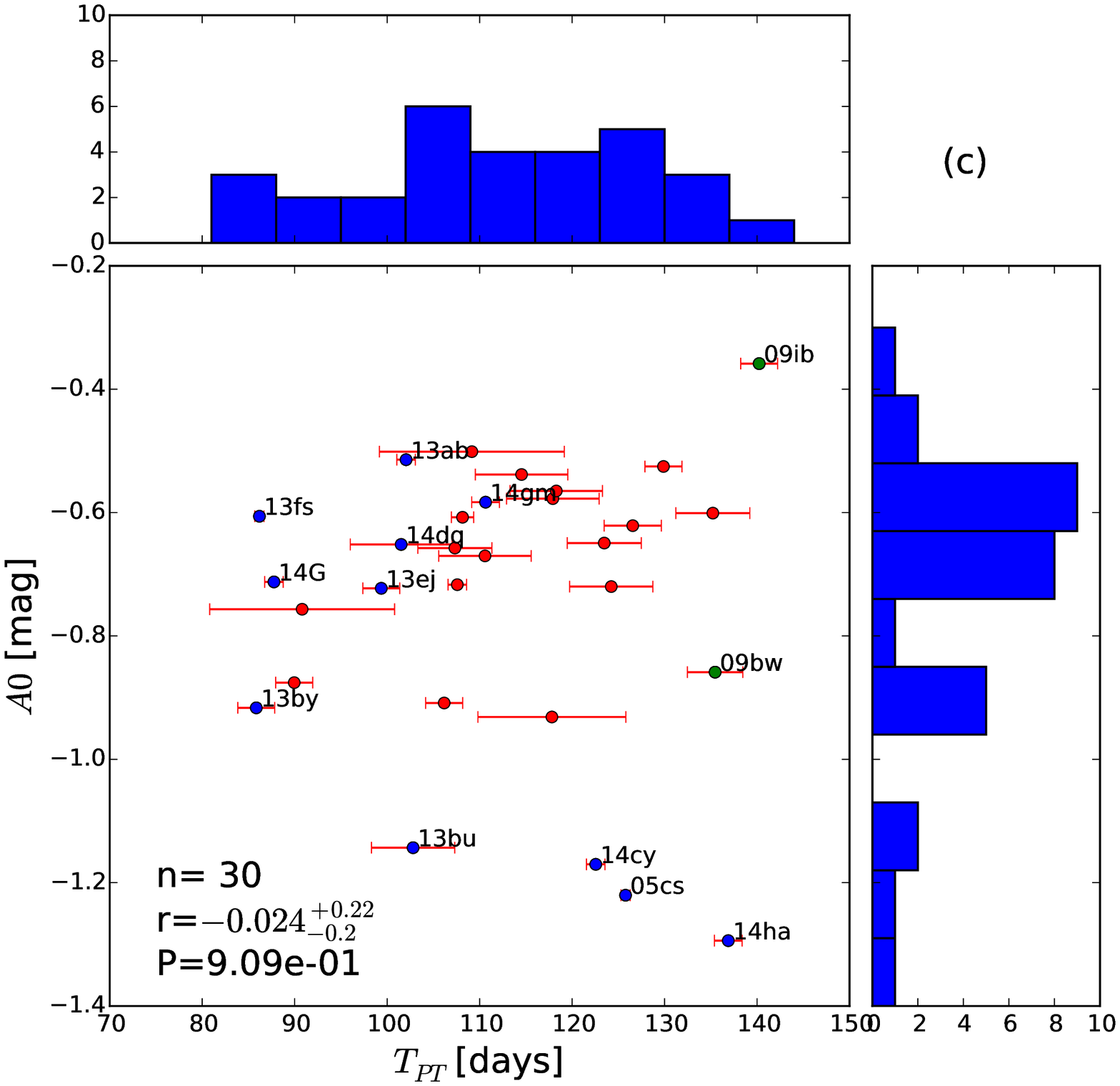}
   \includegraphics[width=8.cm,height=8cm]{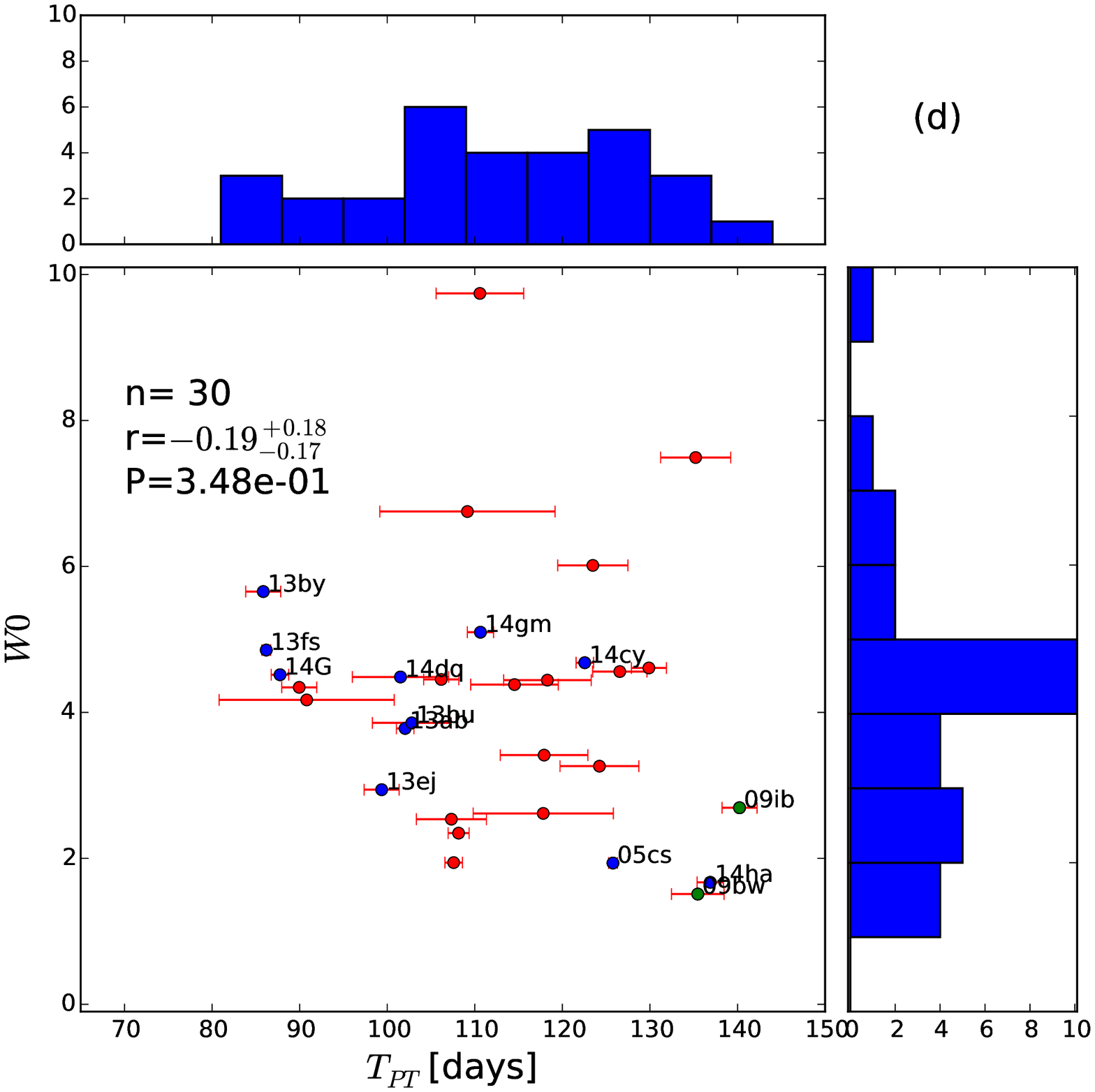}
  \caption{Correlations  among $S1$,  $S1$,  $T_{\rm PT}$,  $W0$, and $A0$  
    as obtained from pseudobolometric light curves. Our sample is marked
    in blue.  Some other SNe are  highlighted in green. For each panel
    $n$ is  the  number  of  events,  $r$ is Pearson's
    correlation coefficient,  and $P$ is  the probability of  detecting a
    correlation by chance. }
  \label{fig:tpt2}
\end{center}
\end{figure*}

\begin{figure*}
\begin{center}
  \includegraphics[width=17.cm,height=20.cm]{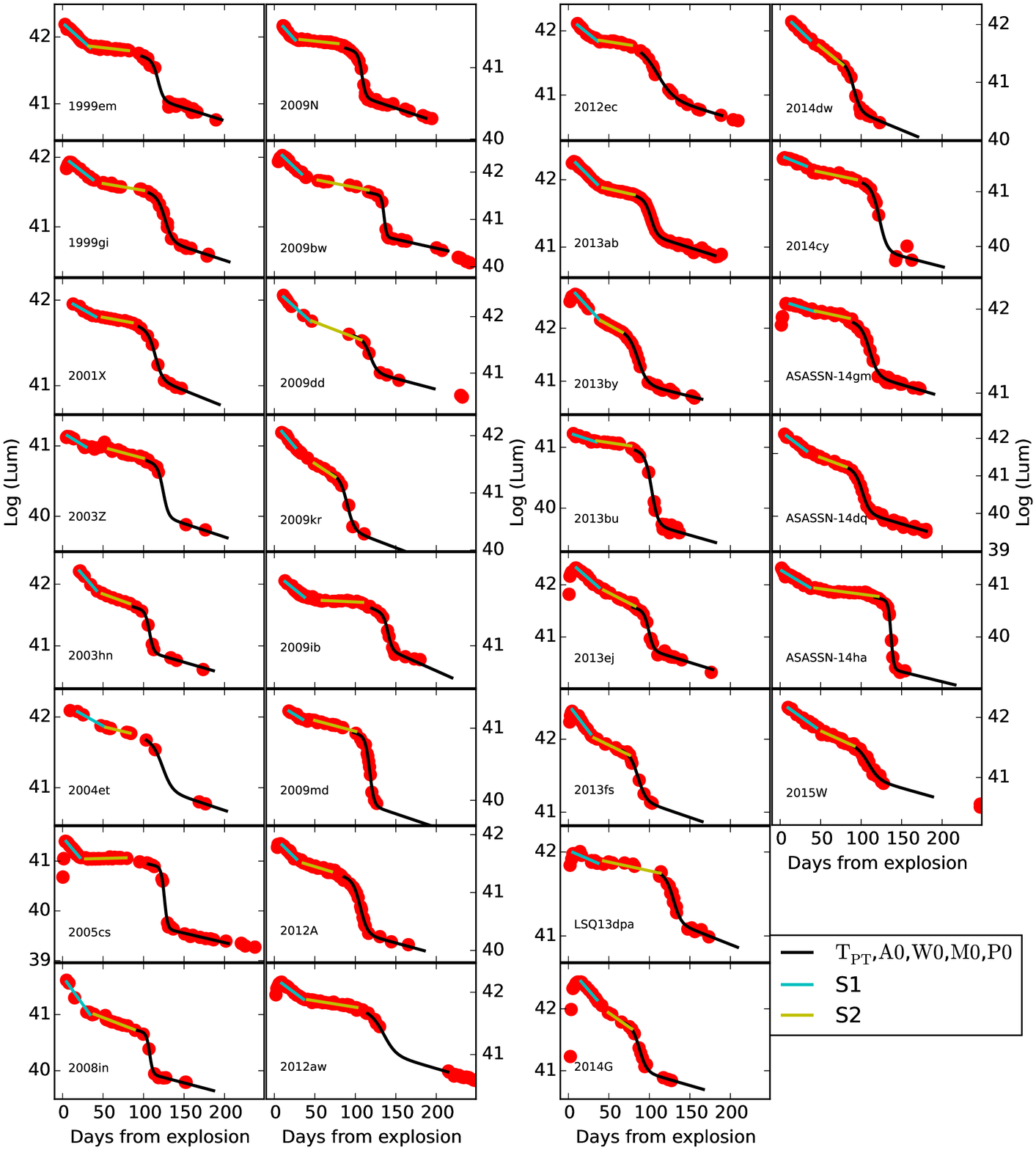}
  \caption{Pseudobolometric light curves and their derived parameters.
  Data: 
    SN 1999em, \protect\cite{2003MNRAS.338..939E}, \protect\cite{Leonard2002};
    SN 1999gi,  \protect\cite{Leonard2002b}, \protect\cite{Faran2014a};
    SN 2001X,  \protect\cite{Faran2014a};
    SN 2003Z,  \protect\cite{Spiro2014};
    SN 2003hn,  \protect\cite{Krisciunas2009};
    SN 2004et,  \protect\cite{2010MNRAS.404..981M};
    SN 2005cs,  \protect\cite{2006MNRAS.370.1752P}, \protect\cite{2009MNRAS.394.2266P};
    SN 2008in,  \protect\cite{Roy2011};
    SN 2009N,  \protect\cite{Takats2013};
    SN 2009bw,  \protect\cite{2012MNRAS.422.1122I};
    SN 2009dd,  \protect\cite{Inserra2013};
    SN 2009kr,  \protect\cite{2010ApJ...714L.280F}, \protect\cite{Elias-Rosa2010};
    SN 2009ib, \protect\cite{Takats2015};
    SN 2009md, \protect\cite{2011MNRAS.417.1417F};
    SN 2012A, \protect\cite{Tomasella2013};
    SN 2012aw, \protect\cite{Dall'Ora2014};
    SN 2012ec, \protect\cite{Barbarino2015};
    SN 2013ab, \protect\cite{Bose2014};
    SN 2013by, \protect\cite{Valenti2015};
    SN 2013bu, this paper;
    SN 2013ej, Yuan et al., in prep.;
    SN 2013fs, this paper;
    LSQ13dpa, this paper;
    SN 2014G, Terreran   et  al., in  prep.;
    SN 2014dw, this paper;
    SN 2014cy, this paper;
    ASASSN-14ha, this paper;
    ASASSN-14gm, this paper;
    ASASSN-14dq, this paper; and
    SN 2015W, this paper.}
  \label{fig:bolofit}
\end{center}
\end{figure*}

No  clear correlation  is found  between  $S1$ and  $T_{\rm PT}$  (see
Fig.  \ref{fig:tpt2}a)   owing  to  the  presence   of  SN~2005cs  and
SN~2009bw,  which  have   long  plateaus  and  steep   $S1$.  A  clear
correlation between  $S2$ and $T_{\rm  PT}$ is found,  confirming that
SNe   with  a   large  slope   also  have   a  shorter   plateau  (see
Fig. \ref{fig:tpt2}b).

\section{Early-Time Data}
\label{sec:early}

The early-phase emission of SNe~II provides clues regarding the nature
of the progenitor  star at the moment of explosion.   For example, the
cooling after the shock breakout  depends on the ejected mass, energy,
and            radius           of            the           progenitor
\citep{Waxman2007,Chevalier2011,Nakar2010}.   Therefore,   a  detailed
study of SNe~II  at these phases, particularly with  UV and $uBg$-band
photometry,  may   reveal  intrinsic   differences  among   the  SN~II
populations.  A large  fraction of SNe observed with  {\it Swift} have
been  published  recently  \citep{Brown2009a,Pritchard2013,Brown2014}.
However,  for SNe~IIP/IIL,  most  of  the UV  published  data are  for
SNe~IIP.  In  the last  two years, several  SNe~II have  been observed
with {\it Swift}.   Seven (out of 16) SNe in  our sample have extended
follow-up observations in the UV. Here  we present the UV light curves
of these SNe, comparing with  others available in the literature, with
a  particular  emphasis on  determining  whether  there are  intrinsic
differences between  SNe~IIL and  IIP.  We  also show  early-time {\it
  Swift} UV spectra  of SN~2013ej and compare to those  of the nearby,
normal Type~IIP SN~2012aw \citep{Bayless2014}.

\subsection{Colour and Temperature}
\label{sec:earlycolor}
The early  photospheric temperature evolution  of a SN is  perhaps the
most    revealing    about    the    nature    of    its    progenitor
\citep{Waxman2007,Chevalier2011,Nakar2010}.     SNe~II     with    RSG
progenitors are  expected to remain  at a relatively  higher effective
temperature for  a longer time than  what is expected for  SNe~II with
blue supergiant  (BSG) progenitors.   A clear example  of this  is the
photospheric  temperature  of  SN~1987A,  which was  close  to  5000~K
\citep{Hamuy1988}  five  days  after  explosion. On  the  other  hand,
SN~2013ab and SN~2013ej exhibited temperatures of 10,000 K at the same
epoch,     and      likely     arose     from      RSG     progenitors
\citep{Valenti2014,Bose2015}.

\cite{Blinnikov1993}  suggested that  the  properties  of the  SNe~IIL
1979C and  1980K may be  explained by  a large progenitor  radius.  If
SNe~IIL  arise  from  progenitors  with   a  large  radius  and  their
temperature follows the simple analytic equations of \cite{Nakar2010},
we  should  expect to  see  SNe~IIL  having higher  temperatures  than
SNe~IIP at early phases.

\begin{figure}
\begin{center}
  \includegraphics[width=8.4cm,height=6.3cm]{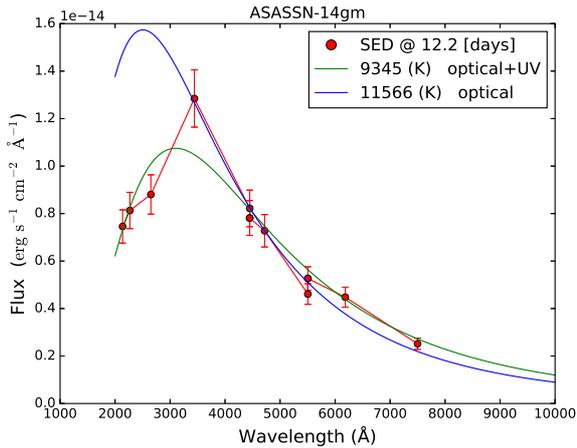}
  \caption{Black-body fit for ASASSN-14gm 12 days after explosion. A different photospheric 
  temperature is obtained with and without the \emph{Swift} UV filters.}
  \label{fig:BB}
\end{center}
\end{figure}

To investigate the  photospheric temperature of our sample  of SNe, we
computed a black-body fit to our multiband photometry
\footnote{The magnitudes were converted to flux. Each flux value was 
assigned to the effective wavelength of each passband.}. 
An example of the fit is reported in Figure \ref{fig:BB}. 
We will show that the photospheric temperature estimated with and without 
the \emph{Swift} data can be different. 
Instead of choosing one or the other, we present the fit both with and
without the UV Swift data for the first 30 days.

\begin{figure}
\begin{center}
  \includegraphics[width=8.4cm,height=11cm]{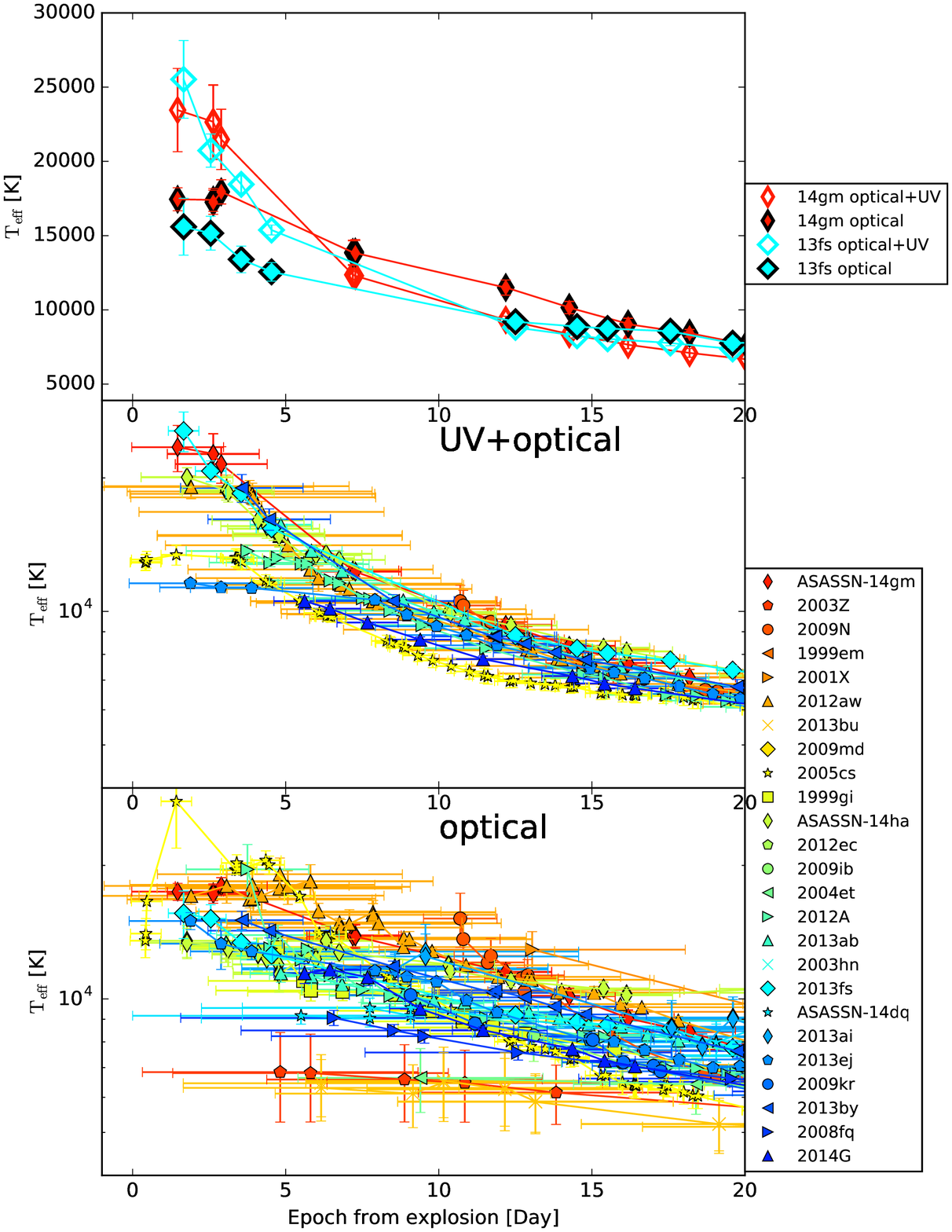}
  \caption{Temperature evolution of a sample  of SNe~II in the first 20 days.
  The objects are  colour-coded with  respect to  the $V$-band  
  decline rate over 50  days ($s_{50V}$).}
  \label{fig:temperature}
\end{center}
\end{figure}

The  temperature  evolution   of  our  SN~II  sample   is  plotted  in
Figure~\ref{fig:temperature}.   Lower  and  central  panels  show  the
temperature  evolution (with  and without  the UV).   The objects  are
colour-coded with  respect to the  $V$-band decline rate over  50 days
($s_{50V}$). Blue  objects are fast-declining SNe~IIL.   Following the
suggestion of \cite{Blinnikov1993}, blue  objects should have a higher
temperature than red objects. Based on our data, SNe~IIL do not always
exhibit higher temperatures than SNe~IIP.

However,  two  caveats  need  to be  mentioned.   First,  the  methods
presented    by    \citet{Waxman2007},   \citet{Chevalier2011},    and
\citet{Nakar2010} are usually  valid only in the first  few days after
explosion  (see  also \citealt{Rubin2015})  when  our  optical and  UV
follow-up observations are limited.

A second caveat is that temperature  estimates at this early phase are
difficult.  \cite{Valenti2014}  noticed that  for SN~2013ej,  a cooler
temperature was inferred when early-time UV data are included with the
optical.  Other SNe~II  show similar  behaviour.  The  upper panel  of
Figure \ref{fig:temperature}  shows the  temperature evolution  of two
SNe~II  with UV  data included  in  the analysis  (empty symbols)  and
without  UV data  (filled  symbols).  Both  objects  show a  different
temperature  estimate depending  on the  spectral energy  distribution
(SED) extension.  In  the first few days after explosion,  they show a
higher temperature  if UV data  are included.  Later,  the temperature
with only  optical bands is  higher than the temperature  including UV
data.   It is  also important  to note  that the  temperature estimate
using only the optical bands is  very sensitive at early phases to the
$U$ band; at this phase, most of  the flux is still emitted in the UV.
Ground-based $U$-band data are also notoriously difficult to calibrate
(sometimes  affected by  the atmosphere  as well),  and may  introduce
large, poorly understood systematic uncertainties into the temperature
estimate.   Considering the  limited number  of SNe~IIL  available and
this    $U$-band    systematic   uncertainty,    more    high-quality,
well-calibrated $U$-band data are  needed to properly investigate this
issue.

\begin{figure}
\begin{center}
  \includegraphics[width=8.4cm,height=6cm]{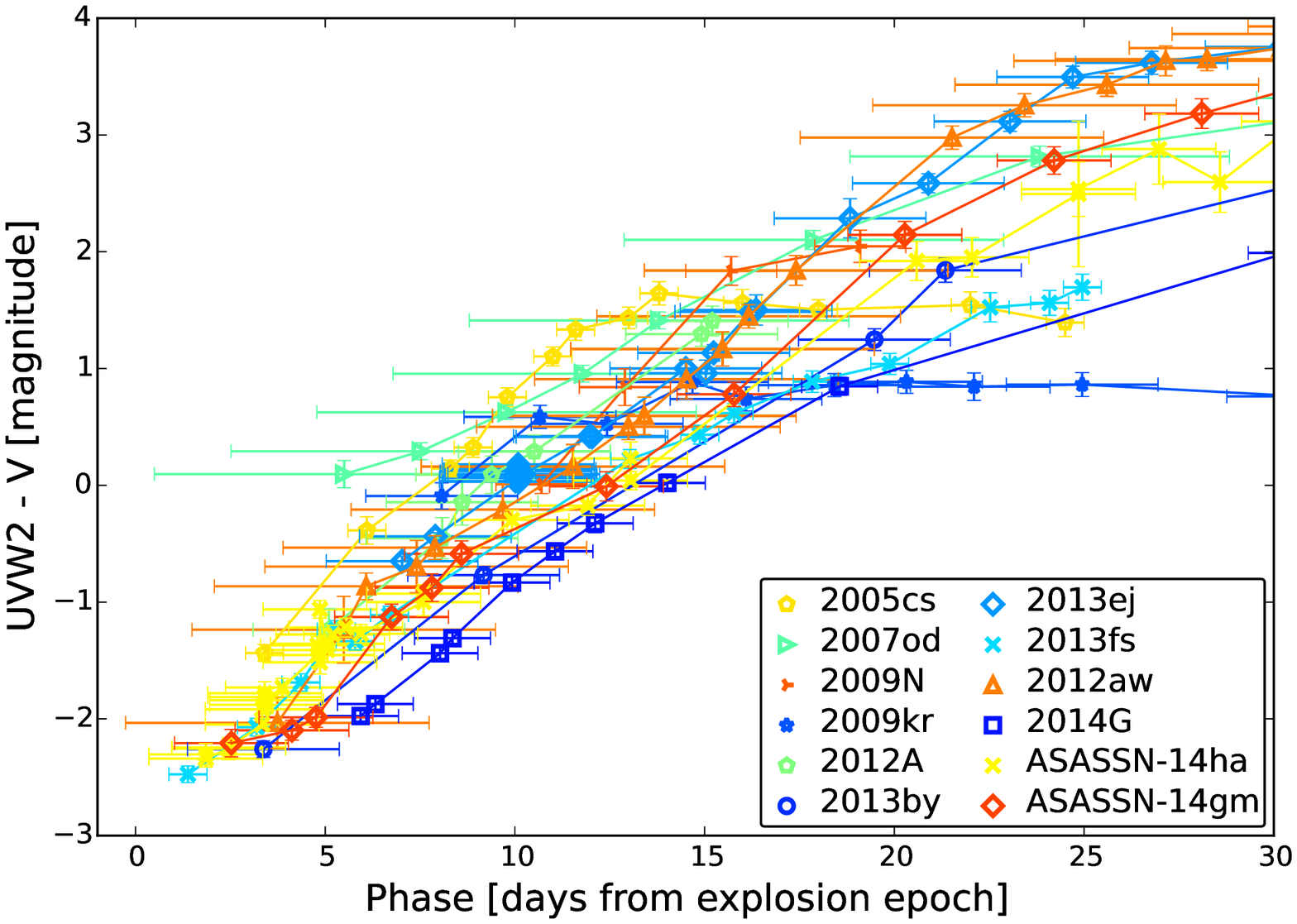}
  \caption{Colour  evolution at  early  phases for  a  sample of  SNe~II.
    Objects   from    the   literature  are  SN~2012A
    \citep{Tomasella2013},                                   SN~2012aw
    \citep{Bayless2013,Bose2013,Dall'Ora2014},               SN~2005cs
    \citep{Pastorello2005},   SN~2007od   \citep{2011MNRAS.417..261I},
    SN~2009N  \citep{Takats2013},  and SN~2006bp \citep{Brown2013a}. 
 The objects are  colour-coded with  respect to  the $V$-band  decline rate
over 50  days ($s_{50V}$).}
  \label{fig:colore}
\end{center}
\end{figure}

Plotted  in  Figure~\ref{fig:colore}  is  the  $UVW2$--$V$  broad-band
colour evolution  for a sample of  SNe~II.  All of these  objects have
been observed  in the UV with  {\it Swift} beginning soon  after their
early discovery.  A  strong change in colour is visible  in all of the
objects regardless of their IIP or IIL classification.  This behaviour
confirms what we have noticed  before: with good photometric precision
and knowledge of  their explosion epochs, SNe~IIP and  SNe~IIL seem to
be indistinguishable based on early-time UV data alone.

\subsection{Ultraviolet Spectra}
\label{sec:earlyspectra}

\begin{figure}
\begin{center}
  \includegraphics[width=8.4cm,height=6cm]{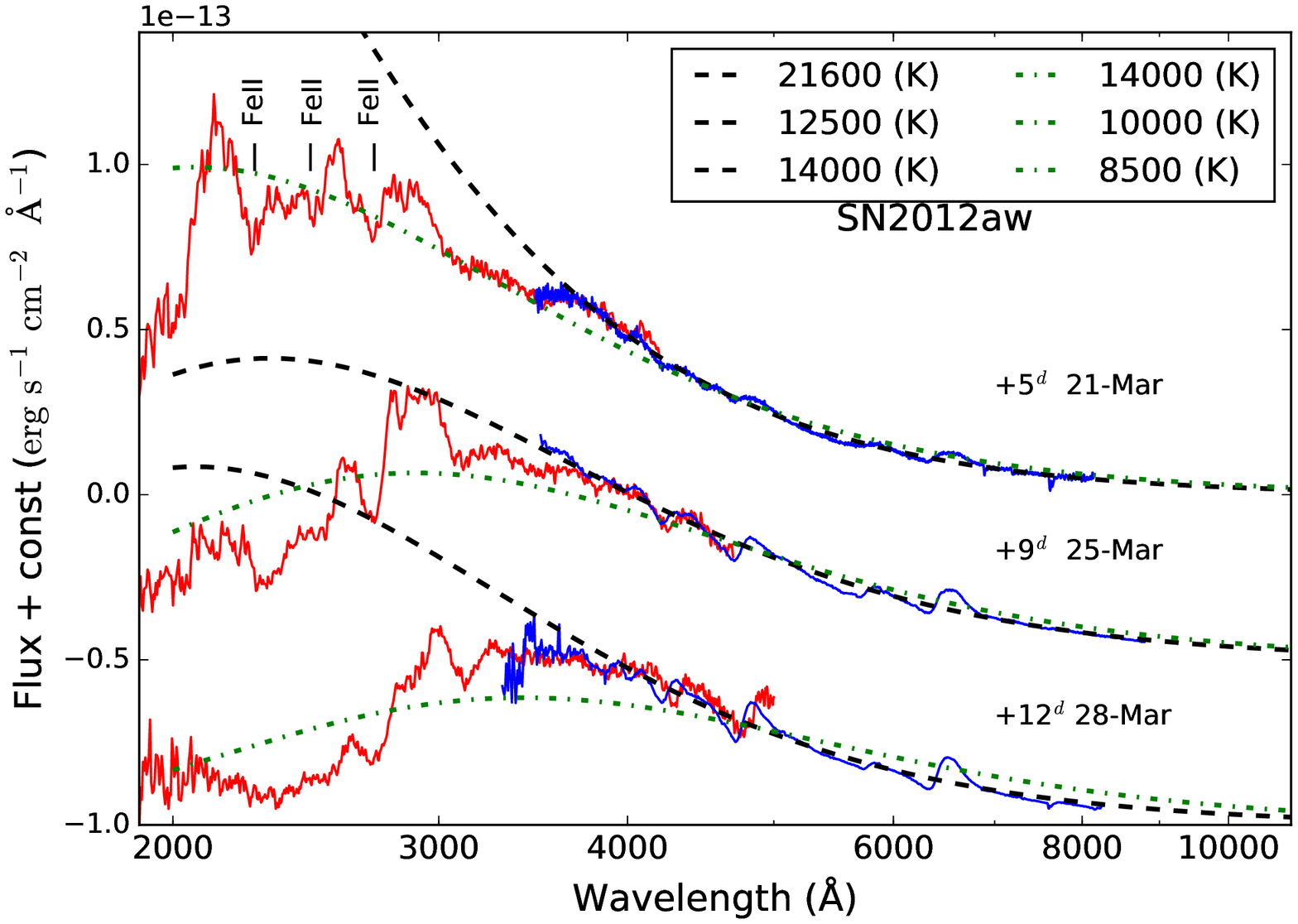}
    \includegraphics[width=8.4cm,height=6cm]{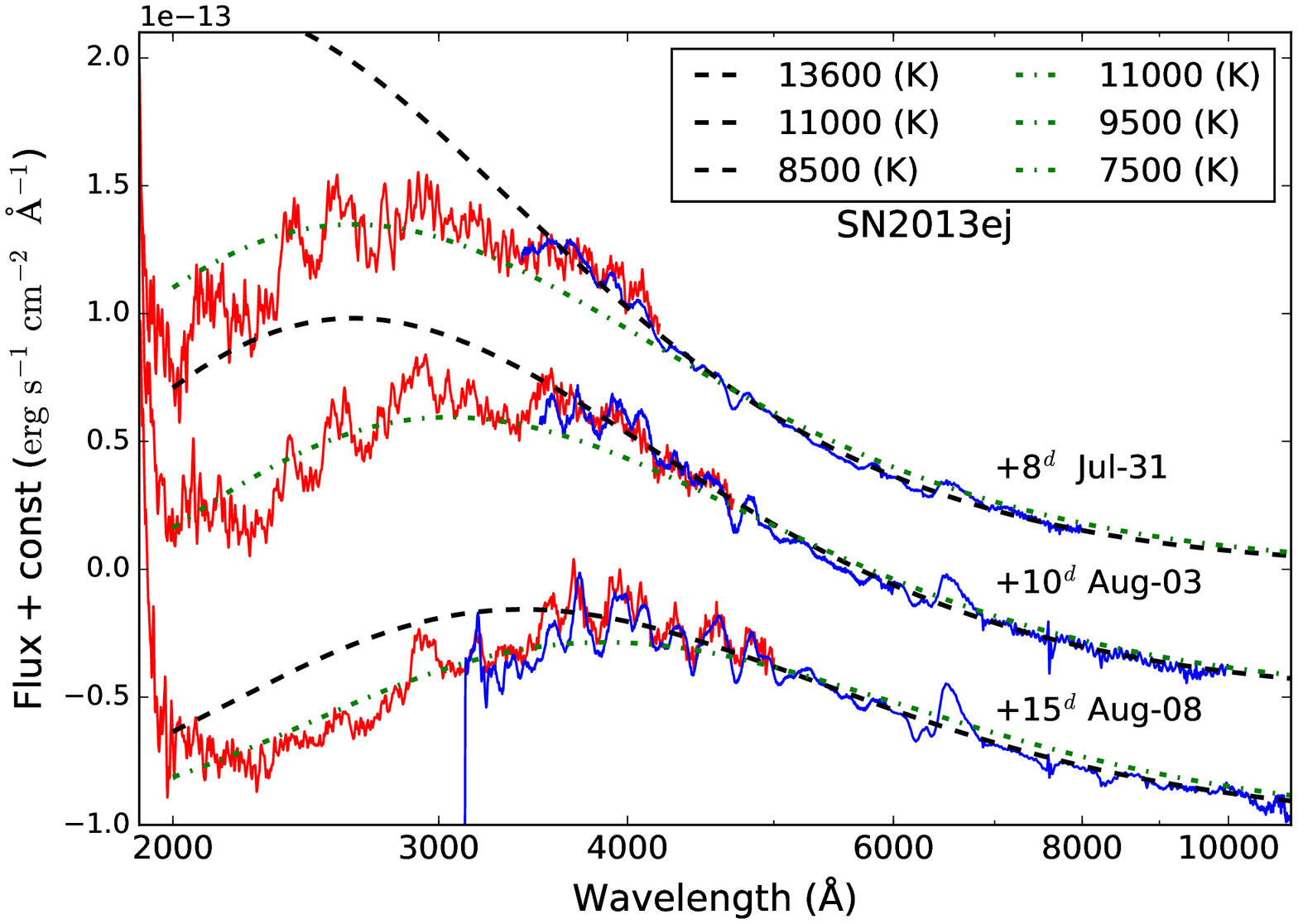}
  \caption{Early-time spectral evolution of  the Type~IIP SN~2012aw and the
    (almost) Type~IIL SN~2013ej. The red  portion of each spectrum
    corresponds  to  UV  coverage  obtained with  {\it  Swift},  while
    visual-wavelength   spectra at similar epochs  are   plotted   in
    blue. Black-body fits are overplotted  as dashed and  dotted lines
    (see text).}
  \label{fig:swiftspectra}
\end{center}
\end{figure}

In Figure~\ref{fig:swiftspectra}  we show {\it Swift}  UVOT UV spectra
of SN~2012aw (top  panel) and SN~2013ej (bottom  panel). The SN~2012aw
spectra have  previously been  published by  \cite{Bayless2013}, while
the    spectra   of    SN~2013ej    were    recently   presented    by
\cite{Dhungana2015}.  As SN~2013ej is  a transitional object between a
SN~IIP  and  a SN~IIL  ($s_{50V}  =  1.12$  mag  per 50  days),  while
SN~2012aw is a  more typical SN~IIP ($s_{50V}=0.26$ mag  per 50 days),
direct comparison between these  objects may highlight the differences
between these SN subtypes.

The  UV spectra  of these  two objects  are plotted  along with  their
corresponding  visual-wavelength spectra  obtained at  similar epochs.
In  both cases,  line  blanketing  is already  apparent  in the  first
spectra.  In  the optical, SNe~II  can be featureless at  early times.
However,  the UV  spectra exhibit  several deep  and broad  absorption
features     between    2000     \AA\     and     3000    \AA.      In
Figure~\ref{fig:swiftspectra}, we also show  the black-body fit to the
optical spectrum (black dashed line) and  to the combination of the UV
and  visual-wavelength spectra  (green  dot-dashed line).   Black-body
fits yield a  higher temperature if we use  only the visual-wavelength
spectra, as in Section~\ref{sec:earlycolor}.  We note that the UV flux
should be  used with  caution to  estimate the  black-body temperature
around  one week  past  explosion, since  line  blanketing is  already
prevalent at these epochs.

\begin{figure}
\begin{center}
  \includegraphics[width=8.4cm,height=6cm]{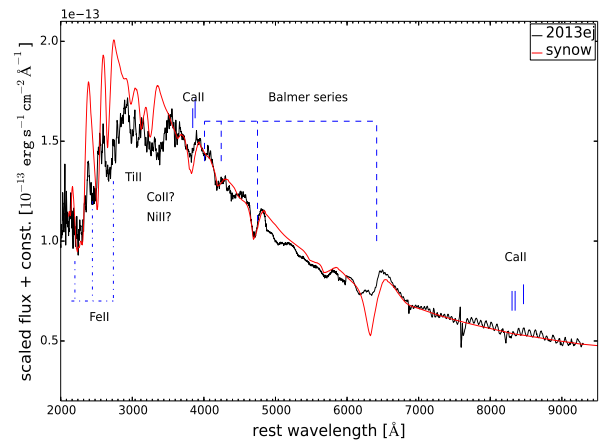}
  \caption{The {\tt SYNOW} synthetic spectrum is compared with the UV and 
   visual-wavelength spectrum of SN~2013ej.}
  \label{fig:synow13ej}
\end{center}
\end{figure}

We made use of {\tt SYNOW} \citep{2000PhDT.........6F} to identify the
lines  in the  early-time UV  spectra (see  Fig.~\ref{fig:synow13ej}).
\FeII{} and \TiII{} are the main ions contributing to the UV features.
\CoII{} or  \NiII{} may  contribute to the  absorption at  $\sim 3500$
\AA, although  it is difficult to  reproduce the depth of  the feature
without introducing  strong \CoII{}  and \NiII{} features  blueward of
2200~\AA.   \FeII{} has  also been  proposed by  \cite{Bayless2013} to
explain  the  line  blanketing  in   the  UV  spectrum  of  SN~2012aw,
confirming that  similar ions characterize  the UV spectra  of SNe~IIP
and IIL.   We note that  the use of  {\tt SYNOW} should  be approached
with  caution  owing to  limitations  in  atomic  line lists  and  its
one-dimensional  non-local-thermodynamic-equilibrium   nature,  making
correct line identifications in this wavelength region difficult.

\subsection{Rise Time}
\label{sec:rise}
The rise  time of  SNe~II has  been a matter  of debate.   Through the
comparison  of early-time  data of  PTF10vdl with  those of  SN~2006bp
\citep{Quimby2007}, \cite{Gal-Yam2011b} suggested that SNe~II may show
a diversity  in their  rise times.  More  recently, \cite{Valenti2014}
suggested a correlation between rise  time and luminosity, with slowly
rising SNe~II having  higher luminosities.  Given that  SNe~IIL are on
average more luminous than SNe~IIP, and several SNe~IIL exhibit a slow
rise, there could be a correlation between the rise time and $s_{50V}$
or between the rise time and the peak luminosity.

From a theoretical point of view, such a correlation may be important.
\cite{Gall2015},  following   the  equations   of  \cite{Rabinak2011},
pointed  out that  the  rise time  of SNe~II  is  proportional to  the
progenitor  radius to  the power  0.55,  while it  is only  marginally
dependent on the explosion energy and the ejected mass.  They measured
the rise  time for a sample  of 19 SNe~II, finding  a weak correlation
between rise time and SN type.  Specifically, SNe~IIL exhibit a longer
rise time than SNe~IIP.  In a recent paper, \cite{Gonzalez-Gaitan2015}
applied a similar analysis to SN~II samples from the Sloan Digital Sky
Survey  II (SDSS-II)  and the  Supernova Legacy  Survey (SNLS).   Such
surveys  offer  the  advantage  of having  a  homogeneous  cadence  of
observations, ideal  for doing rise-time studies.   Surprisingly, they
found  the  opposite correlation:  SNe~II  that  decline slowly  after
maximum light (SNe~IIP)  also exhibit a slower rise  to maximum, while
faster-declining  objects  (SNe~IIL) show  a  faster  rise to  maximum
light.  A  third study,  using a  sample of  SNe~II discovered  by PTF
\citep{2009PASP..121.1334R}, also find a weak correlation between rise
time and luminosity \citep{Rubin2015}.

\begin{figure}
\begin{center}
  \includegraphics[width=8.4cm,height=6.cm]{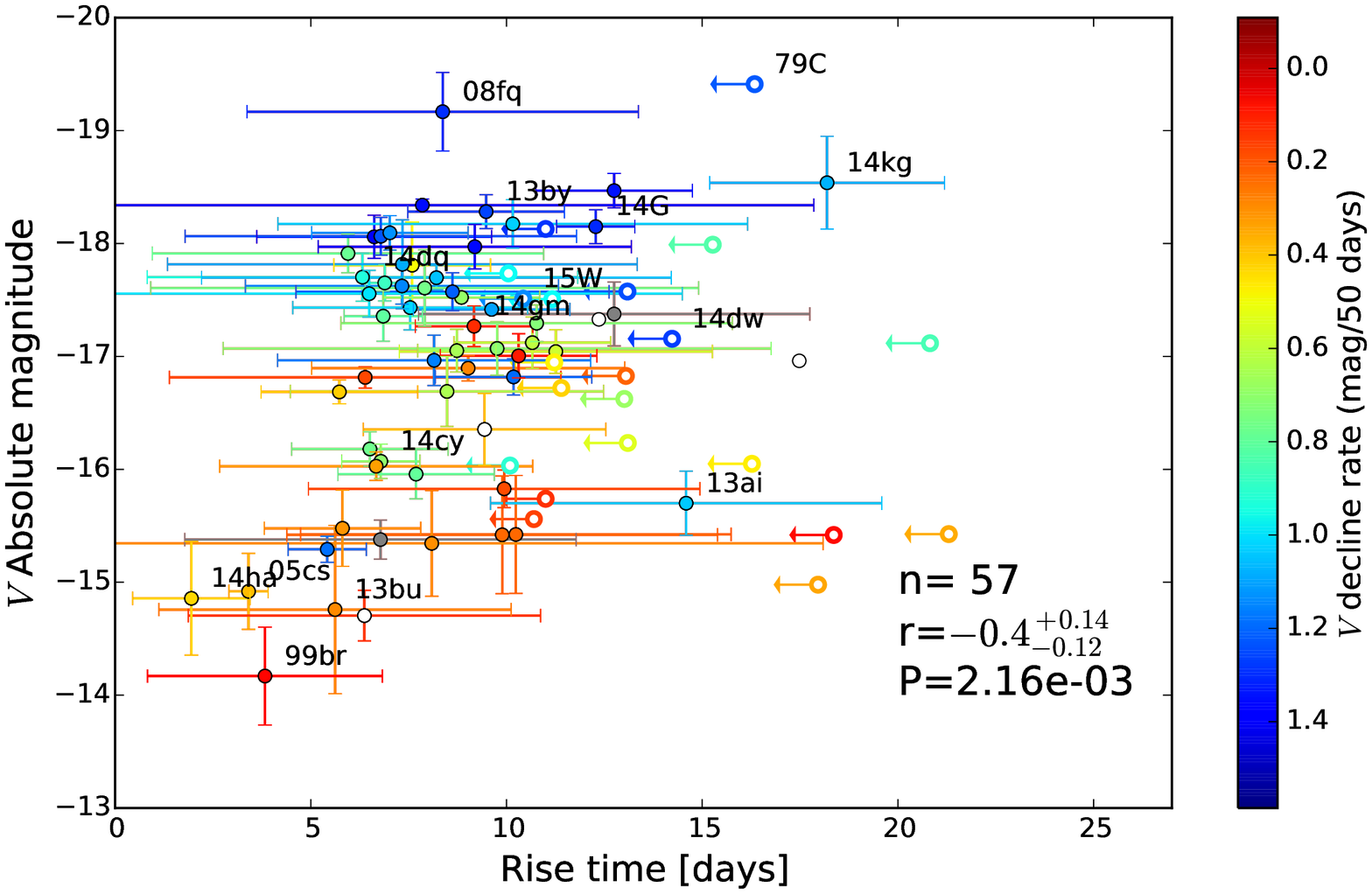}
    \includegraphics[width=8.4cm,height=6.cm]{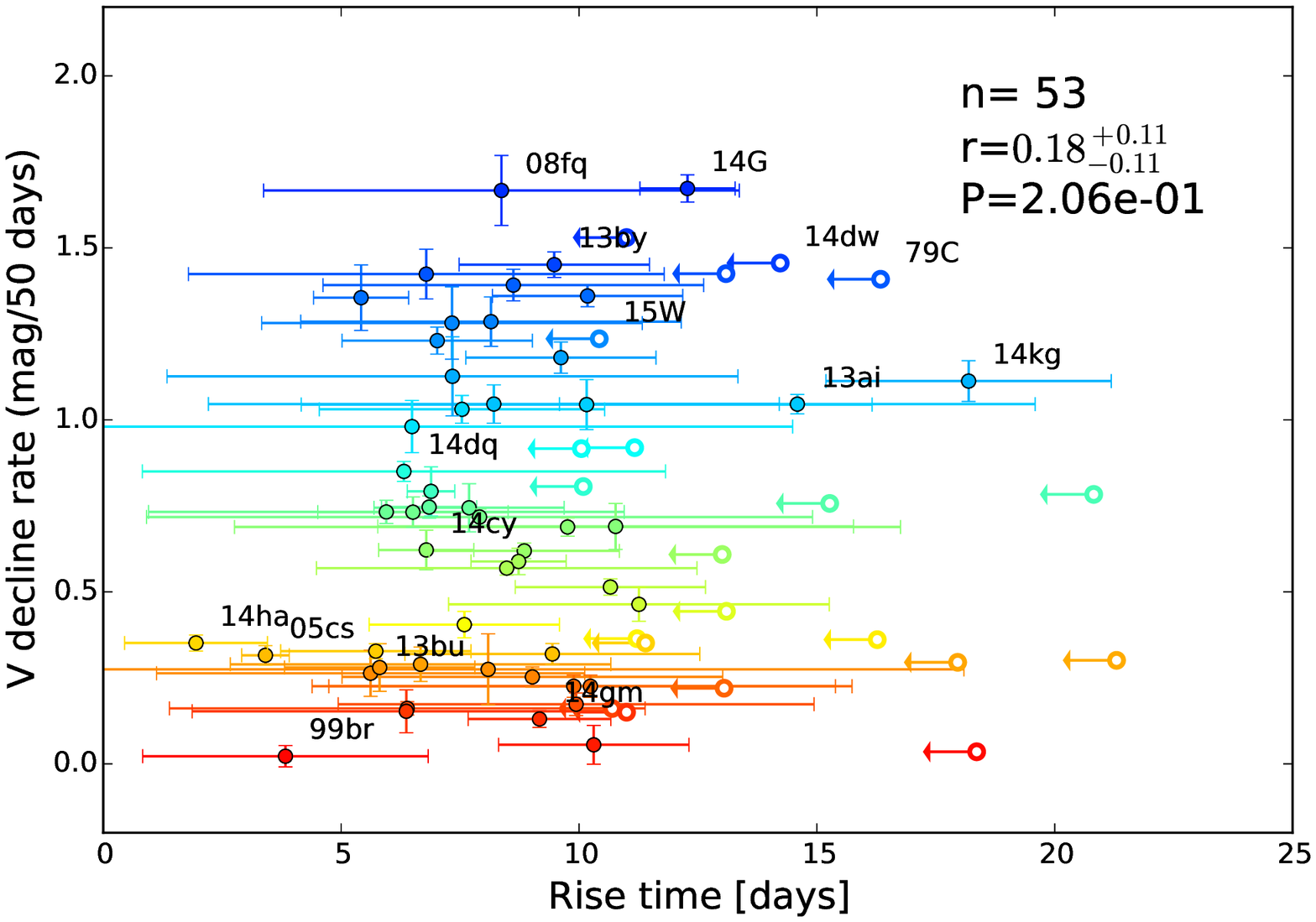}
  \caption{Absolute magnitude as a function  of rise time for a sample
    of SNe~II.  SNe~IIL are blue  and SNe~IIP are red.  For each panel,
    $n$  is  the  number  of  events,  $r$  is  Pearson's
    correlation coefficient,  and $P$ is  the probability of  detecting a
    correlation by chance.  The objects are  colour-coded with  
    respect to  the $V$-band  decline rate over 50  days ($s_{50V}$). }
  \label{fig:risetime}
\end{center}
\end{figure}

A few of  our SNe were discovered quite early  and may add information
to  this  puzzling  question.   We   applied  the  technique  used  by
\cite{Gall2015}, but excluded  SNe with no information  within 10 days
of explosion  (e.g., we excluded  SN~1979C for lack of  information on
the rise  time).  The rise time  is defined as the  difference between
the explosion epoch and the epoch when the light curve does not change
more than 0.01 mag per day.  In Figure \ref{fig:risetime} (top panel),
we show  the peak luminosity as  a function of the  computed rise time
for a  sample of SNe~II.  They are colour-coded by  $s_{50V}$: SNe~IIL
are blue, SNe~IIP are red.  SNe  with no information within 10 days of
explosion are reported as upper limits.

We find that the correlation between rise time and luminosity is weak,
with  a  Pearson's  correlation  coefficient  of  $-0.41$.   The  weak
correlation seems  to be caused  by the lack of  low-luminosity SNe~II
having a long rise.

As  a next  step, we  investigate the  rise time  versus the  $V$-band
decline  rate (Fig.  \ref{fig:risetime},  bottom  panel).  Again,  the
correlation is weak  or absent.  There are a few  objects that clearly
have a  long rise.   \cite{Gonzalez-Gaitan2015} speculated  that these
slow risers may  have a different physical origin. However,  we do not
see a clear separation from other SNe~II.

In summary, using our sample, the correlation found by \cite{Gall2015}
is  not  statistically  significant  (this  is  also  pointed  out  by
\citealt{Rubin2015}).     The    inverse    correlation    found    by
\cite{Gonzalez-Gaitan2015} is  caused by the fact  that their analysis
is computed  using all filters ($g$,  $r$, $i$, $z$) at  the same time
(private  communication).   However,  \cite{Gonzalez-Gaitan2015}  have
also shown  that rise  and decay  times are  dependent on  filters.  A
reanalysis of  the SNLS and  SDSS data  for each filter  is consistent
with  the absence  of  a clear  correlation (Gonzalez-Gaitan,  private
communication).

\section{Type II Supernovae at 50 days }
\label{sec:50}

\begin{figure}
\begin{center}
    \includegraphics[width=8.4cm,height=6.cm]{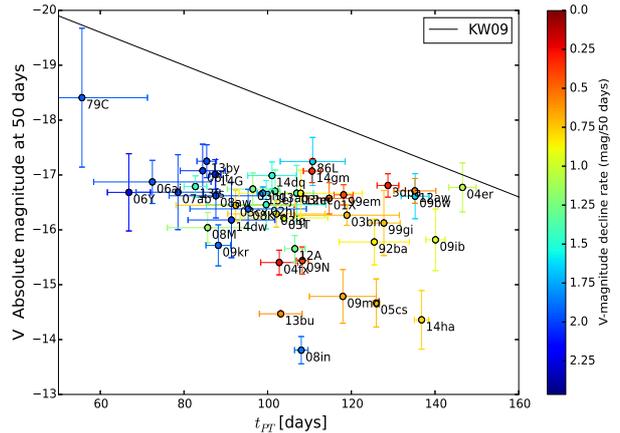}
  \caption{Absolute $V$ magnitude 50 days after explosion vs. plateau length. Objects are colour-coded with the 
$V$-band decline rate in 50 days. The black line is from Eq. 18 of \protect\cite{2009ApJ...703.2205K}.}
  \label{fig:magtpt50}
\end{center}
\end{figure}

At 50  days after  explosion, the photosphere  has receded  within the
outer 0.5  \msun{} of  the ejecta  \citep{Dessart2010}, and  the light
curve is mainly powered by hydrogen recombination.  This phase is used
by  several  authors   to  study  the  SN~II   diversity.   In  Figure
\ref{fig:magtpt50} we  investigate the relation between  luminosity at
50 days after explosion and length  of the plateau, comparing our data
with Eq. 18  of \cite{2009ApJ...703.2205K}. A weak  relation is driven
by  the fact  that faint  SNe  tend to  have  a long  plateau, and  by
SN~1979C  which was  still very  bright 50  days after  explosion. The
data,   all   fall   below   the   line   representing   Eq.   18   of
\cite{2009ApJ...703.2205K}.    However,    the    models    used    in
\cite{2009ApJ...703.2205K}  are  known  to  overestimate  the  plateau
luminosity \citep{Dessart2010}.

\begin{figure}
\begin{center}
    \includegraphics[width=8.4cm,height=6.cm]{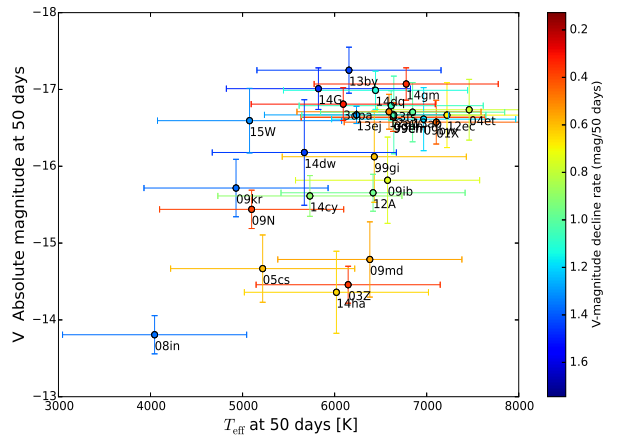}
  \caption{Absolute $V$ magnitude 50 days after explosion vs. $T_{\rm eff}$. Objects are color-coded with the 
$V$-band decline rate in 50 days.}
  \label{fig:teff50}
\end{center}
\end{figure}

At 50  days, the  red part of  the spectrum can  be approximated  by a
black body, while iron-group-element  line blanketing reduces the flux
at  the blue  end \citep{2009ApJ...703.2205K}.   We used  bands redder
than $V$  (inclusive) to measure  an effective temperature at  50 days
after explosion.  We correlate this temperature with the luminosity at
50 days  in Figure \ref{fig:teff50},  color-coding the objects  by the
$V$-magnitude  decline   rate.  Brighter  objects  exhibit   a  higher
temperature    and   IIL-like    SNe    have   systematically    lower
temperatures.  This is  probably expected,  since IIL-like  SNe evolve
faster  and at  50  days are  already  close to  the  end of  hydrogen
recombination.

\section{Nickel Mass in Type II Supernovae}
\label{sec:nickel}

Recent studies of  SNe~II have shown that SNe~IIL are  on average more
luminous       at       peak       brightness       than       SNe~IIP
\citep{Anderson2014,Faran2014,Sanders2015},   though    whether   they
produce more  $^{56}$Ni has not  been demonstrated.  A  clear relation
between  $^{56}$Ni  content  and  plateau luminosity  has  been  found
\citep{Hamuy2003,Spiro2014}.   \cite{Valenti2015} showed  that SNe~IIL
may  follow the  same relation,  although they  increase the  scatter.
However,  \cite{Pejcha2015a}  have  pointed out  that  luminosity  and
$^{56}$Ni content  are strongly correlated, reducing  the significance
of previous claims.

Theoretically,  \cite{Popov1993}  and \cite{2009ApJ...703.2205K}  have
shown that an excess in production of $^{56}$Ni may actually produce a
longer plateau.   Recently, \cite{Nakar2015} suggested  that the typical 
$^{56}$Ni contribution to the time weighted integrated luminosity during 
the photospheric phase is 30 \%. This  may be a  important factor  
in the morphology  of SN~IIP light  curves.  

Most of  our objects were  monitored until they reached  the $^{56}$Co
tail, providing an excellent opportunity to better understand how much
$^{56}$Ni is produced  in SNe~II.  Using SN~1987A as  a reference, the
$^{56}$Ni   mass  is   estimated   using  the   method  described   by
\cite{Spiro2014}.   To do  this, the  pseudobolometric light  curve of
SN~1987A is compared with those of  our sample when the objects are on
the radioactive  tail ($t>  t_{PT}$).  Note that  the pseudobolometric
light curve of  SN~1987A used for comparison was  constructed with the
same limits of integration\footnote{$(U)BVRI$ or $(U)BVgri$, depending
  on the available bands} and over  similar epochs as was done for the
objects in our sample.  The comparison of SN~1987A to any given object
yields a $^{56}$Ni mass estimate following the relation

\begin{equation}
M({^{56}{\rm Ni}}) = 0.075~{\rm M}_{\odot} \times \frac{L_{\rm SN}(t)}{{L_{\rm 87A}(t)}} .
\end{equation}

We next  investigate four  questions: (1) Do  more-massive progenitors
produce SNe~IIL?  (2) Is the  amount of $^{56}$Ni  correlated
with the luminosity  decay rate?  (3)
Is the $^{56}$Ni content correlated with the length of the plateau, as
claimed  by \cite{Popov1993}  and \cite{2009ApJ...703.2205K}?   (4) Do
more-luminous SNe~II produce more $^{56}$Ni?

\begin{figure}
\begin{center}
    \includegraphics[width=8.4cm,height=6.cm]{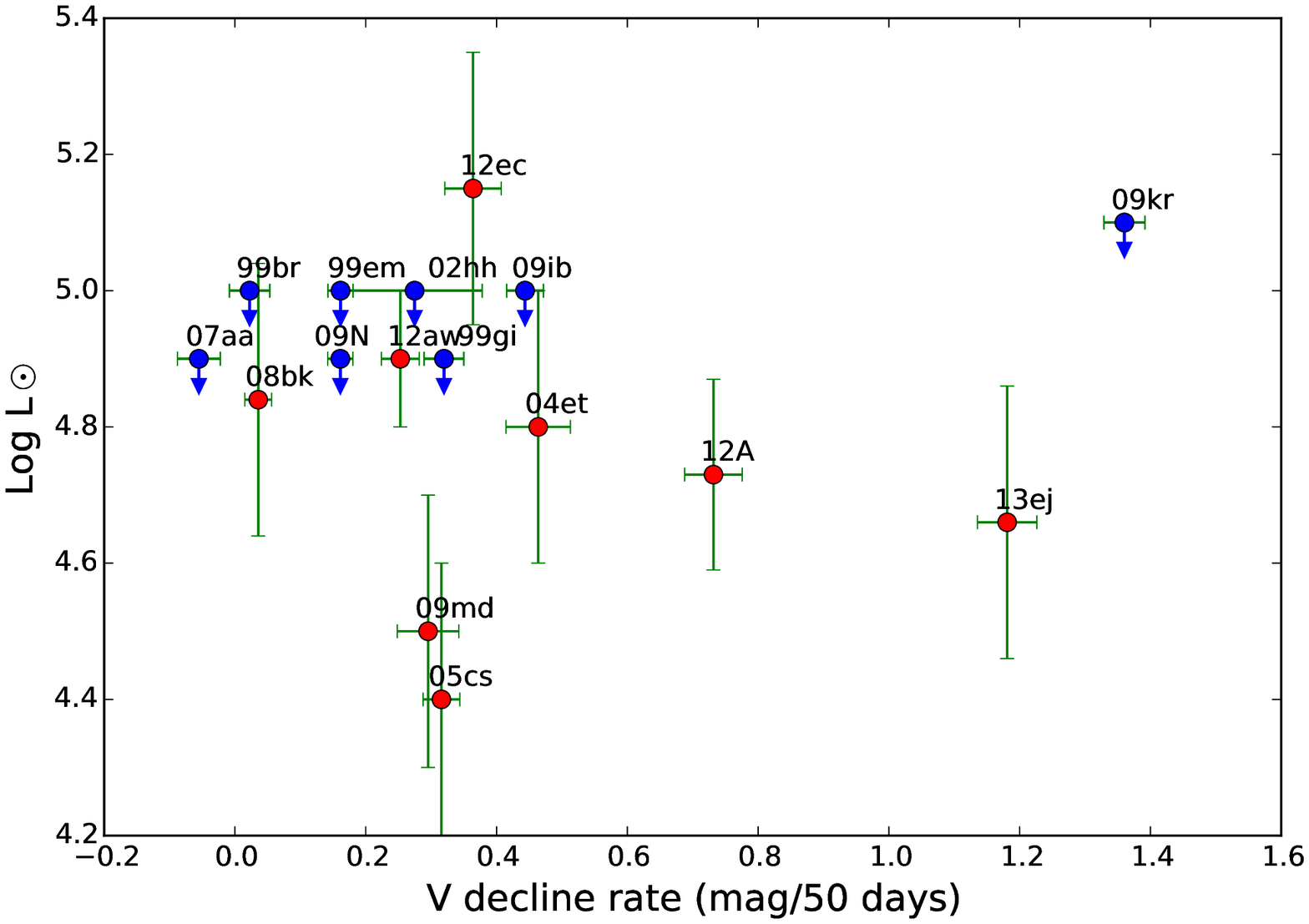}
  \caption{Progenitor luminosity from pre-explosion
    detection vs. the $V$-band decline rate in 50 days. Progenitor data are from
    \protect\cite{Smartt2009}, \protect\cite{Elias-Rosa2011}, \protect\cite{2011MNRAS.417.1417F},
       \protect\cite{Fraser2013a}, \protect\cite{Maund2015}, and   \protect\cite{VanDyk2012}.}
  \label{fig:nickel2}
\end{center}
\end{figure}
\begin{figure}
\begin{center}
  \includegraphics[width=9cm,height=7.cm]{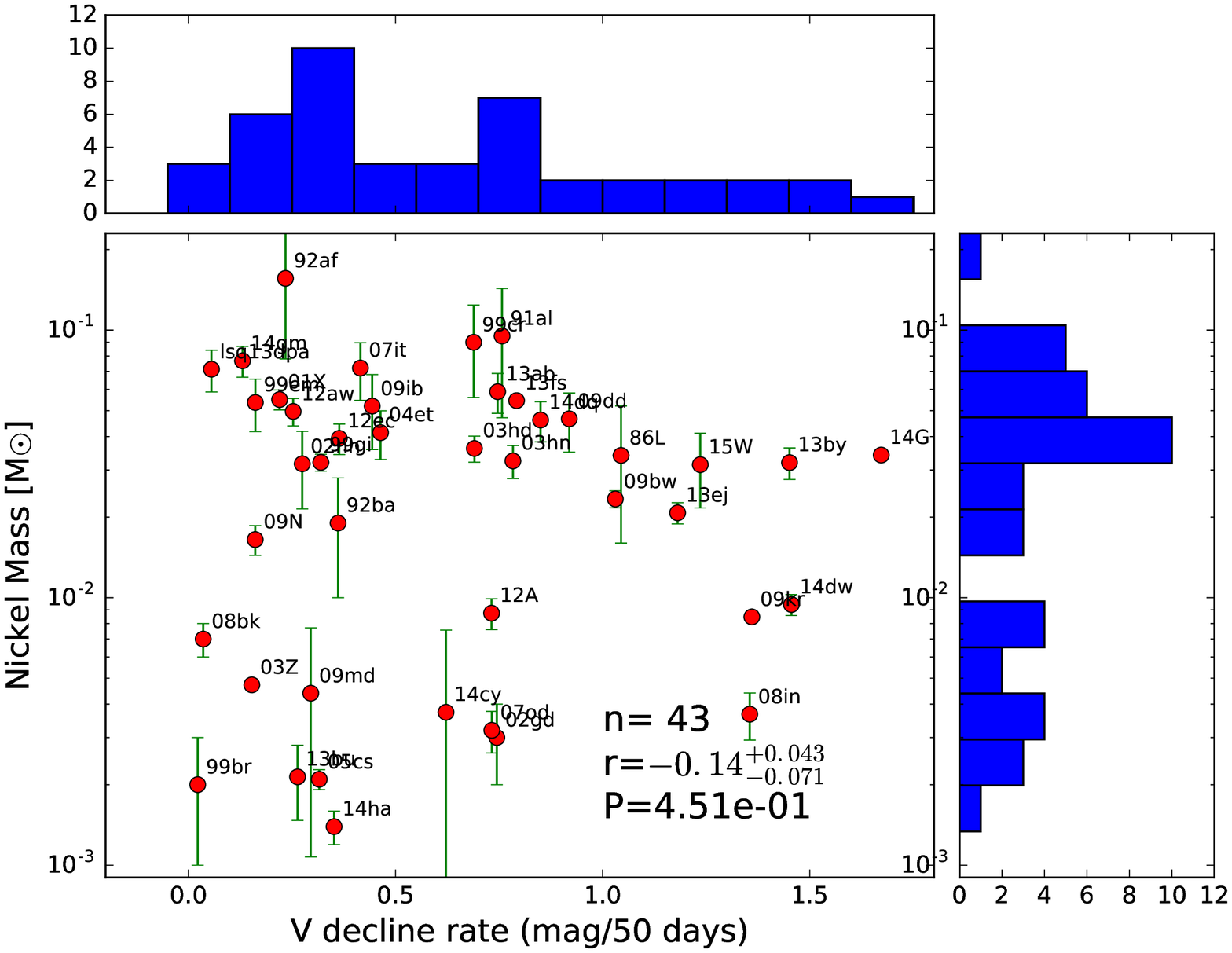}
  \caption{$^{56}$Ni  mass for  a  sample of  SNe~II vs.  the
    $V$-band decline rate in 50 days for a sample of SNe~II.  The data
    for the  sample include  SNe~II from  \protect\cite{Hamuy2003} and
    \protect\cite{Spiro2014}.   For  each panel,  $n$  is  the number  of
    events,  $r$ is  Pearson's correlation  coefficient,
    and $P$ is the probability of detecting a correlation by chance. }
  \label{fig:nickel3}
\end{center}
\end{figure}
\begin{figure}
\begin{center}
    \includegraphics[width=9cm,height=7.cm]{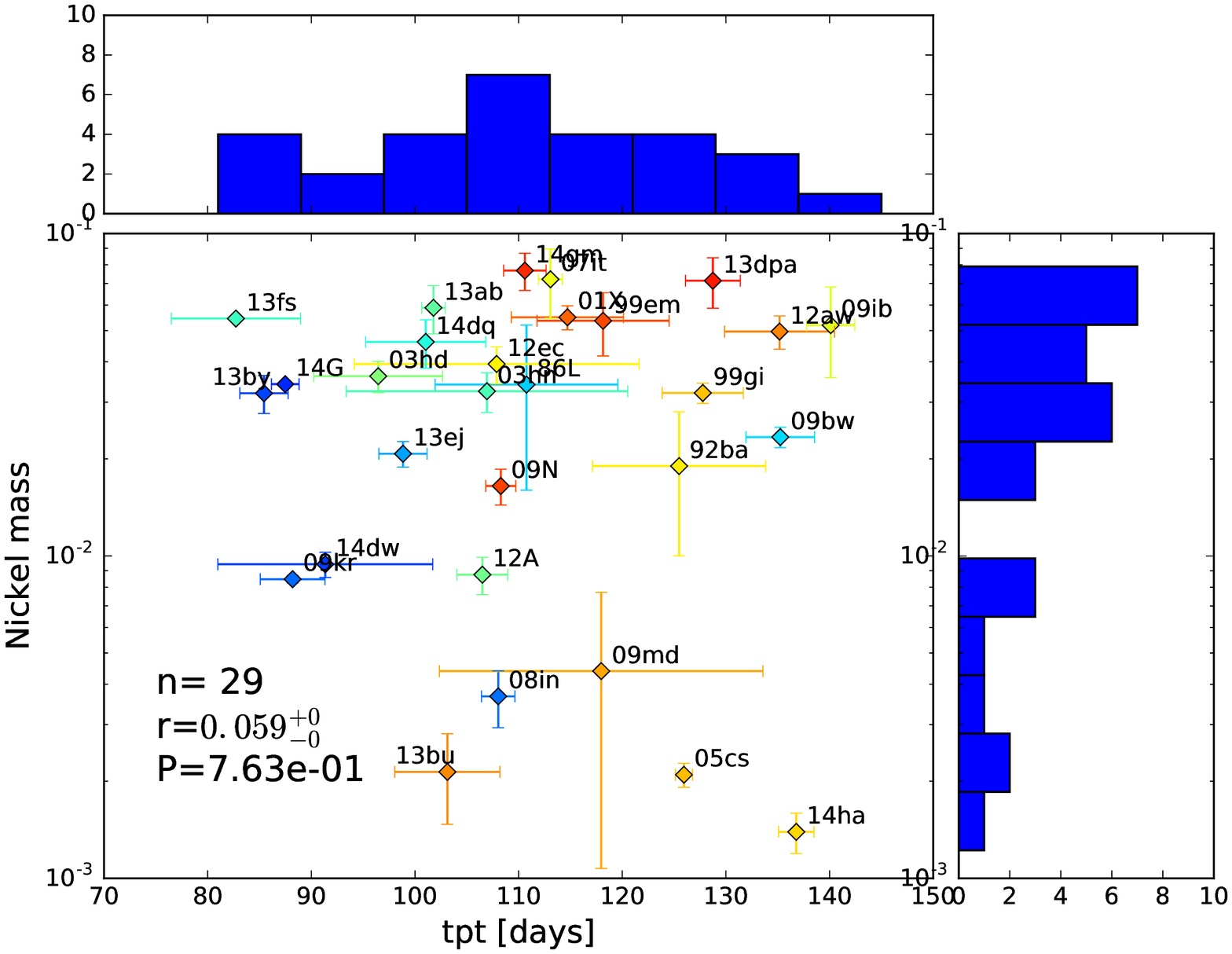}
  \caption{$^{56}$Ni  mass as  a function  of time  of the  transition
    between  the plateau  and radioactive  tail; $n$  is the  number of
    events, $r$ is  Pearson's correlation  coefficient,
    and $P$ is the probability of detecting a correlation by chance.  
    The objects are  colour-coded with respect to  the $V$-band  
    decline rate over 50  days ($s_{50V}$).}
  \label{fig:nickel4}
\end{center}
\end{figure}

\begin{figure}
\begin{center}
    \includegraphics[width=8.4cm,height=6.cm]{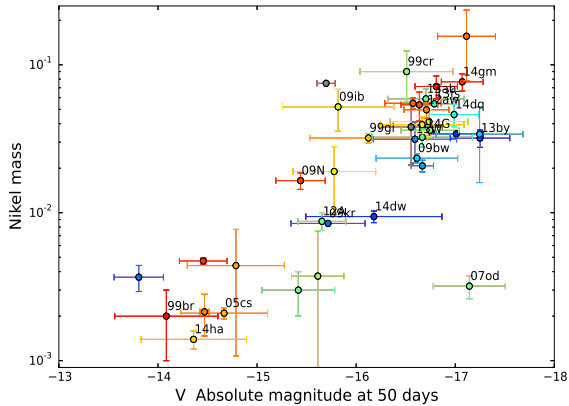} 
  \caption{$^{56}$Ni mass as a function of the absolute $V$-band magnitude at 50 days after maximum
 The objects are  colour-coded with respect to  the $V$-band  decline rate over 50  days ($s_{50V}$).}
  \label{fig:nickel5}
\end{center}
\end{figure}

The current dataset suggests that the  answer to the first question is
negative.  However, the  number of SNe~II with  a progenitor detection
in archival images and a measurement  of the decay rate on the plateau
is still  limited.  With the data  available today, we do  not see any
evidence that  SNe~IIL are related  to a more-massive  progenitor (see
Fig.~\ref{fig:nickel2}). We will  come back to this issue  in the next
section.

For the second question, we have a statistically significant sample of
SNe~II (see Fig.  \ref{fig:nickel3}) and the lack of  a correlation is
robust.  

There is no correlation between $^{56}$Ni mass and the {\it length} of
the  plateau  (see  Fig.~\ref{fig:nickel4}).  A  possible  explanation
comes     from     the      models     of     \cite{Popov1993}     and
\cite{2009ApJ...703.2205K},  which  show   that  $^{56}$Ni  begins  to
contribute to the plateau length  only when large amounts ($\geq 0.06$
M$_{\odot}$) of  $^{56}$Ni are  produced.  For  SNe that  produce less
$^{56}$Ni, the  effect is  usually negligible.   The number  of SNe~II
that produce  more than 0.06  \msun\ of  $^{56}$Ni is very  small.  We
note that the length of the plateau  also depends on the energy of the
explosion   and   the   initial    radius   \cite[see   Eq.    11   of
][]{2009ApJ...703.2205K}.  This  may partially contribute to  the lack
of a correlation between $^{56}$Ni mass and the length of the plateau.
Fig.~\ref{fig:nickel3} and Fig~ \ref{fig:nickel4}  show however a lack
of SNe that produce $10^{-2}$ M$_{\odot}$) of $^{56}$Ni.

Fig.~\ref{fig:nickel5},          confirm          the          finding
\citep{Hamuy2003,Spiro2014,Pejcha2015a}  that   brighter  SNe  produce
larger $^{56}$Ni masses.  However the scatter in the  relation is also
suggesting that  a simplistic two-dimensional parameter  space plateau
luminosity - $^{56}$Ni mass is  probably an over simplification of the
problem.

\section{Nebular Spectra}
\label{sec:nebular}

An alternative  way to constrain  the nature  of SNe~II is  trough the
study  of their  late-phase  ($> 200$  days  after explosion)  nebular
spectra.  At this phase, SN~II  ejecta have sufficiently expanded that
they       become       transparent       to       optical       light
\citep[e.g.,][]{1989ApJ...343..323F,2009MNRAS.397..677T,2012MNRAS.420.3451M}.

The nucleosynthesis  during the last phases  of massive-star evolution
produces a  carbon-oxygen core with  an elemental composition  that is
strongly dependent on mass. For example, the amount of oxygen produced
in  stellar evolutionary  models varies  from 0.2  to 5~M$_\odot$  for
progenitors between 10 and 30~M$_\odot$ (e.g., \citealt{WOOSLEY2007}).
These differences should be  visible (see \citealt{Jerkstrand2014a} for
a discussion), and spectral synthesis modeling has been widely used to
constrain  the   abundance  of   different  elements  in   the  ejecta
\citep{2007ApJ...670..592M,      2010MNRAS.408...87M,     Dessart2010,
  Dessart2013, Jerkstrand2012, Jerkstrand2014,  Jerkstrand2015} and to
elucidate the nature of the progenitor.

\cite{Jerkstrand2011, Jerkstrand2012} developed quantitative method of
taking the ejecta  stratification structure of a stellar  model at the
moment  of  explosion  (e.g., the  models  of  \citealt{WOOSLEY2007}),
carrying  out  a detailed  calculation  of  the energy  deposition  of
gamma-rays/positrons by radioactive isotopes,  and combining this with
radiative-transfer calculations to  produce synthetic nebular spectra.
The computed synthetic spectra can  then be compared with the observed
spectra at different phases.

By  analyzing  the  synthetic   spectra  from  different  progenitors,
\cite{Jerkstrand2014a} found that the intensities of specific lines are
correlated   with  the   progenitor   mass.    In  particular,   [\OI]
$\lambda\lambda$6300, 6364 is a fare indicator of the progenitor mass.
By  measuring the  [\OI] intensity,  \cite{Jerkstrand2015} found  that
most SNe~II  are consistent with  progenitor stars with mass  $\le 17$
\msun.   This is  consistent  with progenitor  studies using  archival
prediscovery images: SNe~II likely come  from progenitors in the range
8--16  \msun\ \citep{Smartt2009,  Smartt2015a}.  As  noted previously,
the lack  of SN~II progenitor  detections with large masses  is called
the RSG  problem. Nebular-phase spectroscopy allows  us to investigate
whether SNe~IIL  come from more-massive progenitors  than SNe~IIP.  We
stress, however, that this type of nebular-spectrum analysis relies on
the assumption  that SNe~II  have roughly  similar values  for certain
parameters (e.g., explosion energy, mixing, mass loss).

We  collected six  spectra for  three of  the objects  in our  sample.
These  spectra  are   shown  in  Figure~\ref{fig:nebularspectra},  and
detailed information is reported in Appendix \ref{sec:specrredu}.

\begin{figure}
\begin{center}
    \includegraphics[width=8.4cm,height=6cm]{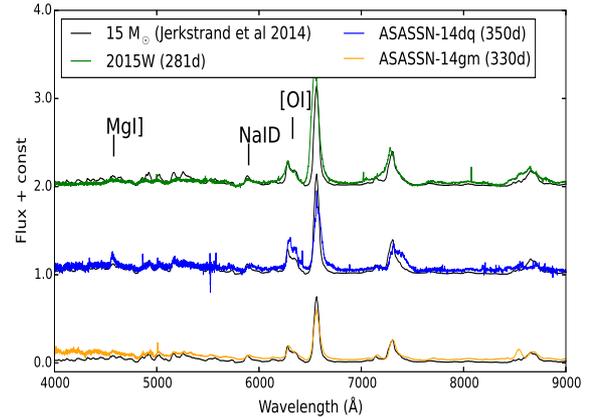}
  \caption{Nebular spectra of three SNe~II from our sample. The intensity of [\OI{}] in all 
  three SNe is consistent with the \citealt{Jerkstrand2015} model of 15 \msun.}
  \label{fig:nebularspectra}
\end{center}
\end{figure} 

\begin{figure}
\begin{center}
    \includegraphics[width=8.4cm,height=6cm]{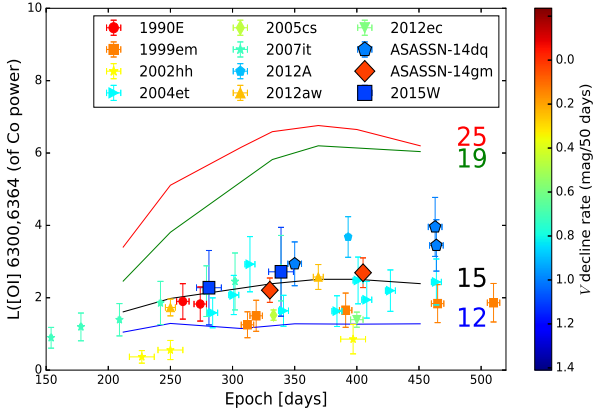}
  \caption{The [\OI] $\lambda\lambda$6300, 6364 luminosity (normalized to the
    $^{56}$Co decay power) for the \protect\cite{Jerkstrand2015}
    sample of SNe and three objects from our sample. Objects are 
    colour-coded with  the  $V$-band  decline rate $s_{50V}$.}
  \label{fig:nebularmass}
\end{center}
\end{figure}

We   show  the   Jerkstrand  et   al.   (2015b)   diagram  in   Figure
\ref{fig:nebularmass} with three more  objects added.  ASASSN-14dq and
    SN~2015W  are  two  fast-declining SNe~II  and  the  oxygen
    profiles in their spectra are consistent with progenitors of $\sim
    15$~\msun,     similar     to      other     SNe~IIP.      Figures
    \ref{fig:nebularspectra} and \ref{fig:nebularmass} show that these
    two SNe~II  (which would be  reasonably classified as  SNe~IIL) do
    not    show    significant    differences    compared    to    the
    15~\msun\  models. In  particular, the  oxygen line  strengths are
    clearly not greater  than either the model spectra or  the bulk of
    the   SNe~IIP   observed   to   date    (see   also   Figure
      \ref{fig:nebular2}).   While a larger sample is required before
    making definitive  conclusions, the  simple interpretation  of our
    data  is that  SNe~IIL come  from progenitors  having core  helium
    masses that  are no more  massive than those of  SNe~IIP.  Whether
    massive RSGs can explode as  SNe~IIL remains an open question, but
    there is no evidence from our  data that they come from stars more
    massive  than $\sim  17$ \msun,  and so  we cannot  solve the  RSG
    problem elucidated by \cite{Smartt2009}.

\section{Discussion and Conclusions}
\label{sec:discussion}
We have  presented extensive multiband  photometry for a sample  of 16
SNe~II.  The light curves follow the  flux evolution of our sample out
to the radioactive  $^{56}$Ni decay tail.  Main  parameters (e.g., the
rise time to peak brightness, the plateau length, and the decay slope)
are estimated from both the $V$-band light curves and pseudobolometric
light curves.  These results are  placed into context with an expanded
sample drawn from the literature.

Our main findings can be summarized as follows.
\begin{itemize}
\item  SNe~II exhibit  some differences  in their  light-curve slopes,
  plateau lengths, and absolute magnitudes.  However, we do not find a
  clear  separation  between these  subtypes  that  would justify  the
  historical  classification  into   distinct  Type~IIP  and  Type~IIL
  subclasses.
\item  SNe~IIP  and SNe~IIL  do  not  exhibit significant  temperature
  differences at  early phases. This suggests  that SN~IIL progenitors
  do  not  have  systematically  larger radii,  as  confirmed  by  the
  early-time colour evolution. We also find that temperature evolution
  studies  at  early  phases  have several  potential  pitfalls.   The
  approximation by a black-body function  is problematic even at a few
  days after explosion,  as strong line blanketing in  the UV strongly
  effects the SED.
\item  We confirm  that there  is  a correlation  between the  plateau
  length and  the slope of the  light curve after maximum  light.  The
  correlation is significant with the $s_{50V}$ parameter (in $V$) and
  with the $S2$ parameter (using the pseudobolometric light curve).
\item SNe~II have a drop from  the plateau to the radioactive decay of
  1.0--2.6 mag, which  appears to be independent of the  length of the
  plateau confirming  that brighter plateau have  more $^{56}$Ni.  The
  three SNe with a large drop (3--4 mag) are low-luminosity SNe.
\item We measure the rise time of SNe~II using the approach adopted by
  \cite{Gall2015}.  We find  a weak correlation between  rise time and
  luminosity,  while no  correlation is  found between  rise time  and
  light-curve decline.
\item  We   investigate  the   possibility  that  SNe~IIL   come  from
  more-massive progenitors. Despite the limited  amount of data on the
  luminosity of the  progenitors and the number of  nebular spectra of
  SNe~IIL, so far there is no conclusive evidence that the mass of the
  progenitors of SNe~II correlates with  the slope of the decline rate
  after maximum light.
\item We  find that the  plateau luminosity  at 50 days  after maximum
  correlates  with   the  plateau   length  and  with   the  effective
  temperature.
\item Despite the fact that SNe~IIL  are on average more luminous than
  SNe~IIP \citep{Patat1994,Li2011},  there is  no clear  evidence that
  the decline rates  of SNe~II correlate with the  amount of $^{56}$Ni
  observed.
\end{itemize}

Finally, our  analysis (in  $V$ and  pseudobolometric light  curve) is
consistent    with    the    result   of    \cite{Anderson2014}    and
\cite{Sanders2015}:   SNe~IIP  and   SNe~IIL  cannot   be  comfortably
separated in two different classes.   Similar investigations in $R$ or
$I$ are, however, still limited by  the small number of SNe published,
but  so  far, are  in  agreement  with the  result  in  $V$ band  (see
Fig. \ref{fig:slope} and \citealt{Galbany2015}).

We find no clear evidence that  SNe~II with a fast light-curve decline
(SNe~IIL) originate from a more massive progenitor.  While SNe~IIP and
SNe~IIL  cannot be  comfortably  separated in  two different  classes,
faint  SNe~II  behave  often  differently  from the  rest  of  SNe  II
(eg. plateau  length and $^{56}$Ni  synthesized). However the  lack of
SNe between \emph{faint} and \emph{normal} SNe~II may still due to the
small sample of SNe.

The construction  of a larger sample  of SN~II progenitors and  a more
thorough statistical study of their  late-time oxygen lines (used as a
progenitor-mass tracer)  is required to increase  our understanding of
SNe~II progenitors.

\section*{Acknowledgements}
The authors acknowledge  the ASASSN, La Silla Quest,  and LOSS surveys
for discovering new SNe that  made this study possible.  This material
is based upon work supported  by the National Science Foundation (NSF)
under  Grant No.  1313484.   M.D.S.  gratefully acknowledges  generous
support provided by  the Danish Agency for Science  and Technology and
Innovation  realized through  a Sapere  Aude Level  2 grant.   M.F. is
supported by the European Union FP7 programme through ERC grant number
320360.   S.J.S.  acknowledges  funding  from  the  European  Research
Council  under  the  European   Union's  Seventh  Framework  Programme
(FP7/2007-2013)/ERC  Grant  agreement  No. [291222]  and  STFC  grants
ST/I001123/1  and  ST/L000709/1.  A.V.F.'s  group at  UC  Berkeley  is
grateful  for financial  assistance  from NSF  grant AST-1211916,  the
TABASGO  Foundation, Gary  and  Cynthia Bengier,  and the  Christopher
R. Redlich  Fund.  This  work was  supported by  the NSF  under grants
PHY-1125915   and  AST-1109174.    M.S.   acknowledges  support   from
EU/FP7-ERC grant  no [615929].   This paper  is based  on observations
made with the  {\it Swift}, LCOGT, Gemini, and  Keck Observatories; we
thank their respective staffs for excellent assistance. The W. M. Keck
Observatory  is  operated  as   a  scientific  partnership  among  the
California Institute of Technology,  the University of California, and
NASA;  the observatory  was made  possible by  the generous  financial
support of the W. M. Keck Foundation.  Based on observations collected
at the European Organisation for Astronomical Research in the Southern
Hemisphere, Chile  as part  of PESSTO,  (the Public  ESO Spectroscopic
Survey for Transient Objects Survey) ESO program ID 188.D-3003.



\appendix

\section[]{Sample}
\label{appsample}
\begin{itemize}
\item \textbf{SN~2013bu}: SN~2013bu was  discovered on 2013 Apr. 21.76
  (UT  dates are  used throughout  this paper)  in NGC  7331 at  J2000
  coordinates (here  and elsewhere) $\alpha  = 22^{h}37^{m}05^{s}.60$,
  $\delta  =   +34\degr{}24'  31.9''$  \citep{Itagaki2013}.    It  was
  classified on 2013 Apr. 24.08 as a  young SN~II a few days after the
  explosion \citep{Ochner2013}.  Our follow-up observations started on
  2013  Apr. 23.45  and continued  with a  few-day cadence  until 2013
  Sep.   1.24.   The   SN  was   not   visible  on   2013  Apr.   12.8
  \citep{Itagaki2013}.

\item \textbf{SN~2013fs}: SN~2013fs was discovered on 2013 Oct.  07.46
  \citep{Nakano2013}  at $\alpha  = 23^{h}19^{m}44^{s}.67$,  $\delta =
  +10\degr{}11'04.5''$.  It was classified on 2013 Oct. 9.2 as a young
  SN~II \citep{Childress2013a}.  The classification spectrum exhibited
  narrow lines leading to a  possible classification as an interacting
  Type IIn supernova (SN~IIn).  However, the narrow lines were visible
  only at early phases, placing SN 2013fs in the sample of SNe~II that
  show signs of  circumstellar material around the  progenitor.  A PTF
  detection  of  SN~2013fs  (iPTF13dqy)  on   2013  Oct.  6.24  and  a
  nondetection the day before places  the explosion epoch of SN~2013fs
  to be less than  24 hr before the first detection  (Yaron et al., in
  preparation).   Our follow-up  observations began  on 2013  Oct. 7.27  and
  continued with several-day observing cadence until 2014 Jan.  22.06.
  {\it Swift} observations  were also obtained in  the month following
  discovery.

\item  \textbf{LSQ13dpa}:  LSQ13dpa   was  discovered  at  coordinates
  $\alpha = 11^{h}01^{m}12^{s}.91$,  $\delta = -05\degr{}50'52.4''$ on
  2013 Dec.  18.28 by the LSQ  survey \citep{Hadjiyska2011,Baltay2013}
  and   classified   on   2013   Dec.    20.3   as   a   young   SN~II
  \citep{Hsiao2013}. Our follow-up observations were initiated on 2013
  Dec. 20.25 and continued with a few-day cadence for $\sim 200$ days.
  A nondetection 4 days prior  to discovery confirms that LSQ13dpa was
  discovered within 4 days after its explosion.
 
\item  \textbf{SN~2013ai}:  SN~2013ai  was discovered  at  coordinates
  $\alpha = 06^{h}16^{m}18^{s}.35$,  $\delta = -21\degr{}22'32.9''$ on
  2013 Mar.  1.66 \citep{Conseil2013a}.   The object was classified by
  the   Public  ESO   Spectroscopic  Survey   for  Transient   Objects
  \citep[PESSTO; ][]{Smartt2015} as a young SN~II \citep{Conseil2013}.
  Our follow-up observations  started on 2013 Mar.  5.06 and continued
  until the  SN disappeared in  the Sun's  glare $\sim 83$  days after
  explosion.

\item \textbf{SN~2014cy}:  SN~2014cy was discovered on  2014 Aug. 31.0
  at   coordinates  $\alpha   =   23^{h}44^{m}16^{s}.03$,  $\delta   =
  +10\degr{}46'12.5''$    \citep{Nishimura2014}.     \cite{Morell2014}
  classified   the  object   as   a  SN~II   around   10  days   after
  explosion.  However,  a  nondetection  on   2013  Aug.  29.3  and  a
  prediscovery LOSS detection on 2013 Aug. 31.3 place the explosion on
  Aug. 30.3, with an uncertainty of 1.0 day\footnote{KAIT nondetection
    and   prediscovery  detection   reported  on   the  TOCP   of  the
    CBAT}. LCOGT follow-up observations started  on 2014 Sep. 3.39 and
  ended on 2015 Feb. 9.06.  However, over the last month of monitoring
  the object was only marginally detected in the $i$ band.

\item \textbf{ASASSN-14ha}:  ASASSN-14ha was  discovered by  ASASSN on
  2014  Sep. 10.29  at coordinates  $\alpha =  04^{h}20^{m}01^{s}.41$,
  $\delta = -54\degr{} 56'  17.0''$ \citep{Kiyota2014}.  We classified
  the object  using the FLOYDS spectrograph  as a young SN  II on 2014
  Sep.  13.68 \citep{Arcavi2014}.   LCOGT follow-up  observations were
  initiated on  2014 Sep. 11.77  and continued with a  few-day cadence
  until the end of Mar. 2015.

\item  \textbf{ASASSN-14gm/SN~2014cx}: ASASSN-14gm  was discovered  on
  2014  Sep.  2.47  at coordinates  $\alpha =  00^{h}59^{m}47^{s}.83$,
  $\delta       =      -07\degr{}34'19.3''$       \citep{Holoien2014}.
  \cite{Elias-Rosa2014} classified the object as a young SN~II the day
  after discovery.   At the same  time (2014  Sep. 3.12) we  began our
  LCOGT follow-up observations, which concluded in Mar. 2015.

\item \textbf{ASASSN-14dq}:  ASASSN-14dq was  discovered on  2014 July
  8.48  at coordinates  $\alpha  =  21^{h}57^{m}59^{s}.97$, $\delta =
  +24\degr{}16'08.1''$.  We classified this transient using the FLOYDS
  spectrograph on  2014 July 9.5  \citep{Arcavi2014a} and observed
  it further  with LCOGT facilities  from 2014  July 10.74  to 2014
  Dec. 30.07.
  
\item  \textbf{SN~2014dw}:  SN~2014dw  was discovered  at  coordinates
  $\alpha = 11^{h}10^{m}48^{s}.41$,  $\delta = -37\degr{}27'02.2''$ on
  2014 Nov.  6.589 \citep{Parker2015}.   We classified  it as  a SN~II
  \citep{Arcavi2014b}.  No recent  prediscovery images  are available.
  However, the classification spectrum was quite blue with conspicuous
  \Ha\  lines, consistent  with  an object  around  a fortnight  after
  explosion.  We adopt  JD = 2,456,958 $\pm$ 10 days  as the explosion
  epoch.

\item \textbf{LSQ14gv}: LSQ14gv was  discovered at coordinates $\alpha
  =  10^{h}54^{m}11^{s}.71$, $\delta  =  -15\degr{}01'30.0''$ on  2014
  Jan.  17.3  by the  LSQ survey  \citep{Hadjiyska2011,Baltay2013} and
  classified  by   PESSTO  on   2014  Jan.   23   as  a   young  SN~II
  \citep{Ergon2014a}.  Our  follow-up  observations  started  on  2014
  Jan. 25.60  and continued  for $\sim 3$  months.  A  nondetection on
  2014 Jan.   13.17 places  the explosion  on 2014  Jan. 15.3  $\pm$ 2
  days.

\item \textbf{ASASSN-14kg}: ASASSN-14kg  was discovered at coordinates
  $\alpha = 01^{h}44^{m}38^{s}.38$,  $\delta = +35\degr{}48'20.5''$ on
  2014 Nov. 17.36 \citep{Nicolas2014} and  classified as a young SN~II
  on  2014 Nov.  18.2 \citep{Falco2014}.   Our follow-up  observations
  started on 2014  Nov. 27.27, 10 days after  discovery.  However, the
  early rise  and a nondetection on  2014 Nov. 11.38 suggest  that the
  explosion epoch  was soon after  the nondetection.  We  followed the
  object until 2015 Mar. 7.09 when  the SN disappeared in the glare of
  the Sun.

\item \textbf{SN~2015W}: SN~2015W was discovered  by LOSS on 2015 Jan.
  12.17  at coordinates  $\alpha =  06^{h}57^{m}43^{s}.03$, $\delta  =
  +13\degr{}34'45.7''$ and was reported to the TOCP pages of the CBAT.
  It  was  classified as  an  SN~II  $\sim  10$ days  after  explosion
  \citep{Tomasella2015}.  Our  follow-up observations started  on 2015
  Jan.  12.92  and   continued  for  four  months   until  the  object
  disappeared in the Sun's glare.
\end{itemize}

\section[]{lcogtsnpipe}
\label{sec:pipe}
In  this  section  we  will briefly  describe  the  {\tt  lcogtsnpipe}
pipeline used to ingest and  automatically reduce the photometric data
presented in  this paper and  all the  data collected under  the LCOGT
Supernova Key Project (``The next-generation sample of supernovae"; PI
D. A.  Howell).  It  is a {\tt PYTHON}  module  based on  a few  main Python
packages (numpy,  MySQLdb, PyRAF,  pyfits, matplotlib) and  uses other
astronomical  software  at  different  stages  of  the  reduction.  In
particular, {\tt lcogtsnpipe}  uses {\tt HOTPANTS} \citep{Becker2015},
{\it SEXTRACTOR}  \citep{Bertin1996}, {\tt  SWARP} \citep{Bertin2010},
and                                                               {\it
  astro-scrappy}\footnote{https://github.com/cmccully/astroscrappy}.
The  instrumental   magnitudes  are   computed  using   {\tt  DAOPHOT}
\citep{Stetson1987}  and the  PSF subtraction  technique.  We  use the
IRAF implementation developed by Enrico Cappellaro in the {\tt snoopy}
package\footnote{http://sngroup.oapd.inaf.it/snoopy.html}.   The  {\tt
  lcogtsnpipe} ``\emph{stages}" are described in detail below.

 \begin{itemize}
\item \textbf{Ingestion:}  The LCOGT network is  a world-wide facility
  that  is   able  to  observe   the  full  sky   nearly  continuously
  \citep{Brown2013a}.  At  the end of  the night for each  LCOGT site,
  the raw  data are preprocessed using  the \emph{lcogt-orac pipeline}
  \citep{Brown2013a,Jenness2015}.  The  prereduced data are  then sent
  to  the  IPAC archive,  where  LCOGT  users  can access  them.   The
  ingestion {\it stage} is triggered within  1~hr after the end of the
  night for each  site.  The images are stored locally  on a dedicated
  machine where  they are  processed.  The  key header  information is
  stored in a database to allow  {\tt lcogtsnpipe} to identify the new
  frames by telescope, filter, instrument, object name or coordinates.
\item  \textbf{Astrometry:} The  ORAC LCOGT  prereduction produces  an
  astrometric solution for each frame. However, this fails on a subset
  of images, so {\tt lcogtsnpipe}  runs a custom astrometry routine to
  solve these  frames.  Starting  from an initial  guess based  on the
  telescope  pointing  and  the  image  pixel  scale,  it  computes  a
  cross-match of extracted sources in the frame with available all-sky
  catalogs (2MASS, UCAC).
\item  \textbf{Point-spread   function:}  The   next  stage   of  {\tt
  lcogtsnpipe} is to compute the PSF of each frame. This is done using
  {\tt DAOPHOT}  within PyRAF.  It also  creates a  FITS table  with a
  catalog  of   $5\sigma$  detections  and  their   aperture  and  PSF
  instrumental  magnitudes. This  table will  be used  to compute  the
  absolute calibration in the next stage of the pipeline.
\item  \textbf{Instrumental  PSF  photometry:} This  stage  uses  {\tt
  DAOPHOT} within PyRAF  to fit the SN. At the  same time, a low-order
  polynomial  fit  to  the  background   is  computed  to  remove  the
  host-galaxy  contamination.  This  implementation was  developed  by
  E. Cappellaro for the {\tt snoopy} package.
\item  \textbf{Zeropoint and  colour-term  calibration:} The  pipeline
  cross-matches the FITS table produced in  the PSF stage with a local
  catalog of  magnitudes.  Since we  are mainly observing  in $BVgri$,
  our first choice is  the APASS catalog \citep{Henden2012}.  However,
  {\tt  lcogtsnpipe} is  also able  to calibrate  using SDSS,  Landolt
  fields, or the  natural LCOGT system. Several  options are available
  to compute the zeropoint and colour  terms. The default option is to
  use  fixed  colour   terms  that  have  been   measured  during  the
  commissioning    of     the    LCOGT    telescopes     (see    Table
  \ref{tab:zeropoint}).
\item \textbf{Apparent  magnitudes of  the SN:} Once  the instrumental
  magnitudes and  zeropoints have been  computed for each  night, this
  stage computes  the apparent magnitude  of the SN  using first-order
  colour corrections.  Our apparent magnitudes are  usually calibrated
  to the Landolt photometric system  (Vega magnitudes) or to the Sloan
  Photometric System (AB magnitudes).
\item \textbf{Difference imaging:}  This stage is applied  when the SN
  is in crowded regions where a background estimate is complicated. We
  use  {\tt HOTPANTS}  to  compute the  convolution  kernel after  the
  target  and  the  reference  images   are  aligned  (with  the  {\tt
    gregister}  task in  PyRAF).  The  previous steps  to compute  the
  apparent magnitudes can be applied to the difference images.
\end{itemize}

Several other stages  are available within {\tt  lcogtsnpipe} that can
be run automatically or in interactive mode to improve our photometric
reduction,  reject  bad-quality  data,  and check  each  step  of  the
reduction. These stages  are usually run manually on each  object as a
sanity  check, to  ensure that  the photometric  product is  ready for
publication.   At  each  stage  the database  is  updated.   Once  the
information is stored in the database,  it is available to all members
of the  Key Project  through the Supernova  Exchange (SNEx;  Arcavi et
al., in preparation), usually within 24~hr of observation.

\begin{table*}
\centering
 \begin{minipage}{140mm}
\caption{LCOGT Colour Terms, Zeropoints, and Extinction Coefficients}
\begin{tabular}{ccccccc}
\hline
Filter\footnote{Zeropoint, colour term, and extinction coefficient 
have been measured during photometric nights using the following equation: 
$m_i-m_c = Z + C_i \times {\rm colour} + K \times {\rm airmass}_i$, where 
$m_i$ is the instrumental magnitude for an observed star, 
$m_c$ is the magnitude from the APASS catalog,
$C_i$ is the colour of this star from the APASS catalog, 
and airmass$_i$ is the airmass at which the star was observed.}
& Colour  & Zeropoint & Colour Term  &   Extinction  & Telescope& Instrument  
\footnote{1m0-08 (McDonald  Observatory,  USA); 1m0-10,  1m0-12,
  1m0-13  (Sutherland, South  Africa), 1m0-04,  1m0-05, 1m0-09  (Cerro
  Tololo, Chile);  1m0-03, 1m0-11  (Siding Spring,  Australia).} \\
  & & 2013 - 2014 - 2015 & & & & \\
\hline
  $g$ &  $gr$ &  23.05 - 22.90 - 22.75  &   0.137 (021)  & 0.21 (04)  & 1m-03  &  SBIG   \\
  $r$ &  $ri$   &  22.80 - 22.65 - 22.50  & -0.005 (011)  & 0.10 (04)  & 1m-03  &  SBIG  \\
  $i$ &  $ri$   &  22.10 - 22.00 - 21.90  &   0.007 (014)  & 0.07 (04)  & 1m-03  &  SBIG  \\
  $B$ &  $BV$ &  22.44 - 22.30 - 22.15  & -0.025 (022)  & 0.20 (06)  & 1m-03  & SBIG  \\
  $V$ &  $BV$ &  22.80 - 22.63 - 22.50  &  0.017  (025)  & 0.14 (07)  & 1m-03  & SBIG  \\
  \hline
  $g$ &  $gr$ &  $--$ -   23.60 - 23.50  &  0.109 (014)  &  0.14 (03)  &1m-04  &  Sinistro   \\
  $r$ &  $ri$   &  $--$ -   23.50 - 23.35  &  0.027 (013)  &  0.08 (02) &1m-04  &  Sinistro  \\
  $i$ &  $ri$   &  $--$ -   23.10 - 22.90  &  0.036 (016)  &  0.06 (03)  & 1m-04  &   Sinistro  \\
  $B$ &  $BV$ &   $--$ -  23.00 - 22.85  & -0.024 (011) & 0.23 (03)  & 1m-04  &  Sinistro  \\
  $V$ &  $BV$ &  $--$ -  23.25 - 23.15  &  -0.014 (030) &  0.12 (05)  &1m-04  &  Sinistro  \\
  \hline
  $g$ &  $gr$ &  23.05 - 22.85 - 22.70      &   0.120 (018) &  0.14 (03)  &1m-05  &  SBIG   \\
  $r$ &  $ri$   &  22.80 - 22.65 - 22.50      & -0.002 (018) & 0.08 (02)  & 1m-05  &  SBIG  \\
  $i$ &  $ri$   &  22.15 - 21.00 - 21.90      &   0.019 (022) & 0.06 (03)  & 1m-05  &  SBIG  \\
  $B$ &  $BV$ &  22.45  -  22.35 - 22.15  &  -0.035 (019)  & 0.23 (03)  &1m-05  &  SBIG  \\
  $V$ &  $BV$ &  22.75  -  22.60 - 22.45  &    0.0     (036) &  0.12 (05)  & 1m-05  &  SBIG  \\
    \hline
  $g$ &  $gr$ &  22.95 - 22.75 - 22.60  &   0.114 (018)  &  0.15 (07)  &1m-08  &  SBIG   \\
  $r$ &  $ri$   &  22.80 - 22.60 - 22.40  & -0.004 (017)  &  0.09 (04)  &1m-08  &  SBIG  \\
  $i$ &  $ri$   &  22.05 - 21.95 - 21.70   &  0.024 (023)  &  0.07 (04)  &1m-08  &  SBIG  \\
  $B$ &  $BV$ &  22.25 - 22.10 - 22.00  & -0.039 (022)  &  0.22 (05)  &1m-08  &  SBIG  \\
  $V$ &  $BV$ &  22.65 - 22.45 - 22.30  & -0.005 (035)  & 0.15 (05)  & 1m-08  &  SBIG  \\
  \hline      
   $g$ &  $gr$ &  $--$ - 23.45 - 23.35  & 0.109 (014)   &  0.14 (03)  & 1m-09  &  Sinistro   \\
  $r$ &  $ri$   &  $--$ - 23.40 - 23.30  & 0.027 (013)    &  0.08 (02)  & 1m-09  &  Sinistro  \\
  $i$ &  $ri$   &  $--$ - 23.10 - 23.00  & 0.036 (016)     &  0.06 (03)  & 1m-09  &   Sinistro  \\
  $B$ &  $BV$ &  $--$ - 22.75 - 22.60  & -0.024 (011)  &  0.23 (03) & 1m-09  &  Sinistro  \\
  $V$ &  $BV$ &  $--$ - 23.15 - 23.05  & -0.014 (030)  &  0.12 (05)  & 1m-09  &  Sinistro  \\  
 \hline
   $g$ &  $gr$ &  23.00 - 22.85 - 22.70  &  0.112  (016)  & 0.16 (05)  & 1m-10  &  SBIG   \\
  $r$ &  $ri$   &  22.75 - 22.60 - 22.50  & -0.001  (014) &  0.10 (04)  &1m-10  &  SBIG  \\
  $i$ &  $ri$   &  22.10 - 22.05 - 21.95  &   0.013  (022) &  0.07 (04)  & 1m-10  &  SBIG  \\
  $B$ &  $BV$ &  22.35 - 22.20 - 22.05  & -0.030  (020) &  0.19 (09)  &1m-10  &  SBIG  \\
  $V$ &  $BV$ &  22.60 - 22.50 - 22.35  & -0.019  (025) &  0.14 (06)  &1m-10  &  SBIG  \\
\hline
  $g$ &  $gr$ &  23.05 - 22.90 - 22.75  &   0.137 (021)  & 0.21 (04)  & 1m-11  &  SBIG   \\
  $r$ &  $ri$   &  22.80 - 22.65 - 22.50  & -0.005 (011)  &  0.10 (04)  & 1m-11  &  SBIG  \\
  $i$ &  $ri$   &  22.10 - 22.00 - 21.90  &   0.007 (014)  &  0.07 (04)  & 1m-11  &  SBIG  \\
  $B$ &  $BV$ &  22.44 - 22.30 - 22.15  & -0.025 (022)  &  0.20 (06)  &  1m-11  & SBIG  \\
  $V$ &  $BV$ &  22.80 - 22.63 - 22.50  &   0.017 (025)  &  0.14 (07)  & 1m-11  & SBIG  \\
\hline
  $g$ &  $gr$ &  23.05 - 22.90 - 22.75  &   0.112 (016)  &  0.16 (05)  & 1m-12  &  SBIG   \\
  $r$ &  $ri$   &  22.80 - 22.65 - 22.50  & -0.001 (014)  &  0.10 (04)  &1m-12  &  SBIG  \\
  $i$ &  $ri$   &  22.10 - 22.00 - 21.90  &   0.013 (022)  & 0.07 (04)  & 1m-12  &  SBIG  \\
  $B$ &  $BV$ &  22.44 - 22.30 - 22.15  & -0.030 (020)  & 0.19 (09)  & 1m-12  & SBIG  \\
  $V$ &  $BV$ &  22.80 - 22.63 - 22.50  & -0.019 (025)  & 0.14 (06)  & 1m-12  & SBIG  \\
\hline
  $g$ &  $gr$ &  23.05 - 22.90 - 22.75  &   0.112 (016)  &  0.16 (05)  &1m-13  &  SBIG   \\
  $r$ &  $ri$   &  22.80 - 22.65 - 22.50  & -0.001 (014)  &  0.10 (04)  &1m-13  &  SBIG  \\
  $i$ &  $ri$   &  22.10 - 22.00 - 21.90  &   0.013 (022)  &  0.07 (04)  &1m-13  &  SBIG  \\
  $B$ &  $BV$ &  22.44 - 22.30 - 22.15  & -0.030 (020)  & 0.19 (09)  &  1m-13  & SBIG  \\
  $V$ &  $BV$ &  22.80 - 22.63 - 22.50  & -0.019 (025)  & 0.14 (06)  & 1m-13  & SBIG  \\
  \hline  
\end{tabular}
\label{tab:zeropoint}
\end{minipage}
\end{table*}

\section[]{Nebular Spectra Reduction}
\label{sec:specrredu}
At Keck,  we obtained  one or  two 1200-s  exposures, and  reduced the
spectrum using routines written specifically  for LRIS in the Carnegie
{\sc Python}  ({\sc CarPy})  package. The two-dimensional  (2D) images
were flat-fielded, corrected for distortion along the $y$ (slit) axis,
wavelength  calibrated with  comparison-lamp spectra,  and cleaned  of
cosmic  rays before  extracting the  1D spectrum  of the  target. This
spectrum was flux calibrated using a sensitivity function derived from
a  standard star  obtained the  same  night with  the same  instrument
configuration. The standard-star spectrum was  also used to remove the
telluric sky absorption features.
 
The Gemini spectra were reduced using the custom pipeline developed by
C. McCully
\footnote{https://github.com/cmccully/lcogtgemini}.\\
Details of the spectra can be found in Table \ref{tab:spectra}.

\begin{table*}
\centering
\tiny
 \begin{minipage}{140mm}
\caption{Nebular spectra of SNe type II}
\begin{tabular}{cccccccc}
\hline
       Object              &     Setup                     &  Exposure &     Epoch        & Phase    &     PI           & Reducer   & Observer
       \footnote{M.L.G. = M. L. Graham; A.F. = A. Filippenko; C.Mc. = C. McCully; W.Z. = W. Zheng; S.V. = S. Valenti}   \\
ASASSN-14dq   & Keck+LRIS (uvir $1.0''$)  & $1 \times 1200$~s  & 2015 June 16.47 & 350  & Filippenko  & M.G.  &  M.G, A.F., W.Z.\\
ASASSN-14dq  &  Keck+LRIS (uvir $1.0''$)  & $1 \times 1200$~s  & 2015 Oct. 10.34 & 464  &  Filippenko  & M.G.  &  M.G. \\
ASASSN-14gm  & Keck+LRIS (uvir $1.0''$)  & $1 \times 1200$~s  & 2015 Oct. 10.39 &  405 &  Filippenko  & M.G.  &  M.G. \\
   SN 2015W           & Keck+LRIS (uvir $1.0''$) & $2 \times 1200$~s  & 2015 Oct. 10.60 &  281  &  Filippenko    & M.G.   & M.G.  \\
ASASSN-14gm &Gemini+GMOS (B600+R400 $1.5''$)& $2 \times 900$~s & 2015 July 27.62 & 330  & Howell &  C. Mc.    & Service \\
ASASSN-14dq  &Gemini+GMOS (B600+R400 $1.0''$ )& $2 \times 1900$~s & 2015 Oct. 09   &  463   & Valenti & S.V. & Service \\
   SN 2015W     & Gemini+GMOS (R400 $1.0''$ )& $2 \times 1900$~s & 2015 Dec. 07   &  339   & Valenti & S.V. & Service \\
  \hline  
\end{tabular}
\label{tab:spectra}
\end{minipage}
\end{table*}

\section[]{Figures and Tables}
\label{apefigures}

\begin{figure}
\begin{center}
  \includegraphics[width=8.cm,height=5cm]{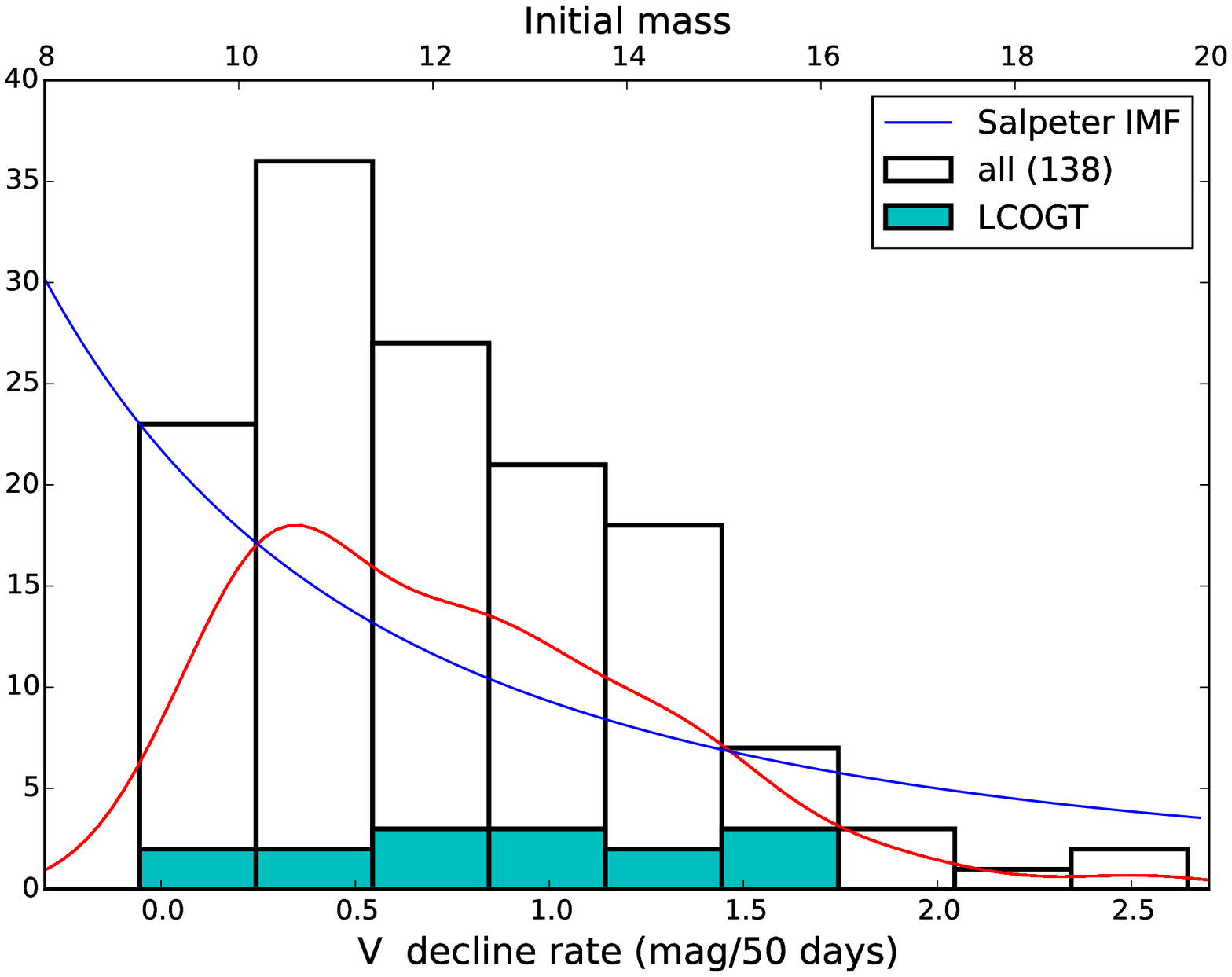}
    \includegraphics[width=8.cm,height=4.5cm]{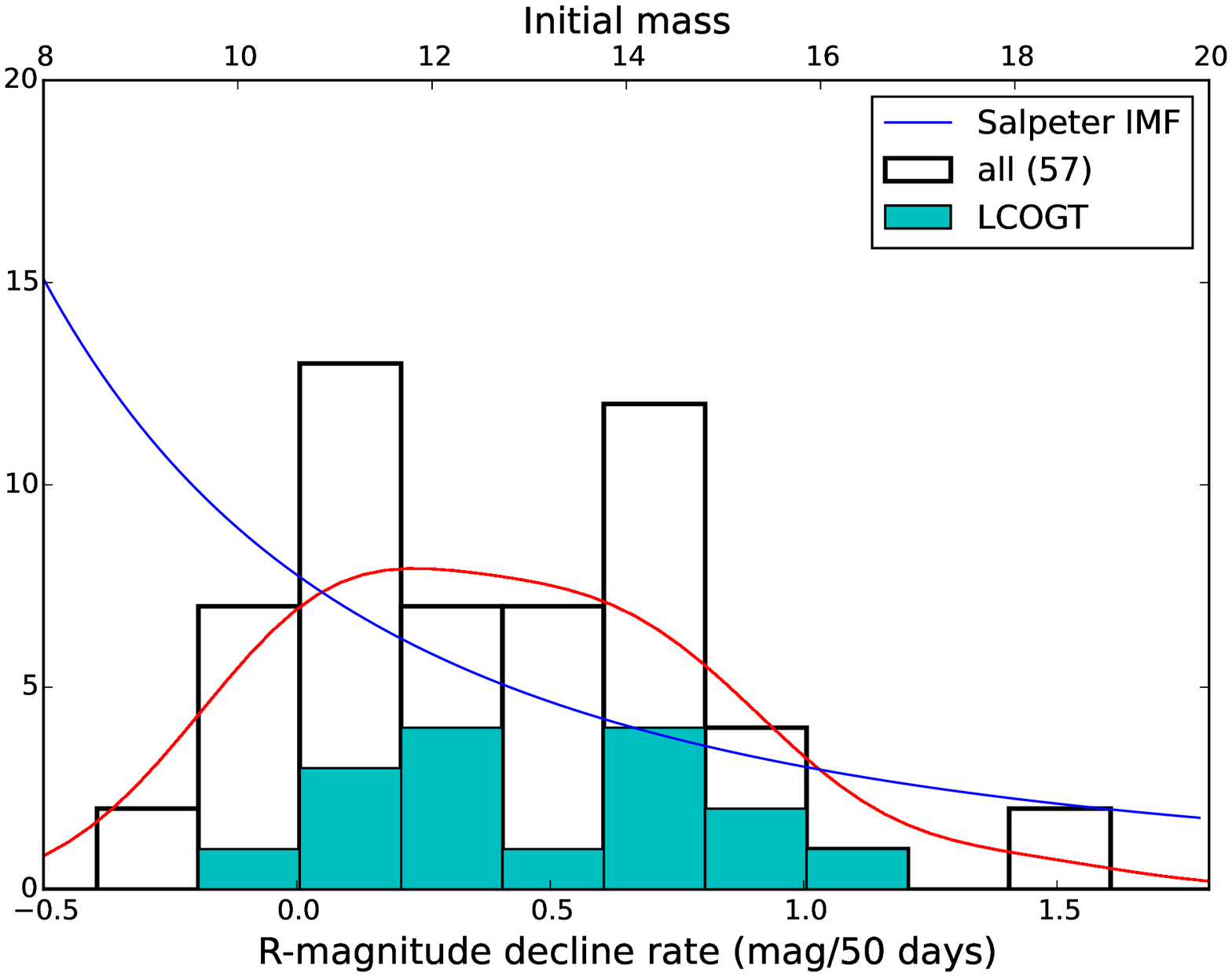}
      \includegraphics[width=8.cm,height=4.5cm]{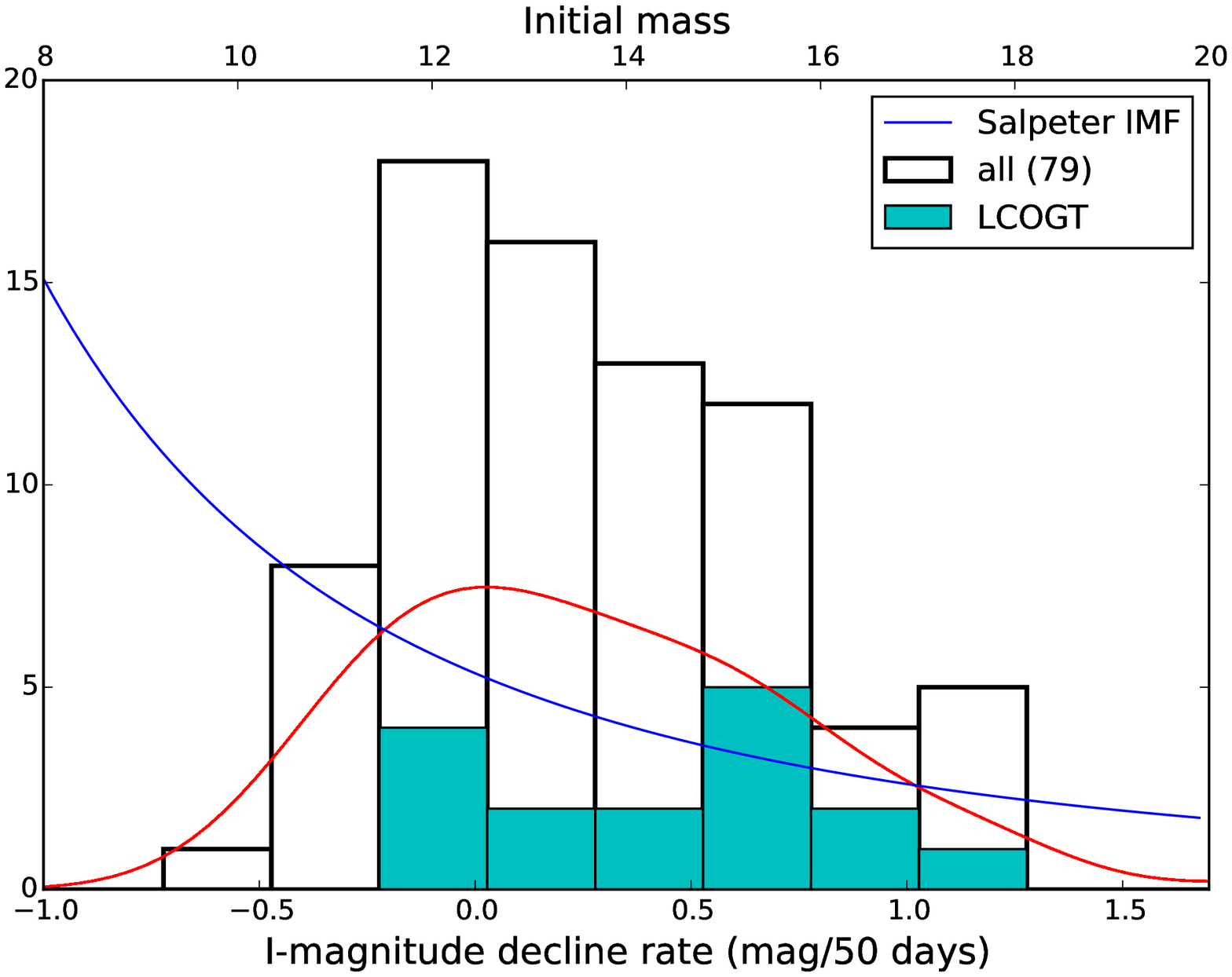}
  \caption{$s_{50}$ distributions for  a sample of SNe~II  in $V$ (top
    panel), $R$  (central panel),  and $I$ (lower  panel). In  blue we
    plot  a  Salpeter initial  mass  function  to emphasize  what  the
    decline-rate   distribution  would   look  like   if  more-massive
    progenitors are related to faster decline  rate SNe. In red is the
    kernel density estimate  using a Gaussian kernel for  the same set
    of data.}
  \label{fig:slope}
\end{center}
\end{figure}

\begin{figure}
\begin{center}
  \includegraphics[width=8.4cm,height=6.cm]{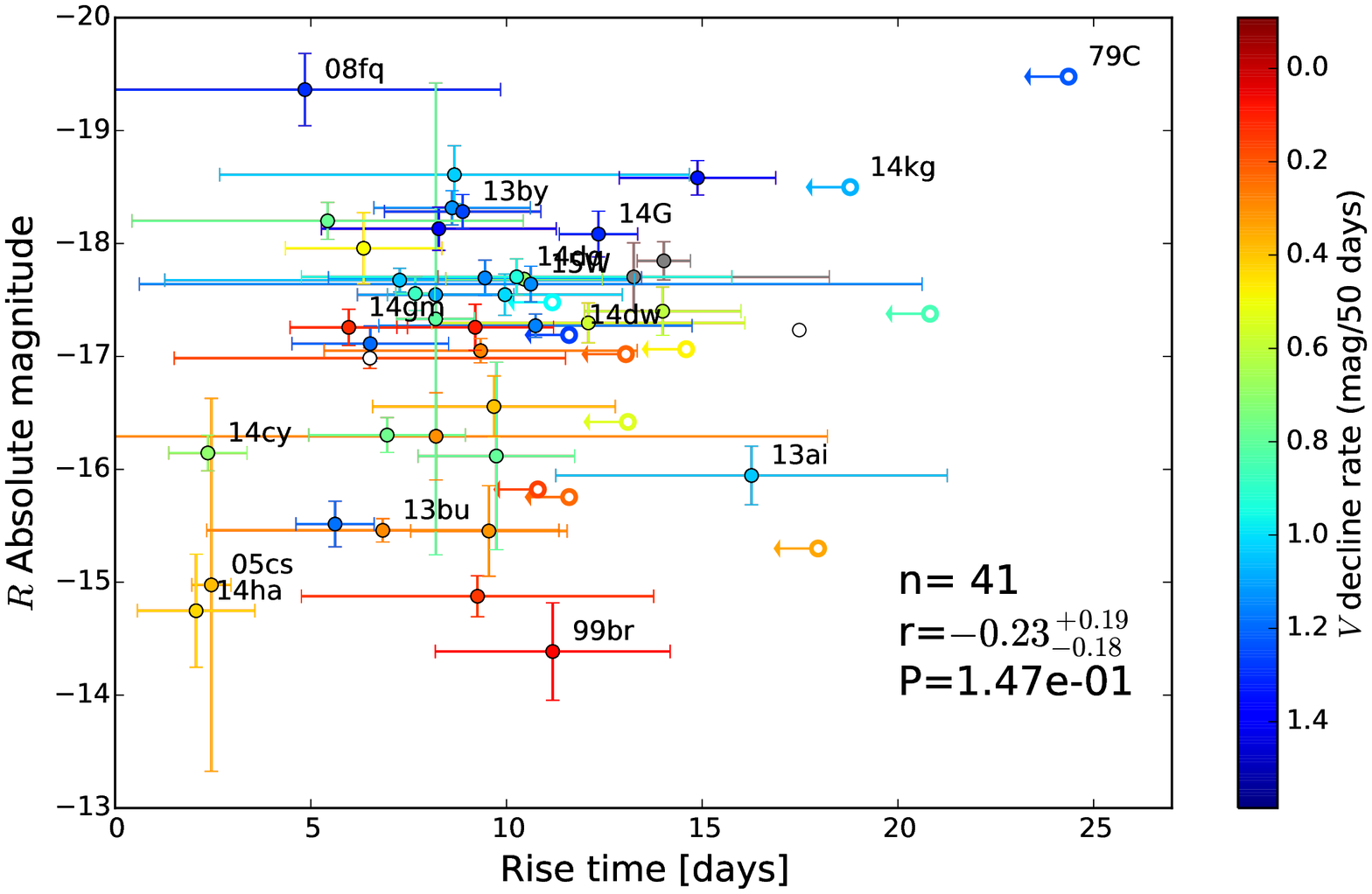}
  \includegraphics[width=8.4cm,height=6.cm]{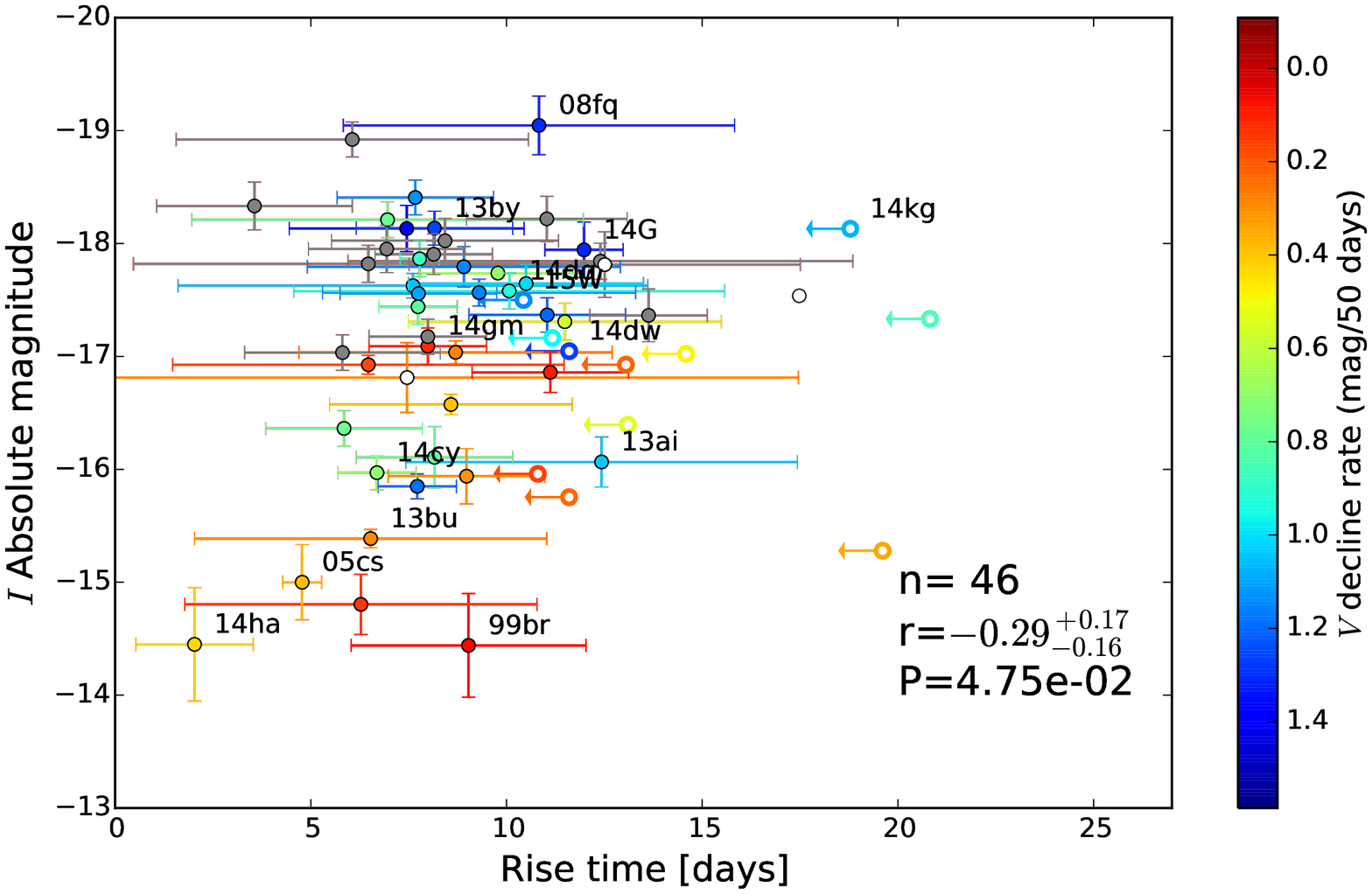}
  \caption{Absolute magnitude as a function  of rise time for a sample
    of  SNe~II.  Objects are  colour-coded with  the  $V$-band  decline rate
    $s_{50V}$. \textbf{Top panel:} $R$ band.  \textbf{Bottom panel:}
    $I$  band. Data  from \protect\cite{Poznanski2015}  have also been
    added in the $I$ band.}
  \label{fig:risetime2}
\end{center}
\end{figure}

\begin{figure}
\begin{center}
  \includegraphics[width=8.cm,height=5cm]{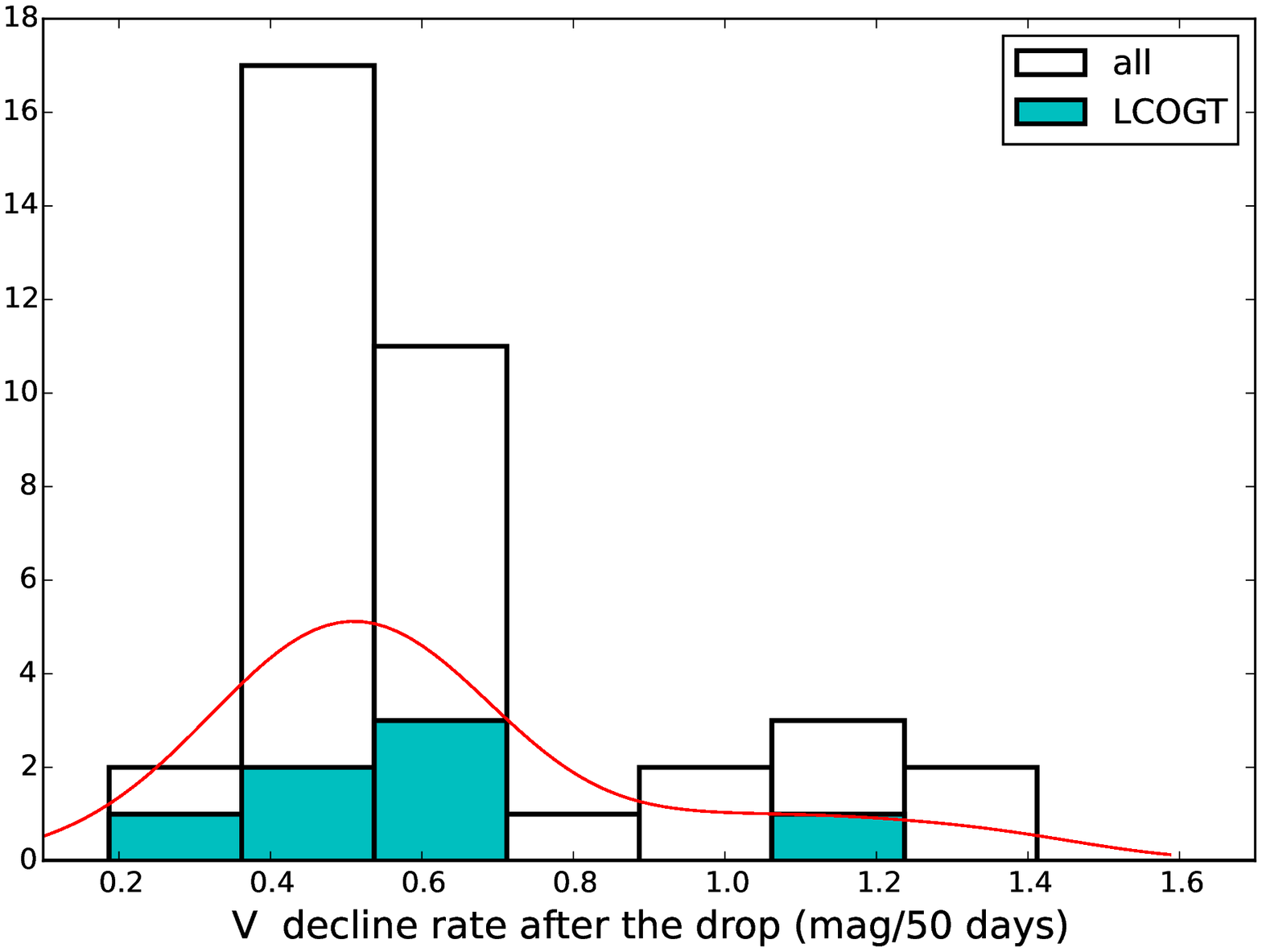}
    \includegraphics[width=8.cm,height=5cm]{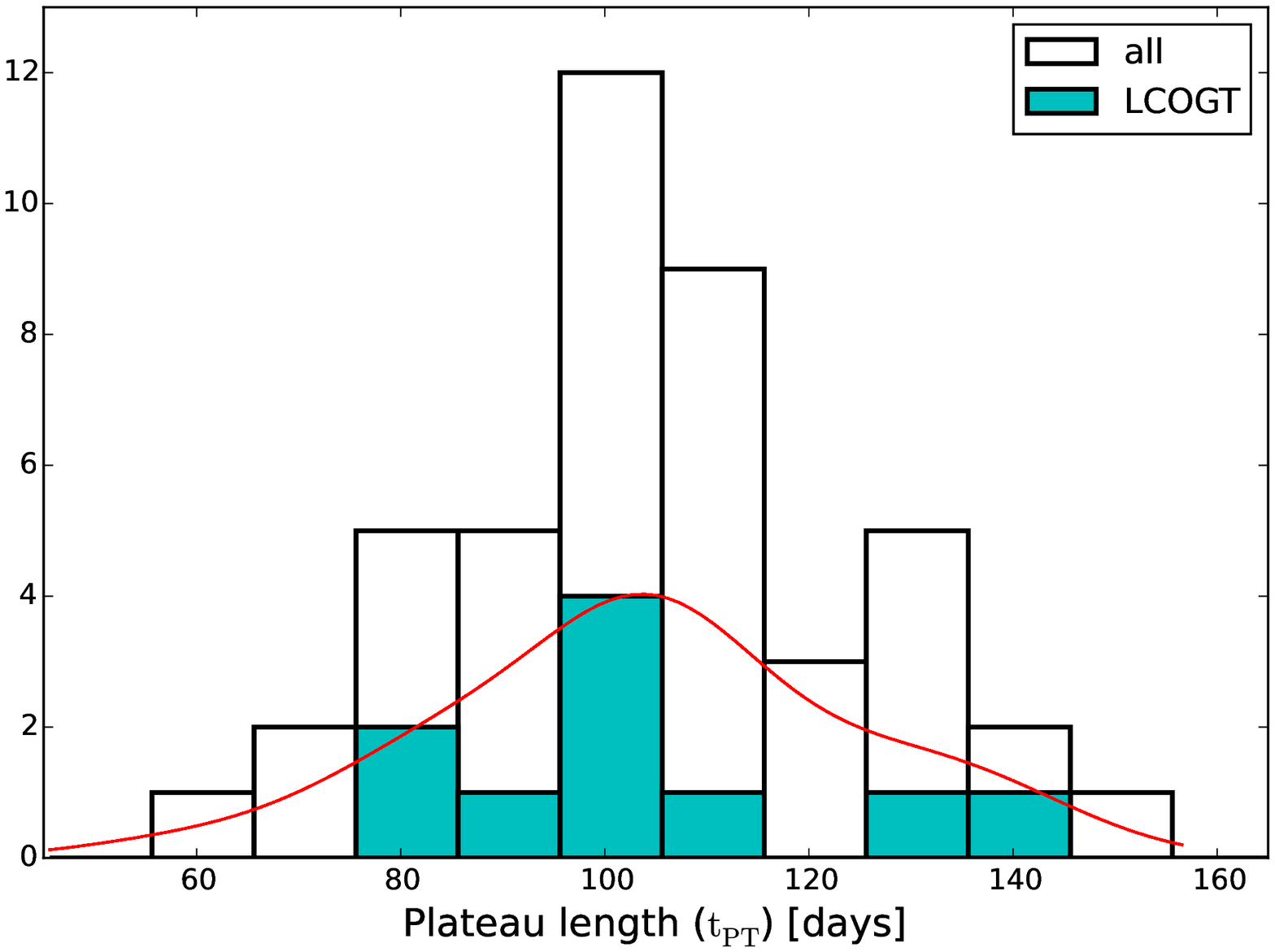}
  \caption{ \textbf{Top panel:} Distribution of light-curve decline 
  rates $s_{50V}$ after the drop for a sample of SNe~II in $V$.
  \textbf{Bottom panel:} 
  Distribution of plateau length $t_{\rm PT}$ for a sample of  
  SNe~II in $V$.  In red is the kernel density estimate using a 
  Gaussian kernel for  the same set of data.
  }
  \label{fig:lateslope}
\end{center}
\end{figure}

\begin{figure}
\begin{center}
  \includegraphics[width=8.cm,height=5cm]{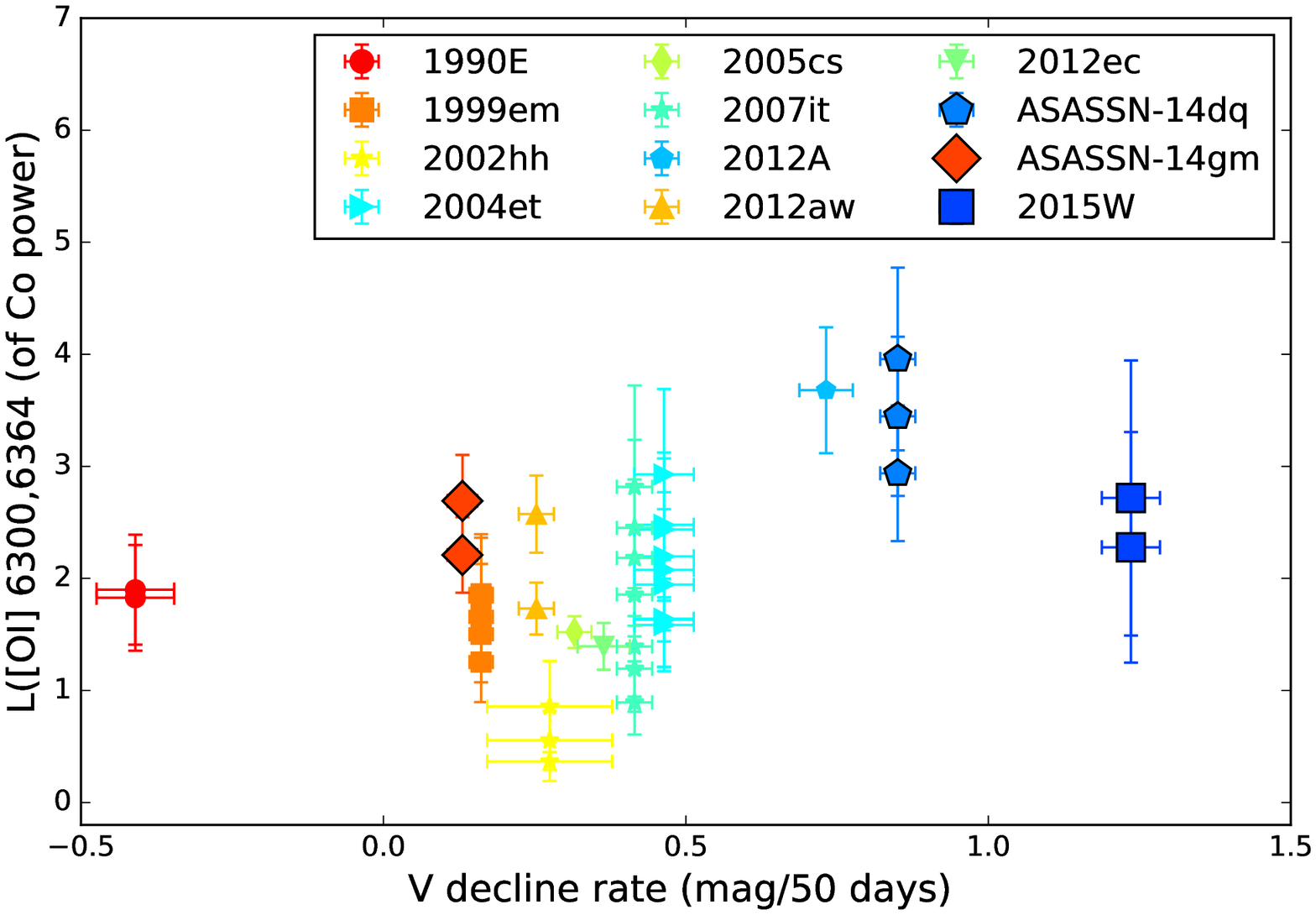}
  \caption{ The [\OI] $\lambda\lambda$6300, 6364 luminosity (normalized to the
    $^{56}$Co decay power) for the \protect\cite{Jerkstrand2015} sample of SNe 
    and three objects from our sample as a function of V band decline rate.
    The [\OI] $\lambda\lambda$6300, 6364 luminosities for all epochs between 
    200 and 500 days after explosion are  plotted and are visible as multiple-points for each object.}
  \label{fig:nebular2}
\end{center}
\end{figure}

\begin{table*}
\centering
\tiny
 \begin{minipage}{140mm}
\caption{Photometric Data (Complete table available in the online version of the paper)}
\begin{tabular}{ccccc|ccccc}
\hline
Date & JD & mag$^{a}$ &  Filter & telescope$^b$ &Date & JD & mag &  Filter & telescope
\footnote{(a) Data have  not been corrected for  extinction; (b) UVOT
  ({\it Swift} Telescope); CSP  (Cerro Tololo, Chile); 1m0-08 (McDonald
  Observatory,  USA);   1m0-10,  1m0-12,  1m0-13   (Sutherland,  South
  Africa), 1m0-04, 1m0-05, 1m0-09 (Cerro Tololo, Chile); 1m0-03, 1m0-11
  (Siding  Spring, Australia).}  \\ 
\hline xx & 2456407.372 & 12.763 0.069  & UW2 & UVOT & x & 2456407.374
& 13.711 0.043 & V & UVOT \\
 \hline
\end{tabular}
\label{tab2}
\end{minipage}
\end{table*}

\newpage 
\begin{landscape}
\begin{table}
\begin{flushleft}
\textbf{Table D2:} Slope Data  \
\end{flushleft}
\centering
\tiny
\begin{tabular}{ccccccccccccccccccc}
\hline 
SN  & $M(V)$ $^{*}$ $\Delta M$ & $s50(V)$ & $\Delta$ s50(V)  &  $ph\_start$ & $ph\_stop$ & $M(R)$ &$\Delta M$ & $s50(R)$ & $\Delta s50(R)$  &  $ph\_start$ & $ph\_stop$ & $M(I)$ &$\Delta M$ &  $s50(I)$ & $\Delta s50(I)$  &  $ph\_start$ & $ph\_stop$  \\
    &        &    $(V)$  &          &                 &  $(V)$        & $(V)$      &        &    $(R)$  &          &                  &  $(R)$       & $(R)$      &        &  $(I)$   &           &      &  $(I)$ & $(I)$  \\
\hline  
2013ai & 16.80 &  0.02 & 0.0209 &  0.0006 &  17.6 &  54.5 & 16.59 &  0.02 & 0.0129 &  0.0004 &  17.6 &  64.5 & 16.29 &  0.05 &  0.0112 &  0.0009 &  19.6 &  63.5 \\ 
2013bu & 16.33 &  0.04 & 0.0053 &  0.0013 &  6.2 &  48.1 & 15.61 &  0.02 & 0.0014 &  0.0005 &  6.2 &  57.1 & 15.63 &  0.01 &  -0.0027 &  0.0004 &  6.2 &  55.1 \\ 
2013fs & 16.03 &  0.04 & 0.0158 &  0.0014 &  4.5 &  58.4 & 15.98 &  0.03 & 0.0075 &  0.0009 &  14.5 &  60.4 & 15.80 &  0.03 &  0.0065 &  0.0008 &  14.5 &  69.6 \\ 
--         &   --   &  --    & --         &  -- &  -- & --  & -- & --  & -- & --& --  & -- & -- & -- & -- & -- & --  & -- \\ 
\hline
\end{tabular}
\label{tabd2}                    
\\
$^{*}$ the slope is computed with the following equation: mag  = M + $s50$ $\times$ t  
\end{table}

\begin{table}
\begin{flushleft}
\textbf{Table D3:} Bolometric light curve parameters  \
\end{flushleft}
\centering
\small
\begin{tabular}{ccccccccccccccc}
\hline 
SN  & $Log10(Lum1(0))$ &  $S1$   & $\Delta S1$ & $ph\_start$ & $ph\_stop$ &  $Log10(Lum2(0))$ &  $S2$ &$\Delta S2$ & $ph\_start$ & $ph\_stop$ &    Ni    & $\Delta$ $Ni$  \\
      &                       &             &                    &   [days]        &     [days]       &                      &           &                  &    [days]                &    [days]      &    \msun         &     \msun     \\ 
\hline  
2013ai & 41.62 &  -0.0081 &  0.0001 &  18.6 &  49.5 & 41.47 & -0.0057 &  0.0005 &  54.5 &  92.2 &    --  &    --    \\ 
2013bu & 41.24 &  -0.0044 &  0.0003 &  6.2 &  35.1 & 41.18 & -0.0020 &  0.0003 &  35.1 &  78.1 & 0.0021 &  0.0007 \\ 
2013fs & 42.47 &  -0.0146 &  0.0009 &  4.5 &  27.6 & 42.19 & -0.0054 &  0.0002 &  31.5 &  75.5 & 0.0545 &  0.0003 \\ 
--         &   --   &  --          & --          &  --  &  --    &  --     & --           & --           & --   & --     &     --     &    --  \\ 
\hline
\end{tabular}
\label{tabD3}                    
\\
$^{*}$ the slope is computed with the following equation: Log10(Lum(t))  = Log10(Lum(0)) + S1 $\times$ t  
\end{table}

\begin{table}
\begin{flushleft}
\textbf{Table D4:} Bolometric MCMC light curve parameters  \
\end{flushleft}
\centering
\small
\begin{tabular}{ccccccccccccccc}
\hline 
SN  &  $T_{pt}$ & $\Delta T_{pt}$   & $A0$ & $A0+$ & $A0-$ & $W0$ & $W0+$ & $W0-$  & $M0$ & $M0+$ & $M0-$  & P0   \\
      &         &           &        &      &        &      &       &        &      &       &        &   \\ 
\hline  
2013bu &   102.8  & 4.5 & -1.14330  &  0.00008 &  0.00011 &  3.85663 & 0.00011 & 0.00011 & 39.73643 & 0.00011 & 0.00010 & -0.0038 \\ 
2013fs &   86.2  & 0.5 & -0.60601  &  0.00010 &  0.00009 &  4.85452 & 0.00010 & 0.00010 & 41.17148 & 0.00010 & 0.00009 & -0.0038 \\ 
LSQ13dpa &   129.9  & 2.0 & -0.52529  &  0.00010 &  0.00009 &  4.60926 & 0.00010 & 0.00010 & 41.16555 & 0.00007 & 0.00010 & -0.0038 \\ 
--           &   --        &  --    & --           &          -- &            -- &          --  &           -- &     --       &     --        &     --       &    --      &   -- \\ 
\hline
\end{tabular}
\label{tabD4}                    
\\
$^{*}$   
\end{table}

\begin{table}
\begin{flushleft}
\textbf{Table D5:} MCMC parameters on V band light curves \
\end{flushleft}
\centering
\small
\begin{tabular}{ccccccccccccccc}
\hline 
SN  &  $t_{pt}$ & $\Delta t_{pt}$   & $a0$ & $a0+$ & $a0-$ & $w0$ & $w0+$ & $w0-$  & $m0$ & $m0+$ & $m0-$  & p0   \\
      &         &           &        &      &        &      &       &        &      &       &        &   \\ 
\hline  
2013bu &   103.1  & 4.5 & 3.08170  &  0.15800 &  0.14293 &  3.51516 & 0.60600 & 0.60607 & 20.27520 & 0.14960 & 0.13669 & 0.0120 \\ 
2013fs &   82.7  & 0.5 & 1.59863  &  0.09900 &  0.07214 &  2.27769 & 2.47320 & 2.47329 & 18.76480 & 0.13160 & 0.08128 & 0.0120 \\ 
2008M &   85.7  & 9.0 & 2.05115  &  0.03200 &  0.03193 &  2.67883 & 0.39460 & 0.39467 & 18.95140 & 0.02770 & 0.02715 & 0.0116 \\ 
--          &   --   &  --    & --         &  --          &    --         & --           & --           & --           & --           & --          & --           &   --  \\ 
\hline
\end{tabular}
\label{tabD5}                    
\\
$^{*}$   
\end{table}
\end{landscape}

\end{document}